\documentclass[longauth]{aa}
\usepackage{graphicx}
\usepackage{gensymb}
\usepackage{txfonts}
\usepackage[flushleft]{threeparttable}
\renewcommand{\arraystretch}{1.1}
\usepackage{hyperref}
\hypersetup{hidelinks,colorlinks=true,linkcolor=blue,citecolor=blue}
\usepackage{longtable,tabu}
\usepackage{lipsum}
\usepackage{adjustbox}
\usepackage{stfloats}
\usepackage{scrextend}
\usepackage{subcaption}
\usepackage{amssymb}
\usepackage{caption}
\usepackage{amsmath}
\usepackage{amssymb}
\usepackage{mathrsfs}
\usepackage{multicol}
\usepackage{supertabular}
\usepackage{tabularx}
\usepackage{float}
\usepackage{footmisc}
\usepackage{hyperref}

\newcommand{\feh}{\mbox{[Fe/H]}\xspace}
\newcommand{\sih}{\mbox{[Si/H]}\xspace}
\newcommand{\mgh}{\mbox{[Mg/H]}\xspace}
\newcommand{\teff}{\ensuremath{T_{\rm eff}}}

\newcommand{\kms}{\mbox{km\,s$^{-1}$}\xspace}
\newcommand{\ms}{\mbox{m\,s$^{-1}$}\xspace}
\newcommand{\gccc}{\mbox{g\,cm$^{-3}$}\xspace}

\newcommand{\me}{\mbox{$\it{M_{\rm \mathrm{\oplus}}}$}\xspace}
\newcommand{\re}{\mbox{$\it{R_{\rm \mathrm{\oplus}}}$}\xspace}

\newcommand{\rstar}{\mbox{$R_{*}$}\xspace}

\newcommand{\msol}{\mbox{$\it{M_\mathrm{\odot}}$}\xspace}
\newcommand{\rsol}{\mbox{$\it{R_\mathrm{\odot}}$}\xspace}
\newcommand{\se}{\mbox{$\it{S_{\rm \mathrm{\oplus}}}$}\xspace}

\newcommand{\denssol}{\mbox{$\it{\rho_\mathrm{\odot}}$}\xspace}

\newcommand{\MonoTools}{{\sc \tt Monotools}\xspace}
\newcommand{\juliet}{{\sc \tt juliet}\xspace}
\newcommand{\pycheops}{{\sc \tt pycheops}\xspace}
\newcommand{\Juliet}{{\sc \tt Juliet}\xspace}

\newcommand{\celerite}{{\sc \tt celerite}\xspace}

\newcommand{\ldcu}{{\sc \tt LDCU}\xspace}

\AtBeginDocument{}
\usepackage{hyperref}
\usepackage{orcidlink}

\begin{document} 

\title{%
  Discovery of two warm mini-Neptunes with contrasting densities orbiting the young K3V star TOI-815\protect\thanks{This study uses TESS data, CHEOPS data observed as part of guaranteed time observation programs CH\_PR10031 and CH\_PR10048, and ESPRESSO data collected with the ESO 3.6 m telescope under programs 105.20P7.001, 109.23DX.001, and 110.2481.001 (PI: Bouchy). The photometric and radial velocity data in this work are only available at the CDS via anonymous ftp to cdsarc.u-strasbg.fr (130.79.128.5) or via \url{http://cdsarc.u-strasbg.fr/viz-bin/qcat?J/A+A/}.}}
\authorrunning{A. Psaridi, et al.}
    \author{Angelica Psaridi\orcidlink{0000-0002-4797-2419},
    \inst{\ref{inst-geneva}}
    Hugh Osborn\orcidlink{0000-0002-4047-4724}, \inst{\ref{inst-bern},\ref{inst-kavli}}
    Fran\c{c}ois Bouchy\orcidlink{0000-0002-7613-393X}, \inst{\ref{inst-geneva}}
    Monika Lendl\orcidlink{0000-0001-9699-1459}, \inst{\ref{inst-geneva}}
    L\'{e}na Parc\orcidlink{0000-0002-7382-1913}, \inst{\ref{inst-geneva}}
    Nicolas Billot\orcidlink{0000-0003-3429-3836}, \inst{\ref{inst-geneva}}
    Christopher Broeg\orcidlink{0000-0001-5132-2614}, \inst{\ref{inst-bern},\ref{inst-bernPlanetologie}}
    Sérgio G. Sousa\orcidlink{0000-0001-9047-2965}, \inst{\ref{inst-caup}}
    Vardan Adibekyan\orcidlink{0000-0002-0601-6199},\inst{\ref{inst-caup},\ref{inst-porto2}}
    Omar Attia\orcidlink{0000-0002-7971-7439}, \inst{\ref{inst-geneva}}
    Andrea Bonfanti\orcidlink{0000-0002-1916-5935}, \inst{\ref{inst-graz}}
    Hritam Chakraborty\orcidlink{0000-0002-5177-1898}, \inst{\ref{inst-geneva}}
    Karen A. Collins\orcidlink{0000-0003-2781-3207}, \inst{\ref{inst-Smithsonian}}
    Jeanne Davoult\orcidlink{0000-0002-6177-2085}, \inst{\ref{inst-bern}}
    Elisa Delgado-Mena\orcidlink{0000-0003-4434-2195}, \inst{\ref{inst-caup}}
    Nolan Grieves\orcidlink{0000-0001-8105-0373}, \inst{\ref{inst-geneva}}
    Tristan Guillot\orcidlink{0000-0002-7188-8428}, \inst{\ref{inst-cotedazur}}
    Alexis Heitzmann\orcidlink{0000-0002-8091-7526}, \inst{\ref{inst-geneva}}
    Ravit Helled\orcidlink{0000-0001-5555-2652}, \inst{\ref{inst-zurich}}
    Coel Hellier\orcidlink{0000-0002-3439-1439}, \inst{\ref{inst-keele}}
    Jon~M.~Jenkins\orcidlink{0000-0002-4715-9460}, \inst{\ref{inst-NASAAmes}}
    Henrik Knierim, \inst{\ref{inst-zurich}}
    Andreas Krenn\orcidlink{0000-0003-3615-4725}, \inst{\ref{inst-graz}}
    Jack J.~Lissauer\orcidlink{0000-0001-6513-1659}, \inst{\ref{inst-NASAAmes}}
    Rafael Luque\orcidlink{0000-0002-4671-2957}, \inst{\ref{inst-Chicago}}
    David Rapetti\orcidlink{0000-0003-2196-6675}, \inst{\ref{inst-NASAAmes},\ref{inst-Washington}}
    Nuno C.~Santos\orcidlink{0000-0003-4422-2919}, \inst{\ref{inst-caup},\ref{inst-porto2}}
    Olga Su\'{a}rez\orcidlink{0000-0002-3503-3617},\inst{\ref{inst-cotedazur}}
    Julia Venturini\orcidlink{0000-0001-9527-2903}, \inst{\ref{inst-geneva}}
    Francis P. Wilkin\orcidlink{0000-0003-2127-8952}, \inst{\ref{inst-Union}}
    Thomas G.~Wilson\orcidlink{0000-0001-8749-1962}, \inst{\ref{inst-Warwick}}
    Joshua N. Winn\orcidlink{0000-0002-4265-047X},\inst{\ref{inst-Princeton}}
    Carl Ziegler\orcidlink{0000-0002-0619-7639}, \inst{\ref{inst-Austin}}
    Tiziano Zingales    \orcidlink{0000-0001-6880-5356}, \inst{\ref{inst-padova},\ref{inst-inaf}}
    Yann Alibert\orcidlink{0000-0002-4644-8818}, \inst{\ref{inst-bern},\ref{inst-bernPlanetologie}}
    Alexis Brandeker\orcidlink{0000-0002-7201-7536}, \inst{\ref{inst-AlbaNova}}
    Jo Ann Egger\orcidlink{0000-0003-1628-4231}, \inst{\ref{inst-bernPlanetologie}}
    Davide Gandolfi\orcidlink{0000-0001-8627-9628}, \inst{\ref{inst-torino}}
    Matthew J. Hooton\orcidlink{0000-0003-0030-332X}, \inst{\ref{inst-Cavendish}}
    Amy Tuson\orcidlink{0000-0002-2830-9064}, \inst{\ref{inst-Cambridge}}
    Solène Ulmer-Moll\orcidlink{0000-0003-2417-7006}, \inst{\ref{inst-geneva},\ref{inst-bernPlanetologie}}
    Lyu Abe, \inst{\ref{inst-cotedazur}}
    Romain Allart\orcidlink{0000-0002-1199-9759}, \inst{\ref{inst-montreal}}
    Roi Alonso\orcidlink{0000-0001-8462-8126},\inst{\ref{inst-canarias},\ref{inst-laguna}}
    David R.~Anderson\orcidlink{0000-0001-7416-7522}, \inst{\ref{inst-Warwick}}
    Guillem Anglada\orcidlink{0000-0002-3645-5977}, \inst{\ref{inst-CSIC},\ref{inst-IEEC}}
    Tamas Bárczy\orcidlink{0000-0002-7822-4413}, \inst{\ref{inst-Hungary}}
    David Barrado\orcidlink{0000-0002-5971-9242}, \inst{\ref{inst-cabmadrid}}
    Susana C. C.~Barros\orcidlink{0000-0003-2434-3625}, \inst{\ref{inst-caup},\ref{inst-porto2}}
    Wolfgang Baumjohann\orcidlink{0000-0001-6271-0110}, \inst{\ref{inst-graz}}
    Mathias Beck\orcidlink{0000-0003-3926-0275}, \inst{\ref{inst-geneva}}
    Thomas Beck, \inst{\ref{inst-bernPlanetologie}}
    Willy Benz\orcidlink{0000-0001-7896-6479}, \inst{\ref{inst-bernPlanetologie},\ref{inst-bern}}
    Xavier Bonfils\orcidlink{0000-0001-9003-8894}, \inst{\ref{inst-grenoble}}
    Luca Borsato\orcidlink{0000-0003-0066-9268}, \inst{\ref{inst-inaf}}
    Vincent Bourrier\orcidlink{0000-0002-9148-034X}, \inst{\ref{inst-geneva}}
    David~ R.~Ciardi\orcidlink{0000-0002-5741-3047}, \inst{\ref{inst-IPAC}}
    Andrew Collier Cameron\orcidlink{0000-0002-8863-7828}, \inst{\ref{inst-SUPA}}
    S\'{e}bastien Charnoz\orcidlink{0000-0002-7442-491X}, \inst{\ref{inst-paris}}
    Marion Cointepas, \inst{\ref{inst-geneva},\ref{inst-grenoble}}
    Szil\'ard Csizmadia\orcidlink{0000-0001-6803-9698}, \inst{\ref{inst-DLR}}
    Patricio Cubillos\orcidlink{0000-0002-1347-2600},  \inst{\ref{inst-inaf},\ref{inst-graz}}
    Gaspare Lo Curto\orcidlink{0000-0002-1158-9354}, \inst{\ref{inst-ESOchile}}
    Melvyn B. Davies\orcidlink{0000-0001-6080-1190}, \inst{\ref{inst-lund}}
    Tansu Daylan\orcidlink{0000-0002-6939-9211}, \inst{\ref{inst-WashU}}
    Magali Deleuil\orcidlink{0000-0001-6036-0225}, \inst{\ref{inst-Marseille}}
    Adrien Deline, \inst{\ref{inst-geneva}}
    Laetitia Delrez\orcidlink{0000-0001-6108-4808}, \inst{\ref{inst-liege},\ref{inst-star}}
    Olivier D. S. Demangeon\orcidlink{0000-0001-7918-0355}, \inst{\ref{inst-caup}}
    Brice-Olivier Demory\orcidlink{0000-0002-9355-5165}, \inst{\ref{inst-bern},\ref{inst-bernPlanetologie}}
    Caroline Dorn\orcidlink{0000-0001-6110-4610}, \inst{\ref{inst-ETH}}
    Xavier Dumusque\orcidlink{0000-0002-9332-2011}, \inst{\ref{inst-geneva}}
    David Ehrenreich\orcidlink{0000-0001-9704-5405}, \inst{\ref{inst-geneva}}
    Anders Erikson, \inst{\ref{inst-DLR}}
    Alain Lecavelier des Etangs\orcidlink{0000-0002-5637-5253}, \inst{\ref{inst-pierre}}
    Elena Diana de Miguel Ferreras, \inst{\ref{inst-sauc}}
    Andrea Fortier\orcidlink{0000-0001-8450-3374}, \inst{\ref{inst-bernPlanetologie},\ref{inst-bern}}
    Luca Fossati\orcidlink{0000-0003-4426-9530}, \inst{\ref{inst-graz}}
    Yolanda G.~C.~Frensch\orcidlink{0000-0003-4009-0330}, \inst{\ref{inst-geneva}}
    Malcolm Fridlund\orcidlink{0000-0002-0855-8426},\inst{\ref{inst-leiden},\ref{inst-onsala}}
    Michaël Gillon\orcidlink{0000-0003-1462-7739}, \inst{\ref{inst-liege}}
    Manuel Güdel, \inst{\ref{inst-Vienna}}
    Maximilian N. Günther\orcidlink{0000-0002-3164-9086}, \inst{\ref{inst-estec}}
    Janis Hagelberg\orcidlink{0000-0002-1096-1433}, \inst{\ref{inst-geneva}}
    Christiane Helling\orcidlink{0000-0002-8275-1371}, \inst{\ref{inst-graz},\ref{inst-Petersgasse}}
    Sergio Hoyer\orcidlink{0000-0003-3477-2466}, \inst{\ref{inst-Marseille}}
    Kate G. Isaak\orcidlink{0000-0001-8585-1717}, \inst{\ref{inst-estec}}
    Laszlo L. Kiss, \inst{\ref{inst-Konkoly},\ref{inst-elte}}
    Kristine Lam\orcidlink{0000-0002-9910-6088}, \inst{\ref{inst-DLR}}
    Jacques Laskar\orcidlink{0000-0003-2634-789X}, \inst{\ref{inst-sorbonne}}
    Baptiste Lavie\orcidlink{0000-0001-8884-9276}, \inst{\ref{inst-geneva}}
    Christophe Lovis\orcidlink{0000-0001-7120-5837}, \inst{\ref{inst-geneva}}
    Demetrio Magrin\orcidlink{0000-0003-0312-313X}, \inst{\ref{inst-inaf}}
    Luca Marafatto\orcidlink{0000-0002-8822-6834}, \inst{\ref{inst-inaf}}
    Pierre Maxted\orcidlink{0000-0003-3794-1317}, \inst{\ref{inst-keele}}
    Scott McDermott, \inst{\ref{inst-protologic}}
    Djamel Mékarnia\orcidlink{0000-0001-5000-7292}, \inst{\ref{inst-cotedazur}}
    Christoph Mordasini\orcidlink{0000-0002-1013-2811}, \inst{\ref{inst-bernPlanetologie},\ref{inst-bern}}
    Felipe Murgas\orcidlink{0000-0001-9087-1245}, \inst{\ref{inst-laguna}}
    Valerio Nascimbeni\orcidlink{0000-0001-9770-1214}, \inst{\ref{inst-inaf}}
    Louise D.~Nielsen\orcidlink{0000-0002-5254-2499}, \inst{\ref{inst-ESOgarching}}
    Göran Olofsson\orcidlink{0000-0003-3747-7120}, \inst{\ref{inst-AlbaNova}}
    Roland Ottensamer\orcidlink{0000-0001-5684-5836}, \inst{\ref{inst-Vienna}}
    Isabella Pagano     \orcidlink{0000-0001-9573-4928}, \inst{\ref{inst-INAFcatania}}
    Enric Pallé        \orcidlink{0000-0003-0987-1593}, \inst{\ref{inst-canarias},\ref{inst-laguna}}
    Gisbert Peter\orcidlink{0000-0001-6101-2513}, \inst{\ref{inst-DLR}}
    Giampaolo Piotto\orcidlink{0000-0002-9937-6387},  \inst{\ref{inst-inaf},\ref{inst-padova}}
    Don Pollacco\orcidlink{0000-0001-9850-9697}, \inst{\ref{inst-Warwick}}
    Didier Queloz\orcidlink{0000-0002-3012-0316}, \inst{\ref{inst-ETH},\ref{inst-Cavendish}}
    Roberto Ragazzoni\orcidlink{0000-0002-7697-5555}, \inst{\ref{inst-inaf},\ref{inst-padova}}
    Devin Ramos\orcidlink{0009-0002-2478-4308}, \inst{\ref{inst-Union}}
    Nicola Rando, \inst{\ref{inst-estec}}
    Heike Rauer\orcidlink{0000-0002-6510-1828},  \inst{\ref{inst-DLR},\ref{inst-tuberlin},\ref{inst-freie}}
    Christian Reimers\orcidlink{0000-0002-2334-1620}, \inst{\ref{inst-Vienna}}
    Ignasi Ribas\orcidlink{0000-0002-6689-0312},\inst{\ref{inst-CSIC},\ref{inst-IEEC}}
    Sara~Seager\orcidlink{0000-0002-6892-6948},\inst{\ref{inst-kavli},\ref{inst-mitAtmospheric},\ref{inst-mitAstronautics}}
    Damien Ségransan\orcidlink{0000-0003-2355-8034}, \inst{\ref{inst-geneva}}
    Gaetano Scandariato\orcidlink{0000-0003-2029-0626}, \inst{\ref{inst-INAFcatania}}
    Attila Simon\orcidlink{0000-0001-9773-2600}, \inst{\ref{inst-bernPlanetologie},\ref{inst-bern}}
    Alexis~M.~S.~Smith\orcidlink{0000-0002-2386-4341}, \inst{\ref{inst-DLR}}
    Manu Stalport\orcidlink{0000-0003-0996-6402},  \inst{\ref{inst-star},\ref{inst-liege}}
    Manfred Steller\orcidlink{0000-0003-2459-6155}, \inst{\ref{inst-graz}}
    Gyula Szabó\orcidlink{0000-0002-0606-7930}, \inst{\ref{inst-Szent},\ref{inst-Szombathely}}
    Nicolas Thomas, \inst{\ref{inst-bernPlanetologie}}
    Tyler A.~Pritchard\orcidlink{0000-0001-9227-8349}, \inst{\ref{inst-Goddard}}
    Stéphane Udry\orcidlink{0000-0001-7576-6236}, \inst{\ref{inst-geneva}}
    Carlos Corral Van Damme, \inst{\ref{inst-estec}}
    Valérie Van Grootel\orcidlink{0000-0003-2144-4316}, \inst{\ref{inst-star}}
    Eva Villaver\orcidlink{0000-0003-4936-9418},  \inst{\ref{inst-canarias},\ref{inst-laguna}}
    Ingo Walter\orcidlink{0000-0002-5839-1521}, \inst{\ref{inst-DLR}}
    Nicholas Walton\orcidlink{0000-0003-3983-8778}, \inst{\ref{inst-Cambridge}}
    Cristilyn N. Watkins, \inst{\ref{inst-Smithsonian}}
    Richard G.~West\orcidlink{0000-0001-6604-5533}\inst{\ref{inst-Warwick},\ref{inst-warwickcenter}}
}
 \institute{
    Observatoire de Gen{\`e}ve, Universit{\'e} de Gen{\`e}ve, Chemin Pegasi, 51, 1290 Versoix, Switzerland \label{inst-geneva}
    \and
    Center for Space and Habitability, University of Bern, Gesellschaftsstrasse 6, 3012 Bern, Switzerland \label{inst-bern}
    \and
    Department of Physics and Kavli Institute for Astrophysics and Space Research, Massachusetts Institute of Technology, Cambridge, MA 02139, USA  \label{inst-kavli}
    \and
    Weltraumforschung und Planetologie, Physikalisches Institut, University of Bern, Gesellschaftsstrasse 6, 3012 Bern, Switzerland  \label{inst-bernPlanetologie}
    \and
    Instituto de Astrof\'isica e Ci\^encias do Espa\c{c}o, Universidade do Porto, CAUP, Rua das Estrelas, 4150-762 Porto, Portugal \label{inst-caup}
    \and
    Departamento de F\'isica e Astronomia, Faculdade de Ci\^encias, Universidade do Porto, Rua do Campo Alegre, 4169-007 Porto, Portugal\label{inst-porto2}
    \and
    Space Research Institute, Austrian Academy of Sciences, Schmiedlstraße 6, 8042 Graz, Austria \label{inst-graz}
    \and
    Center for Astrophysics \textbar \ Harvard \& Smithsonian, 60 Garden Street, Cambridge, MA 02138, USA \label{inst-Smithsonian}
    \and
    Universit\'e C\^ote d'Azur, Observatoire de la C\^ote d'Azur, CNRS, Laboratoire Lagrange, Bd de l'Observatoire, CS 34229, 06304 Nice cedex 4, France \label{inst-cotedazur}
    \and
    Institute for Computational Science, University of Zurich, Winterthurerstr. 90, 8057 Zurich, Switzerland \label{inst-zurich}
    \and
    Astrophysics Group, Keele University, Staffordshire ST5 5BG, UK\label{inst-keele}
    \and
    Department of Astronomy \& Astrophysics, University of Chicago, Chicago, IL 60637, USA\label{inst-Chicago} 
    \and
    NASA Ames Research Center, Moffett Field, CA 94035, USA \label{inst-NASAAmes} 
    \and
    Research Institute for Advanced Computer Science, Universities Space Research Association, Washington, DC 20024, USA \label{inst-Washington}
    \and
    Department of Physics and Astronomy, Union College, 807 Union St., Schenectady, NY 12308, USA \label{inst-Union}
    \and
    Department of Physics, University of Warwick, Gibbet Hill Road, Coventry CV4 7AL, UK\label{inst-Warwick}
    \and
    Department of Astrophysical Sciences, Princeton University, Princeton, NJ 08544, USA \label{inst-Princeton}
    \and
    Department of Physics, Engineering and Astronomy, Stephen F. Austin State University, 1936 North St, Nacogdoches, TX 75962, USA \label{inst-Austin} 
    \and
    Dipartimento di Fisica e Astronomia ""Galileo Galilei"", Università degli Studi di Padova, Vicolo dell'Osservatorio 3, 35122 Padova, Italy\label{inst-padova}
    \and
    INAF, Osservatorio Astronomico di Padova, Vicolo dell'Osservatorio 5, 35122 Padova, Italy\label{inst-inaf}
    \and
    Department of Astronomy, Stockholm University, AlbaNova University Center, 10691 Stockholm, Sweden \label{inst-AlbaNova}
    \and
    Dipartimento di Fisica, Università degli Studi di Torino, via Pietro Giuria 1, I-10125, Torino, Italy\label{inst-torino}
    \and
    Cavendish Laboratory, JJ Thomson Avenue, Cambridge CB3 0HE, UK\label{inst-Cavendish}
    \and
    Institute of Astronomy, University of Cambridge, Madingley Road, Cambridge, CB3 0HA, United Kingdom\label{inst-Cambridge}
     D\'epartement de Physique, Institut Trottier de Recherche sur les Exoplan\`etes, Universit\'e de Montr\'eal, Montr\'eal, Qu\'ebec, H3T 1J4, Canada\label{inst-montreal}
     \and
     Instituto de Astrofisica de Canarias, Via Lactea s/n, 38200 La Laguna, Tenerife, Spain\label{inst-canarias}
     \and
     Departamento de Astrofisica, Universidad de La Laguna, Astrofísico Francisco Sanchez s/n, 38206 La Laguna, Tenerife, Spain\label{inst-laguna}
     \and
     Institut de Ciencies de l'Espai (ICE, CSIC), Campus UAB, Can Magrans s/n, 08193 Bellaterra, Spain\label{inst-CSIC}
     \and
     Institut d’Estudis Espacials de Catalunya (IEEC), Gran Capità 2-4, 08034 Barcelona, Spain\label{inst-IEEC}
     \and
     Admatis, 5. Kandó Kálmán Street, 3534 Miskolc, Hungary \label{inst-Hungary}
     \and
     Depto. de Astrofisica, Centro de Astrobiologia (CSIC-INTA), ESAC campus, 28692 Villanueva de la Cañada (Madrid), Spain\label{inst-cabmadrid}
     \and
     Université Grenoble Alpes, CNRS, IPAG, 38000 Grenoble, France\label{inst-grenoble}
    \and
    Caltech/IPAC-NASA Exoplanet Science Institute, 770 S. Wilson Avenue, Pasadena, CA 91106, USA\label{inst-IPAC}
    \and
    Centre for Exoplanet Science, SUPA School of Physics and Astronomy, University of St Andrews, North Haugh, St Andrews KY16 9SS, UK \label{inst-SUPA}
    \and
    Universit\'{e} de Paris Cit\'{e}, Institut de physique du globe de Paris, CNRS, 1 Rue Jussieu, F-75005 Paris, France \label{inst-paris}
    \and
    Institute of Planetary Research, German Aerospace Center (DLR), Rutherfordstrasse 2, 12489 Berlin, Germany\label{inst-DLR}  
    \and
    European Southern Observatory, Av. Alonso de Cordova 3107, Casilla 19001, Santiago de Chile, Chile \label{inst-ESOchile}
    \and
    Centre for Mathematical Sciences, Lund University, Box 118, 221 00 Lund, Sweden \label{inst-lund}
    \and
    Department of Physics and McDonnell Center for the Space Sciences, Washington University, St. Louis, MO 63130, USA\label{inst-WashU}
    \and
    Aix Marseille Univ, CNRS, CNES, LAM, 38 rue Frédéric Joliot-Curie, 13388 Marseille, France \label{inst-Marseille}
    \and
    Astrobiology Research Unit, Université de Liège, Allée du 6 Août 19C, B-4000 Liège, Belgium \label{inst-liege}
    \and
    Space sciences, Technologies and Astrophysics Research (STAR) Institute, Université de Liège, Allée du 6 Août 19C, 4000 Liège, Belgium\label{inst-star}
    \and
    ETH Zurich, Institute for Particle Physics and Astrophysics, Wolfgang-Pauli-Strasse 27, CH-8093 Zurich Switzerland\label{inst-ETH}
    \and
    Institut d'astrophysique de Paris, UMR7095 CNRS, Université Pierre \& Marie Curie, 98bis blvd. Arago, 75014 Paris, France\label{inst-pierre}
    \and
    Airbus Defence and Space SAU, C/ Aviocar 2,  28906 Getafe, Madrid, Spain\label{inst-sauc}
    \and
    Leiden Observatory, University of Leiden, PO Box 9513, 2300 RA Leiden, The Netherlands \label{inst-leiden}
    \and
    Department of Space, Earth and Environment, Chalmers University of Technology, Onsala Space Observatory, 439 92 Onsala, Sweden\label{inst-onsala}
    \and
    Department of Astrophysics, University of Vienna, Türkenschanzstrasse 17, 1180 Vienna, Austria \label{inst-Vienna}
    \and
    European Space Agency (ESA), European Space Research and Technology Centre (ESTEC), Keplerlaan 1, 2201 AZ Noordwijk, The Netherlands\label{inst-estec}
    \and
    Institute for Theoretical Physics and Computational Physics, Graz University of Technology, Petersgasse 16, 8010 Graz, Austria\label{inst-Petersgasse}
    \and
    Konkoly Observatory, Research Centre for Astronomy and Earth Sciences, 1121 Budapest, Konkoly Thege Miklós út 15-17, Hungary\label{inst-Konkoly}
    \and
    ELTE E\"otv\"os Lor\'and University, Institute of Physics, P\'azm\'any P\'eter s\'et\'any 1/A, 1117 Budapest, Hungary\label{inst-elte}
    \and
    IMCCE, UMR8028 CNRS, Observatoire de Paris, PSL Univ., Sorbonne Univ., 77 av. Denfert-Rochereau, 75014 Paris, France\label{inst-sorbonne}
    \and
    Proto-Logic LLC, 1718 Euclid Street NW, Washington, DC 20009, USA\label{inst-protologic}
    \and
    European Southern Observatory, Karl-Schwarzschild-Straße 2, 85748 Garching bei München, Germany\label{inst-ESOgarching}
    \and
    INAF Osservatorio Astrofisico di Catania, via S. Sofia 78, 95123 Catania, Italy\label{inst-INAFcatania}
    \and
    Zentrum für Astronomie und Astrophysik, Technische Universität Berlin, Hardenbergstr. 36, D-10623 Berlin, Germany\label{inst-tuberlin}
    \and
    Institut fuer Geologische Wissenschaften, Freie Universitaet Berlin, Maltheserstrasse 74-100,12249 Berlin, Germany\label{inst-freie}
    \and
    Department of Earth, Atmospheric and Planetary Sciences, Massachusetts Institute of Technology, Cambridge, MA 02139, USA\label{inst-mitAtmospheric}
    \and
    Department of Aeronautics and Astronautics, MIT, 77 Massachusetts Avenue, Cambridge, MA 02139, USA\label{inst-mitAstronautics}
    \and
    ELTE E\"otv\"os Lor\'and University, Gothard Astrophysical Observatory, 9700 Szombathely, Szent Imre h. u. 112, Hungary\label{inst-Szent}
    \and
    HUN-REN–ELTE Exoplanet Research Group, Szent Imre h. u. 112., Szombathely, H-9700, Hungary\label{inst-Szombathely}
    \and
    NASA Goddard Space Flight Center, 8800 Greenbelt Road, Greenbelt, MD 20771, USA\label{inst-Goddard}
    \and
    Centre for Exoplanets and Habitability, University of Warwick, Gibbet Hill Road, Coventry CV4 7AL, UK \label{inst-warwickcenter}
}
   \date{}

   \abstract{We present the discovery and characterization of two warm mini-Neptunes transiting the K3V star TOI-815 in a K-M binary system. Analysis of its spectra and rotation period reveal the star to be young, with an age of $200 ^{+400} _{-200}$~Myr. TOI-815b has a 11.2-day period and a radius of 2.94 $\pm$ 0.05 \re with transits observed by TESS, CHEOPS, ASTEP, and LCOGT. The outer planet, TOI-815c, has a radius of 2.62 $\pm$ 0.10 \re, based on observations of three nonconsecutive transits with TESS; targeted CHEOPS photometry and radial velocity follow-up with ESPRESSO were required to confirm the 35-day period. ESPRESSO confirmed the planetary nature of both planets and measured masses of 7.6 $\pm$ 1.5 \me ($\rho_\mathrm{P}$ = 1.64$ ^{+0.33} _{-0.31}$~\gccc) and 23.5 $\pm$ 2.4 \me ($\rho_\mathrm{P}$ = 7.2$ ^{+1.1} _{-1.0}$~\gccc), respectively. Thus, the planets have very different masses, which is unusual for compact multi-planet systems. Moreover, our statistical analysis of mini-Neptunes orbiting FGK stars suggests that weakly irradiated planets tend to have higher bulk densities compared to those undergoing strong irradiation. This could be ascribed to their cooler atmospheres, which are more compressed and denser. Internal structure modeling of TOI-815b suggests it likely has a H-He atmosphere that constitutes a few percent of the total planet mass, or higher if the planet is assumed to have no water. In contrast, the measured mass and radius of TOI-815c can be explained without invoking any atmosphere, challenging planetary formation theories. Finally, we infer from our measurements that the star is viewed close to pole-on, which implies a spin-orbit misalignment at the 3$\sigma$ level. This emphasizes the peculiarity of the system's orbital architecture, and probably hints at an eventful dynamical history.}


    \maketitle
%

\section{Introduction}\label{sec:Introduction}
The Transiting Exoplanet Survey Satellite (TESS; \citealt{Ricker2015}) and the \textit{Kepler} space mission have profoundly transformed our understanding of planets by detecting thousands of exoplanets using the transit method, particularly those with radii below 4~\re and relatively short orbital periods. \citet{Fulton2017}, using precise planet radius measurements, identified a clear bimodal distribution in the radii of small planets. It appears that close-orbiting planets predominantly fall into two categories: super-Earths (below 1.8~\re) or sub-Neptunes (2 to 4~\re), with very few planets occupying the intermediate range (the so-called radius valley). Four mechanisms have been proposed to explain the origin of these two populations: photoevaporation of the H-He envelopes due to high-energy radiation from the host star (\citealt{Lopez2013}; \citealt{Owen2017}; \citealt{Jin2018}; \citealt{Rogers2021}; \citealt{Attia2021}), atmospheric mass-loss fueled by the cooling luminosity of a planet's core (\citealt{Ginzburg2018}; \citealt{Gupta2019}), a combination of ex situ formation and photoevaporation \citep{Venturini2020}, and an in situ gas-poor formation scenario (\citealt{Lee_2022}). These mechanisms predict different compositions for the mini-Neptunes, ranging from dry planets composed of a H-He envelope atop a rocky core \citep{Owen2017, Jin2018, Gupta2019, Lee_2022} to water worlds with a thin or no H-He atmosphere \citep{Venturini2020}. The mechanisms also predict distinct relationships between stellar and planetary mass, as well as differing locations of the valley (in the period-radius-mass space) and varying occurrence rates relative to the age of the systems \citep{Berger2020A, Petigura2022}.

Numerous ongoing follow-up programs are dedicated to characterizing the masses of super-Earth and mini-Neptune planets located close to the ``radius valley'' of short-period orbits. However, there is limited coverage of mini-Neptunes with low stellar irradiation (less than 30 \se) and long orbital periods (greater than 15 days). We started a radial velocity (RV) follow-up campaign with the
Échelle SPectrograph for Rocky Exoplanets and Stable Spectroscopic Observations (ESPRESSO; \citealt{pepe2021}) dedicated to the characterizations and precise density measurements of warm mini-Neptune planets that transit FGK dwarfs. Currently, there are known 17 exoplanets in this parameter range with stars brighter than V=12 and well-constrained 
densities\footnote{$\sigma_\mathrm{M}/M$ $\leq$ 25$\%$ and $\sigma_\mathrm{R}/R$ $\leq$8$\%$\label{precisedensity}} according to the PlanetS catalog\footnote{Available on the Data \& Analysis Center for Exoplanets (DACE) platform (\url{https://dace.unige.ch})\label{dacefootnote}} (\citealt{otegi2020}).

Multi-planet systems provide the opportunity to study planets of different densities and compositions within a single system, and to compare such planets with each other. One observed trend in the diversity of planetary systems is that planets seem to resemble ``peas in the pod,'' that is to say, multi-transiting systems tend to have planets with similar sizes and geometrically spaced orbital distances \citep{Weiss2018A,Otegi2022,Mishra2023}. Understanding these trends is crucial since they can elucidate underlying physical processes and constrain formation theories. 

In this work we present the discovery of two mini-Neptune planets transiting TOI-815, a V=11.2 mag K3V star. The innermost planet, designated TOI-815b, has an orbital period of 11.2 days. This planet's transits were captured by TESS during Sectors 6, 36, and 63, as well as by the CHaracterising ExOPlanet Satellite (CHEOPS; \citealt{Benz2021}), the Antarctic Search for Transiting ExoPlanets (ASTEP) telescope (\citealt{Guillot2015}), and the Las Cumbres Observatory Global Telescope (LCOGT; \citealt{Brown:2013}). As for the outer planet, TOI-815c, TESS detected three transits during the aforementioned sectors; however, a unique period could not be determined. Additionally, targeted CHEOPS photometry was performed to investigate potential period aliases, leading to the confirmation of a 35-day orbital period for TOI-815c. The planetary nature of both objects has been confirmed via RV follow-up using ESPRESSO.

The paper is structured as follows: In Section \ref{sec:Observations} we provide a detailed description of the space- and ground-based observations with TESS, CHEOPS, ASTEP, and LCOGT, as well as the ESPRESSO RV observations. Section \ref{sec:Stellar characterization} details how we determined the host star parameters by combining high-resolution spectra. In Section \ref{sec:Joint transit and RV analysis} we present our global photometric and RV analysis and its results.  Finally, in Section \ref{sec:Discussion} we present a discussion of the system, and we conclude in Section \ref{sec:Conclusions}.

\section{Observations}\label{sec:Observations}
\subsection{TOI-815 in a binary system}\label{sec:binary}
\citet{Behmard2022} searched for stellar companions to TESS objects of interest (TOIs; \citealt{Guerrero2021}) by cross-matching them with the \textit{Gaia} Early Data Release 3 (EDR3) catalog. TOI-815 was identified to be in a binary system with the nearest known neighbor in the TESS Input Catalog Version 8 (TICv8; \citealt{Stassun2019}), TIC 102840237 (\textit{Gaia} DR3 5415648821874435584) that is located approximately 359 AU ($\sim6\farcs4$) north of TOI-815 (Figure \ref{fig:TPF}). According to the TICv8 catalog, TIC 102840237 is an M-dwarf with $T_{\mathrm{eff}}$ = 3766 $\pm$ 157 K and is 2.5 magnitudes dimmer than TOI-815 in the G band. To compensate for flux contamination in the photometry, we included a dilution factor for the light curves, as presented below.

\subsection{TESS photometry}\label{sec:TESS_photometry}
The TOI-815 system (TIC 102840239) was observed in TESS Sector 9 (2018 December 12 to 2019 January 06) in 30-minute cadence and Sectors 36 (2021 March 07 to 2021 April 01) and 63 (2023 March 10 to 2023 April 06) in 20-second cadence.  In Sectors 9 and 36, the target was imaged on CCD 2 of camera 2, and in Sector 63, on CCD 1 of camera 2. The 30-minute full frame image light curves were extracted by the MIT Quick Look Pipeline (QLP; \citealt{QLP}), resulting in the simple aperture photometry (SAP) flux and the QLP \textit{Kepler} Spline SAP (KSPSAP; \citealt{Vanderburg2014}) flux. The 2-minute and 20-second cadence photometry was produced by the Science Processing Operations Center (SPOC; \citealt{Jenkins2016}) pipeline at the NASA Ames Research Center, resulting in the SAP \citep{twicken2010,morris2020} flux and the Presearch Data Conditioning SAP (PDCSAP; \citealt{Smith2012}; \citealt{Stumpe2012}, \citeyear{Stumpe2014}) flux.

On 2019 June 21, the TESS data public website\protect\footnote{\url{ https://tev.mit.edu/data/}\label{tessweb}} announced the discovery of a 11.2-day TESS TOI (\citealt{Guerrero2021}), TOI-815b. The SPOC detected the transit signature of TOI-815b in searches of Sector 36, and a search of Sectors 36 and 63 with a noise-compensating matched filter (\citealt{Jenkins2002}, \citeyear{Jenkins2010}, \citeyear{Jenkins2020}), and the signature was fitted with an initial limb-darkened transit model (\citealt{Li2019}) and passed all the diagnostic tests presented in the Data Validation report (\citealt{Twicken2018}). In particular, the difference image centroiding test for the multi-sector search located the host star within 1.4 $\pm$ 3.4 arcsec of the transit source. Seven transits with a depth of 1301 $\pm$ 54 ppm and a duration of 3.06 $\pm$ 0.43 hours were identified in Sectors 9, 36, and 63 (Figure \ref{fig:fullTESSLCs}). However, TOI-815 showed one additional transit in each sector with a transit depth of 1004 $\pm$ 73 ppm and a duration of 2.53 $\pm$ 0.59 hours, suggesting a second transiting companion. As presented in this work, with follow-up observations using ESPRESSO and CHEOPS, we confirmed the planetary nature and characterized the second planet.

For the analysis of the 30-minute cadence data we used the KSPSAP flux. For the 2-minute and 20-second cadence light curves, we used the uncorrected SAP flux and accounted for the systematic noise with the \juliet package (\citealt{Espinoza2019}) using Gaussian process (GP) models with an approximate Mat\'ern 3/2 kernel. 

\begin{figure}
  \centering
    \includegraphics[width=0.45\textwidth]{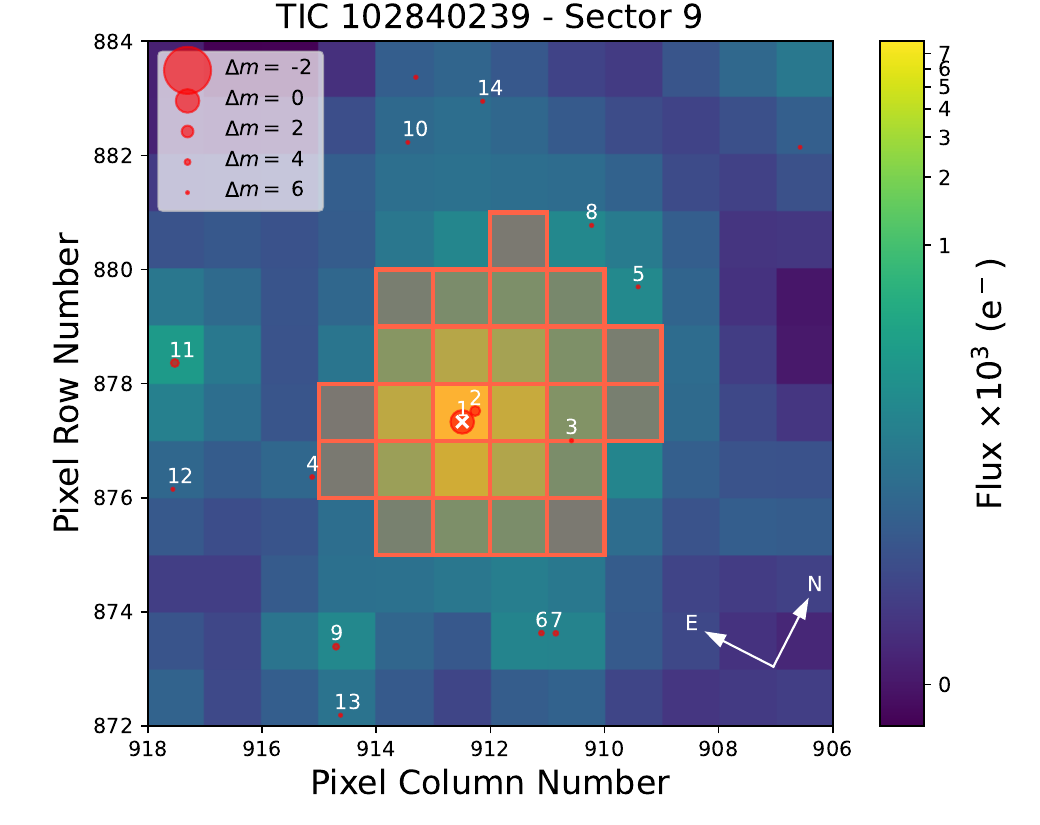}
  \caption{TESS TPF of TOI-815 created with \texttt{tpfplotter} (\citealt{TPFplotter}). The orange pixels define the aperture mask used for extracting the photometry. Additionally, the red circles indicate neighboring objects from the \textit{Gaia} DR2 catalog, with the circle size corresponding to the brightness difference compared to the targets (as indicated in the legend). Our target is marked with a white cross. Pixel scale is 21$\arcsec$/pixel.} 
  \label{fig:TPF}
\end{figure}

In order to check the existence of contaminant sources, we plotted the target pixel file (TPF; Figure \ref{fig:TPF}) along with the aperture mask used for the SAP flux. The plot was generated using \texttt{tpfplotter} (\citealt{TPFplotter}). The apertures used for extracting the light curves in all three sectors were contaminated by TIC 102840237 (Section \ref{sec:binary}) and therefore we included a dilution coefficient during the modeling of the TESS light curves. Our TESS light curve analysis is described in more detail in Section $\ref{sec:joint}$.

\subsection{CHEOPS observations}\label{sec:CHEOPS_photometry}
CHEOPS is a 32~cm aperture ESA telescope dedicated to the characterization of transiting exoplanets. It performs high-precision photometry in a broad optical passband (330 to 1100 nm). The instrument is optimized for bright stars, making it an ideal facility to observe small-amplitude transits created by low-mass planets (e.g., \citealt{Delrez2021}; \citealt{Leleu2021}), or characterize the minute emission signals of hot Jupiters at optical wavelengths \citep{Lendl2020, Krenn2023}. Among various scientific goals, the CHEOPS space mission is also suited to improve the radius precision of known exoplanets discovered by TESS \citep{bonfanti2021} by observing additional transits and to recover the true period of double transit candidates detected by TESS (e.g., \citealt{Osborn2023}; \citealt{Ulmer2023}; \citealt{Garai2023}). TOI-815 was jointly observed by two programs within the CHEOPS guaranteed time observation (GTO): one targeting improved radius precision for multi-planet systems detected by NASA TESS (CHESS; PR0031) and the other targeted on confirming the orbital periods of long-period candidates found by TESS (DUOS; PR0048). The observations had an exposure time of 60~s, typical duration of 5-7 orbits (8.2-11.4~hours), and achieved efficiencies between 51 and 73\% (e.g., due to Earth occultations, Earth's stray light contamination and South Atlantic Anomaly crossings). CHEOPS data are automatically processed with an automatic aperture photometry pipeline \citep{Hoyer2020}. We however re-extracted the photometry using point spread function (PSF) photometry with the \texttt{PIPE} package\footnote{\url{https://github.com/alphapsa/pipe}} (\citealt{Brandeker2022}), which, due to the large asymmetric PSF of CHEOPS, typically results in photometry with less influenced by external factors such as companion stars, roll-angle trends, etc. \texttt{PIPE} takes into account the flux contamination from nearby stars by subtracting a synthetic image using the PSF and \textit{Gaia} DR2 parameters for the neighbor stars. Therefore, the dilution factor for the CHEOPS light curves was fixed to one.

\subsection{Ground-based photometry}\label{sec:ground_photometry}
The TESS pixel scale is $\sim 21\arcsec$ pixel$^{-1}$ and photometric apertures typically extend out to roughly 1$\arcmin$, generally causing multiple stars to blend in the TESS aperture. To determine the true source of transit signals in the TESS data, improve the transit ephemerides, and check the SPOC pipeline transit depth after accounting for the crowding metric, we conducted photometric ground-based follow-up observations with ASTEP and LCOGT of the field around TOI-815 as part of the TESS Follow-up Observing Program (TFOP)\protect\footnote{\url{https://tess.mit.edu/followup/}\label{tessmit}} Sub Group 1 \citep{collins:2019}. We used the {\tt TESS Transit Finder}, which is a customized version of the {\tt Tapir} software package \citep{Jensen:2013}, to schedule our transit observations.

\subsubsection{ASTEP}\label{sec:ASTEP}
\begin{figure}[h]
  \centering
    \includegraphics[width=0.43\textwidth]{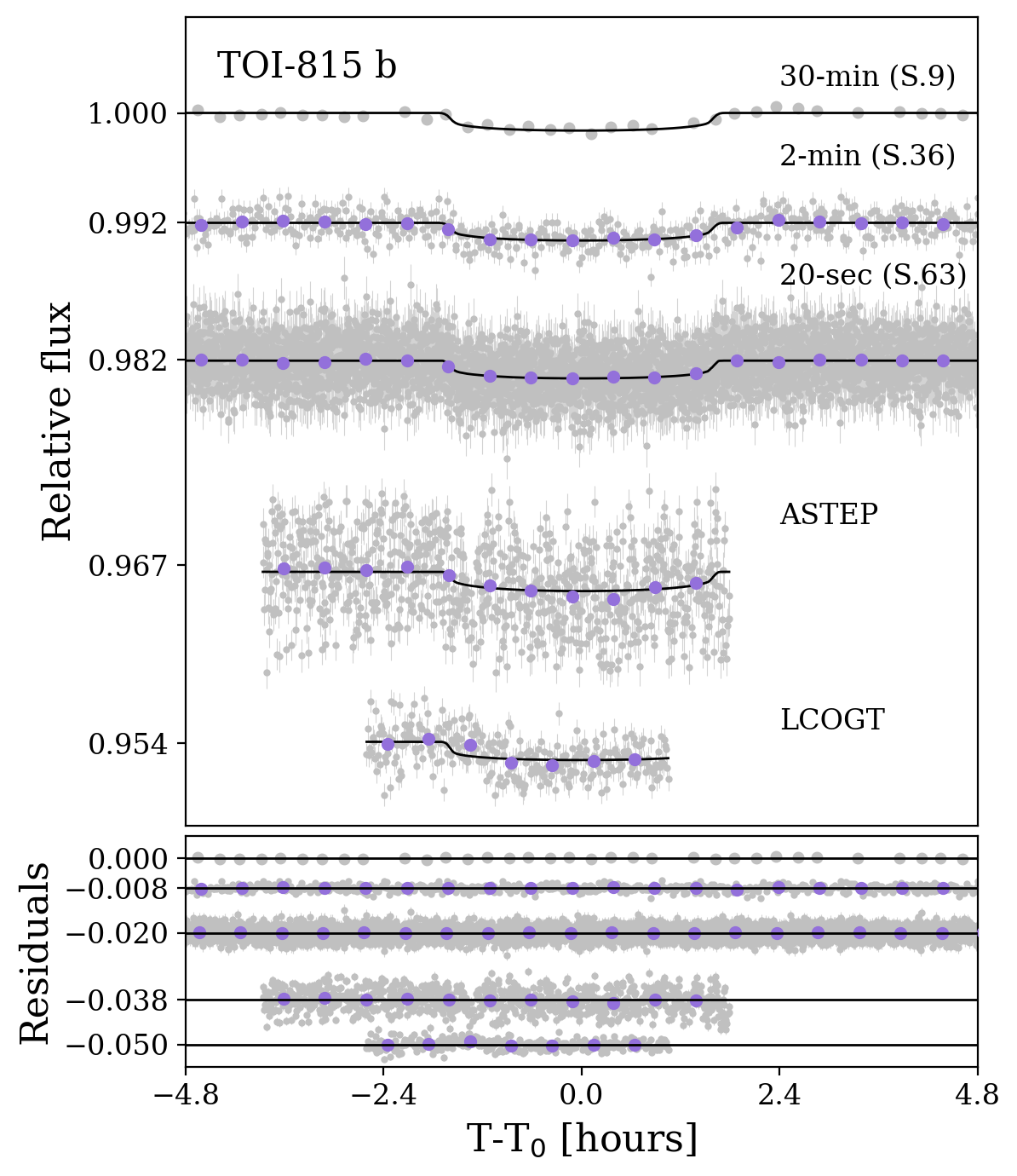}
    \includegraphics[width=0.43\textwidth]{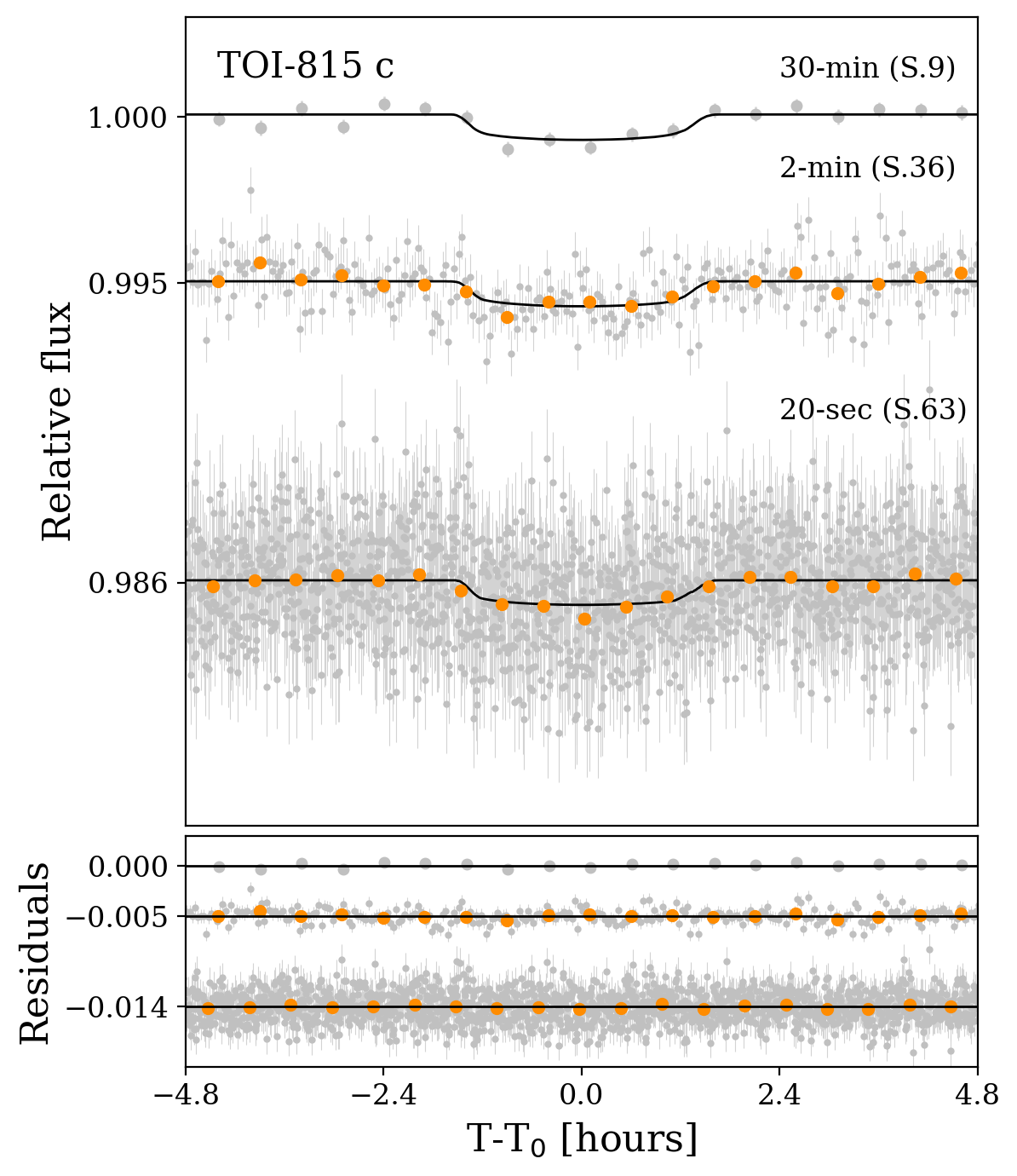}
  \caption{Phase-folded TESS, ASTEP, and LCOGT light curves of TOI-815b (top panel) and TESS light curves of TOI-815c (bottom panel). The data are binned to 30 minutes (purple and orange circles).} 
  \label{fig:TESSp1}
\end{figure}

We observed TOI-815 with ASTEP on 2021 April 04 but could only rule out an ingress of TOI-815b that was, at the time, predicted to occur at BJD 2459309.02. Subsequent analysis (see Section \ref{sec:TTVs}) shows that as a result of transit timing variations (TTVs), the prediction was incorrect, and that the ingress should have started at $2459309.93 \pm 0.0285$, later than the end of this first series of observations. On 2022 September 13, using an improved camera system \citep{Dransfield2022, Schmider2022} we observed the target again (Figure \ref{fig:TESSp1}), with 15 second exposures, at wavelengths ranging from about 700 to 1000 nm (\citealt{Schmider2022}). The data were processed on site with an IDL-based aperture photometry pipeline \citep{Abe2013, Merkania2016}. We detected an on-time 2.0 ppt transit of TOI-815b, using a 13.9 arcsec target aperture. While the ingress was clearly evident, the egress was affected by sky brightening, with a Sun at about $12^\circ$ below the horizon at end of the observation sequence.  

\subsubsection{LCOGT}\label{sec:LCOGT}
We observed an intended full transit window of TOI-815b on 2021 February 18 (Figure \ref{fig:TESSp1}) in Pan-STARRS $z$-short band using the LCOGT 1.0\,m network node at South Africa Astronomical Observatory (SAAO). However, after updating the ephemeris from TESS Sector 36 data, we determined that the follow-up observations were out of transit by approximately one day. Therefore, these data are not considered further in this work. We observed a partial transit window of TOI-815b on 2023 March 11 using the same band-pass and LCOGT network node. The images were calibrated by the standard LCOGT {\tt BANZAI} pipeline \citep{McCully:2018} and differential photometric data were extracted using {\tt AstroImageJ} \citep{Collins:2017}. We used circular photometric apertures with radius $3\farcs9$. The target star aperture was diluted by 0.85$\%$ by the neighbor star (Section \ref{sec:binary}). The best zero, one, or two detrending vectors were retained if they decreased the Bayesian information criterion (BIC) for a fit by at least two per detrend parameter. We found that a proxy for sky transparency and the Y-centroid of the target star on the detector detrending pair provided the best improvement to the light curve fit that was justified by the BIC values.

\subsection{Spectroscopic follow-up with ESPRESSO}\label{sec:Spectroscopic follow-up}

We acquired 34 high-resolution spectroscopic observations of TOI-815 using ESPRESSO on the 8.2 m Very Large Telescope (VLT) located in Paranal, Chile. The observations were carried out between 2021 April 04 and 2022 December 04 as part of the observing programs 105.20P7.001, 109.23DX.001 and 110.2481.001 (PI: Bouchy), dedicated to the characterization of warm mini-Neptune transiting exoplanets. The observations have an integration time ranging from 600 to 700 s, a median resolving power of 140,000 using 2 $\times$ 1 binning, and a wavelength range of 380-788 nm. The RVs and activity indicators were extracted using version 3.0.0. of the ESPRESSO pipeline, and we computed the RVs by cross-correlating the Echelle spectra with a G2 numerical mask.

\begin{figure*}[t]
  \centering
  \begin{minipage}[t]{\textwidth}
  \centering
    \includegraphics[width=\textwidth]{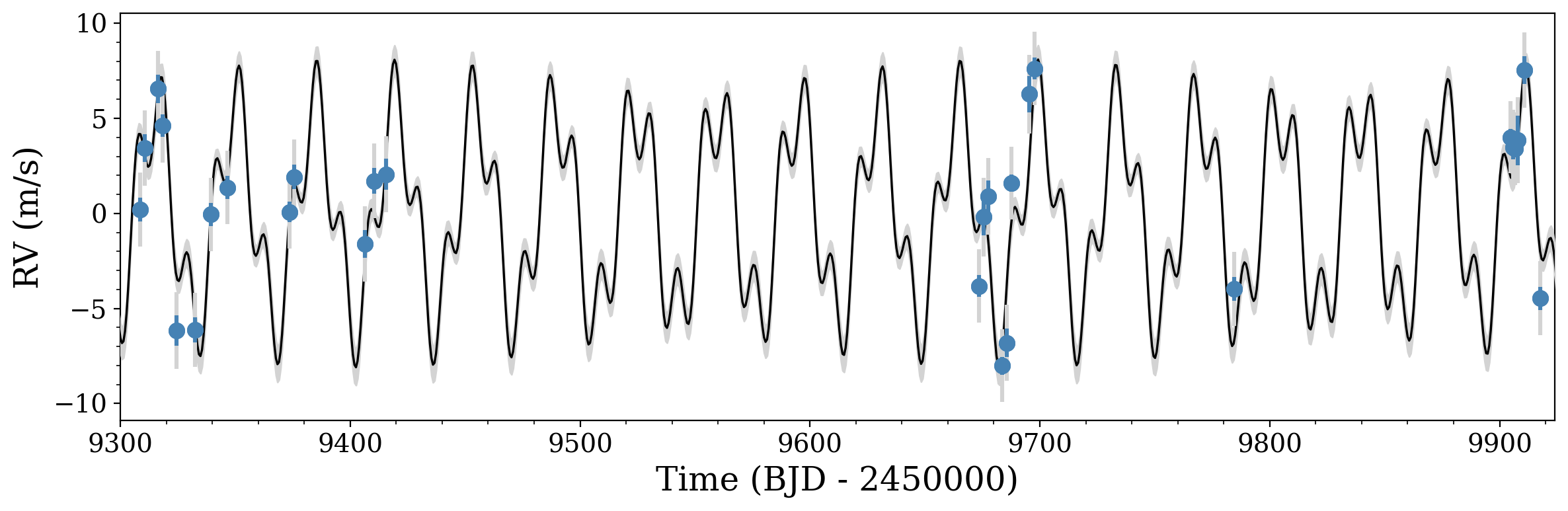}
  \end{minipage}
  \begin{minipage}{\textwidth}
    \hspace{0.04\textwidth} 
    \includegraphics[width=0.45\textwidth]{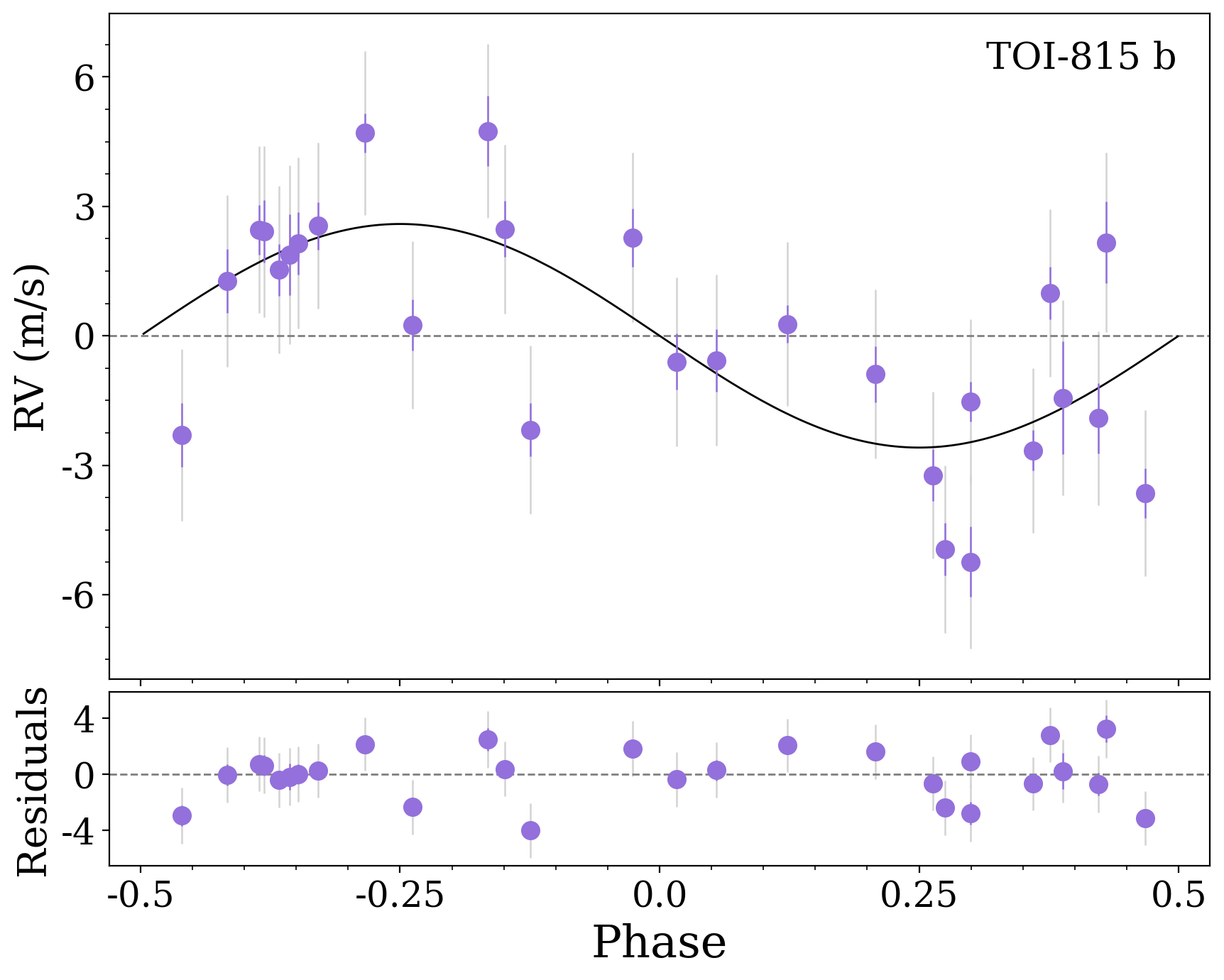}
    \includegraphics[width=0.45\textwidth]{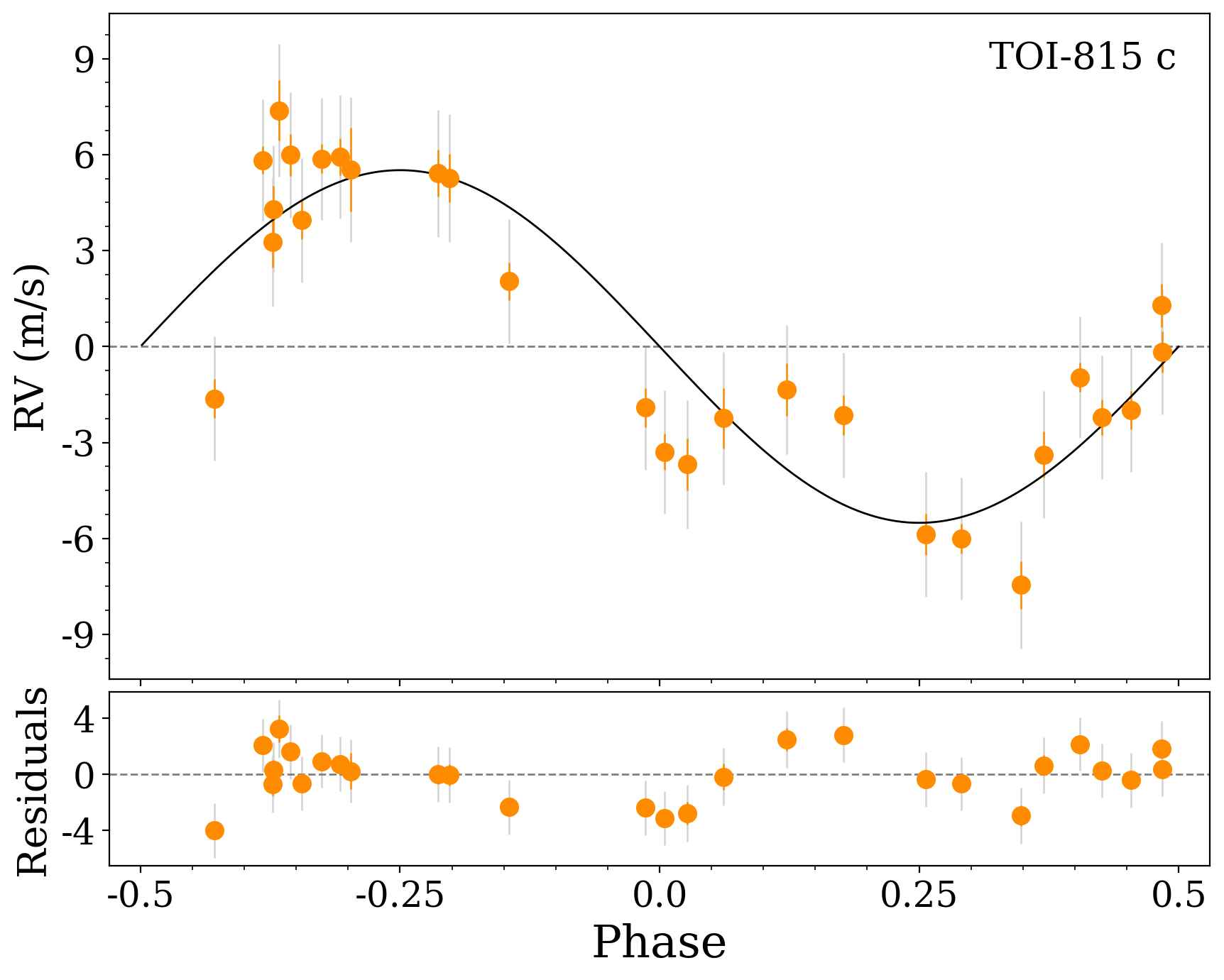}
  \end{minipage}
  \vspace{0.0cm}
  \caption{Relative RV measurements with ESPRESSO and best-fit models for TOI-815. $\it{Top}$: Time series of ESPRESSO RV measurements (in blue). The complete inferred model, which comprises signals from two planets along with the activity-induced signal, is represented by a continuous solid line. $\it{Bottom}$: Phase-folded RVs for TOI-815b (left) and TOI-815c (right). In all plots, the error bars in gray account for the estimated jitter ($\sigma_{w,\mathrm{ESPRESSO}}$).} 
  \label{fig:RVs} 
\end{figure*}

The ESPRESSO pipeline was devised to correct the residual atmosphere dispersion, that is, after atmospheric dispersion corrector correction. This reduces significantly the chromatic effects on the data. Out of the total observations, six were identified as unreliable since the flux correction was not applied and therefore were excluded from the global modeling. The average uncertainty of the RV data is 0.68~\ms and the RMS is 0.74~\ms. We report the ESPRESSO RV measurements and their uncertainties, along with the full width at half maximum (FWHM), bisector, contrast, S-index and H$\alpha$-index in Table \ref{tab:rvs}. The excluded data points are noted in the same Table. The RV time series and the phase-folded RVs for TOI-815b and TOI-815c are shown in Figure $\ref{fig:RVs}$.

\subsection{High-resolution imaging}\label{sec:High resolution imaging}
\begin{figure}
  \centering
  \includegraphics[width=0.45\textwidth]{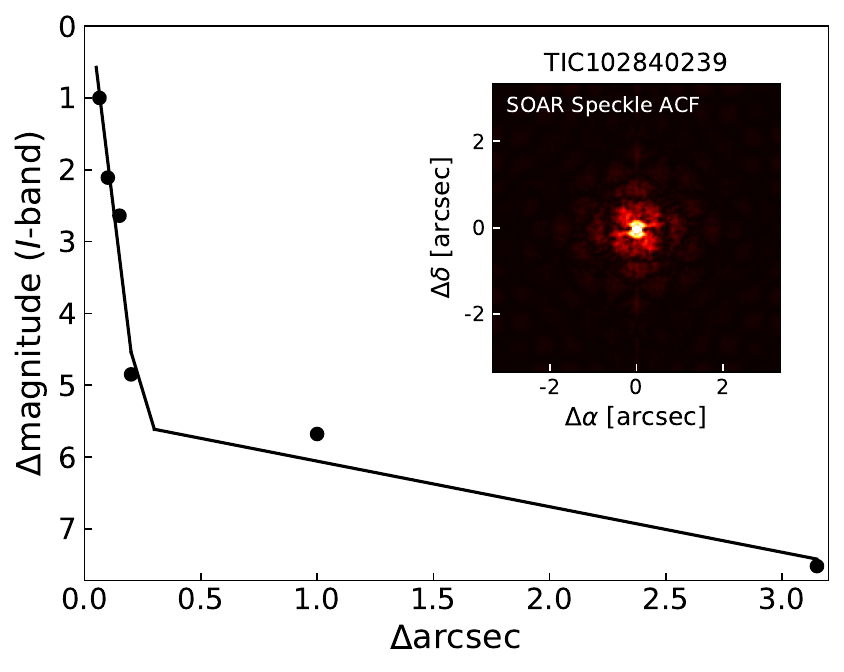}
  \hspace{0.2cm}
  \caption{SOAR speckle imaging (5$\sigma$ upper limits) of TOI-815 that rules out the presence of stellar companions within 3$\arcsec$. The image inset illustrates the speckle autocorrelation function.} 
  \label{fig:SOAR}
\end{figure}

High-angular-resolution imaging is needed to search for nearby sources that can contaminate the TESS photometry, resulting in an underestimated planetary radius, or be the source of astrophysical false positives, such as background eclipsing binaries. We searched for stellar companions to TOI-815 with speckle imaging on the 4.1 m Southern Astrophysical Research (SOAR) telescope (\citealt{Tokovinin2018}) on 2019 July 14, observing in Cousins I-band, a similar visible band-pass as TESS. This observation was sensitive to a 5.8-magnitude fainter star at an angular distance of 1 arcsec from the target. More details of the observations within the SOAR TESS survey are available in \citet{Ziegler2020}. The 5$\sigma$ detection sensitivity and speckle autocorrelation functions from the observations are shown in Figure \ref{fig:SOAR}. No nearby stars were detected within 3\arcsec of TOI-815 in the SOAR observations. The binary star mentioned in Section \ref{sec:binary} has a separation of 6$\farcs4$, which places it outside SOAR's field of view; as a result, it was not detected.

\section{Stellar characterization}\label{sec:Stellar characterization}
\subsection{Spectroscopic parameters}\label{sec:Stellar parameters}

The stellar spectroscopic parameters ($T_{\mathrm{eff}}$, $\log g$, microturbulence, [Fe/H]) were estimated using the ARES+MOOG methodology. The methodology is described in detail in \citet[][]{Sousa-21, Sousa-14, Santos-13, Sousa-08}. For this, we used the latest version of ARES\footnote{The last version, ARES v2, can be downloaded at \url{https://github.com/sousasag/ARES}} \citep{Sousa-07, Sousa-15} to consistently measure the equivalent widths of selected iron lines on the combined ESPRESSO spectrum of TOI-815. Because the star has an effective temperature below 5200 K, we used the appropriate list of iron lines presented in \citet[][]{Tsantaki-13}. In this analysis we used a minimization process to find the ionization and excitation equilibrium to converge for the best set of spectroscopic parameters. This process makes use of a grid of Kurucz model atmospheres \citep{Kurucz-93} and the radiative transfer code MOOG \citep{Sneden-73}. We also derived a more accurate trigonometric surface gravity using recent \textit{Gaia} data following the same procedure as described in \citet[][]{Sousa-21}. 

Using the stellar atmospheric parameters, we determined the abundances of Mg ([Mg/H] = $-0.03$ $\pm$ 0.09 dex) and Si ([Si/H] = $-0.03$ $\pm$ 0.06 dex), closely following the classical curve-of-growth analysis method described in, for example, \citet{Adibekyan-12,Adibekyan-15}. As for the stellar parameter determination, we used ARES to measure the equivalent widths of the spectral lines of these elements, and used a grid of Kurucz model atmospheres \citep{Kurucz-93} and the radiative transfer code MOOG \citep{Sneden-73} to convert the equivalent widths into abundances under assumption of local thermodynamic equilibrium.

Finally, we estimated the projected rotational velocity, $v \sin i_{*}$, by performing spectral synthesis around isolated iron lines in the region of Li line at 6708\,\AA~(discussed in Section \ref{sec:age}). We used the code MOOG and the same model atmosphere used for the abundance determination. The spectral lines are very narrow and, despite the very high resolution of ESPRESSO spectra, we are only able to set an upper limit of $v \sin i_{*}$ < 1 \kms. All the parameters are presented in Table \ref{tab:Stellar parameters}.

\subsection{Mass, radius, and age}\label{sec:age}

We determined the stellar radius of TOI-815 via a modified Markov chain Monte Carlo (MCMC) infrared flux method \citep{Blackwell1977,Schanche2020} by constructing spectral energy distributions using stellar atmospheric models from two catalogs \citep{Kurucz1993,Castelli2003} and the results of our spectral analysis as priors. By comparing synthetic derived from these spectral energy distribution and the observed broadband photometry in the following bandpasses: {\it Gaia} $G$, $G_\mathrm{BP}$, and $G_\mathrm{RP}$, 2MASS $J$, $H$, and $K$, and WISE $W1$ and $W2$ \citep{Skrutskie2006,Wright2010,GaiaCollaboration2022} we derived the bolometric flux of TOI-815. This is converted into effective temperature and angular diameter that is subsequently combined with the offset-corrected \textit{Gaia} parallax \cite{Lindegren2021} to determine the stellar radius. To allow for model uncertainties to propagate to our stellar radius, we used a Bayesian modeling averaging of the radius posterior distributions produced using the aforementioned stellar catalogs.

We determined the stellar mass $M_{\star}$ and age $t_{\star}$ by inputting $T_{\mathrm{eff}}$, [Fe/H], and $R_{\star}$ (Table \ref{tab:Stellar parameters}) along with their uncertainties in the isochrone placement algorithm \citep{bonfanti2015,bonfanti2016}. Following interpolation of the input parameters within precomputed grids of PARSEC\footnote{\textsl{PA}dova and T\textsl{R}ieste \textsl{S}tellar \textsl{E}volutionary \textsl{C}ode: \url{http://stev.oapd.inaf.it/cgi-bin/cmd}} v1.2S \citep{marigo2017} isochrones and evolutionary tracks, we obtained $M_{\star}=0.776\pm0.036\,M_{\odot}$ and $t_{\star}=200_{-200}^{+400}$ Myr, where the error bars were computed by adding 4\% and 20\% in quadrature to the internal uncertainties coming from the interpolation scheme \citep[as justified in][]{bonfanti2021} to account for the isochrone precision. On the one hand, the young age we derived is apparently in disagreement with the low $v\sin{i}$ value. On the other hand, an older main sequence star would have a smaller stellar radius differing more than 3$\sigma$ from our $R_{\star}$ estimate.

\begin{figure}
  \centering
  \includegraphics[width=0.5\textwidth]{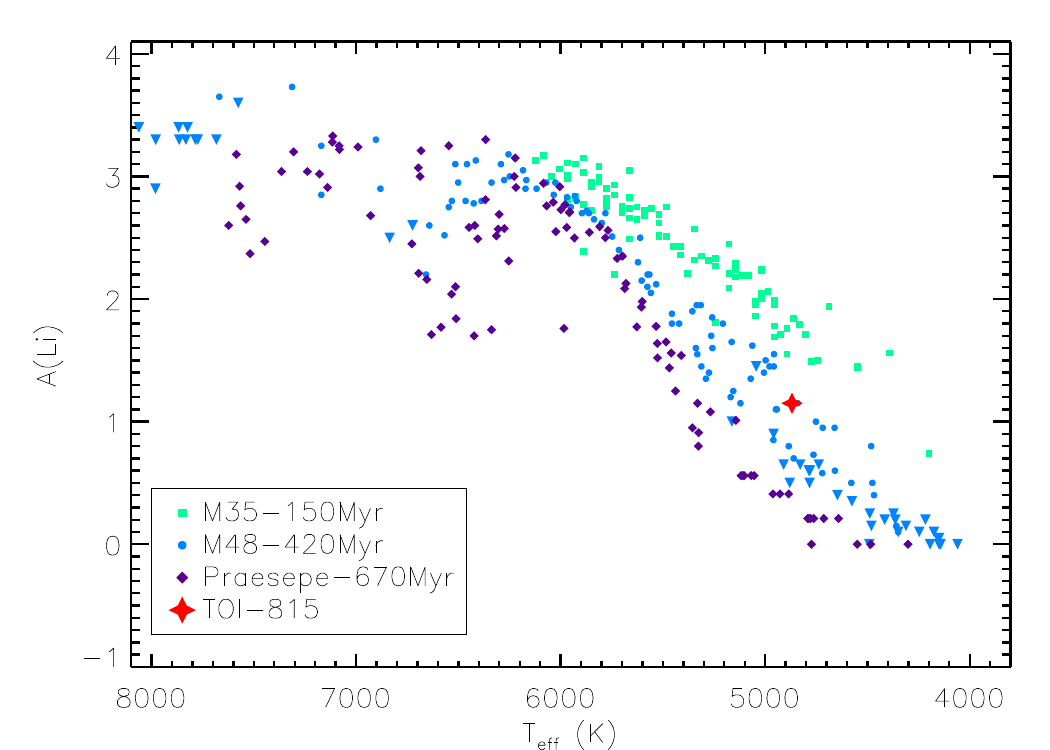}
  \caption{Lithium abundance as a function of \teff. The green, blue, and purple points represent the position of targets in the M35 \citep{sestito05,anthony-twarog18}, M48 (\citealt{sun23}), and Praesepe (\citealt{cummings17}) clusters, respectively. TOI-815 is marked in red.} 
  \label{fig:liclusters}
\end{figure}

To further assess the evolutionary stage of the star, we derived the Li abundance by performing a spectral synthesis (Figure \ref{fig:spectralsynthesis}) using the same model atmospheres and radiative transfer code as to obtain the spectroscopic parameters and abundances \citep[see][for further details]{delgado14}. We obtain a value of A(Li)\footnote{${\rm A(Li)}=\log[N({\rm Li})/N({\rm H})]+12$}\,=\,1.15\,dex, which is very high for stars with such cool temperatures; their thick convective envelopes allow for the mixing of material into interior regions hot enough to burn Li. As a consequence, Li is depleted at early ages for cool stars. The comparison of this Li abundance to that of stars of similar $T_{\mathrm{eff}}$ in open clusters also points to a very young age. In particular, the stars around 4900\,K in the Hyades (with an age of $\sim$600 Myr) have A(Li) below 1\,dex \citep[see Fig. 5 in][]{sestito05}. Similar values are observed in Praesepe, with an age of $\sim$670 Myr \citep{cummings17}. Therefore, TOI-815 should be younger than the Hyades and Praesepe. On the other hand, for clusters of 150-250 Myr (e.g., M35, NGC 6475, and M34) analyzed in \citep{sestito05,anthony-twarog18}, stars with $T_{\mathrm{eff}}\sim$4900~K have Li around 1.5\,dex. The recent work on M48, a 420\,Myr old cluster \citep{sun23} shows that stars with effective temperatures near 4900\,K have lower Li values (around 1\,dex) than TOI-815, adding an additional constrain for an age younger than $\sim$400\,Myr. The Li abundance of TOI-815 is shown with those clusters as comparison in Figure \ref{fig:liclusters}.

To gain a more precise understanding of TOI-815's age, we utilized a recently developed kinematic-based method \citep{Maciel2011,AlmeidaFernandes2018} that relies on a higher probability of kinematic disturbances and thus stellar Galactic velocity relative to known kinematic families over the lifetimes of stars \citep{Wielen1977,Nordstrom2004,Casagrande2011}. We compared the Galactic $U$, $V$, and $W$ velocities of TOI-815 to kinematic-age probability distributions bench-marked using a sample of 9000 stars with well-known isochronal ages \cite{AlmeidaFernandes2018} and determine a nominal age of $\sim$470\,Myr.
Finally, the gyrochronological relations of \citet{Barnes2007} and the rotation period of $P_{\rm rot}$ = 15.3 $\pm$ 1.2 days constrained from the photometry (Section \ref{sec: Stellar rotation period}) suggest an age of 660 $\pm$ 150 Myr.

\subsection{Stellar rotation}\label{sec: Stellar rotation period}
Using the ESPRESSO RVs, we measured a median logR$'_{\mathrm{HK}}$ = -4.67$\pm$ 0.03, which suggests that TOI-815 is a quite active \citep[see Fig. 7 in][]{Henry1996} star. Visual inspection of the raw TESS transit light curves shows a sinusoidal variation that is likely attributed to the stellar activity (Figure \ref{fig:fullTESSLCs}). For our analysis, we used the generalized Lomb-Scargle (GLS; \citealt{Zechmeister2009}) on the TESS SAP photometry of Sector 63 and we identified a significant periodic signal with a period of $\sim$16.5 days, which could be associated with the stellar rotation (Figure \ref{fig:gls}). TOI-815 was also observed by the Wide Angle Search for Planets (WASP; \citealt{2006PASP..118.1407P}) transit survey in multiple years between 2006 and 2014, with the observing season in each year spanning $\sim$\,120 nights. A total of 86\,000 photometric data points was obtained using 200-mm, f/1.8 Canon lenses until 2012 and 85 mm f/1.2 Canon lenses after that.  TOI-815 is by far the brightest star in the extraction aperture. In the years 2008, 2012, and\ 2013 we detect a significant periodicity with a period of 14.2-15.7 days.  In each of those years the estimated false alarm probability (FAP) is below 1\%. By combining the TESS Sector 63 and WASP photometric data, we estimate the stellar rotation to be 15.3~$\pm$~1.2 days.

We repeated the frequency analysis for the ESPRESSO RV measurements, as well as for the Ca\hspace{0.04cm}II\hspace{0.04cm}H\hspace{0.04cm}\&\hspace{0.04cm}K activity index (logR$'_{\mathrm{HK}}$) and the cross-correlation-function asymmetry indicators (FWHM and bisector). The GLS periodogram of the RV measurements shows a strong peak at the orbital frequency of TOI-815c ($P_{\mathrm{orb,c}}$ $\sim$ 35 days) with a FAP < 1\%. Following the subtraction of the outer planet's signal, we found a significant peak at the orbital period of TOI-815b ($P_{\mathrm{orb,b}}$ $\sim$ 11.2 days) with a FAP $\approx$ 1\%. Moreover, the stellar rotation period has no counterpart in the GLS periodograms of the RVs or the activity indicators.


The most straightforward way to estimate the stellar inclination is to assume solid body rotation \citep[e.g.,][]{Borucki1984,Doyle1984}:
\begin{equation}
    \label{eq:sinis}
    \sin i_\star = \frac{v \sin i_\star}{v} = \frac{v \sin i_\star}{2 \pi R_\star/P_\mathrm{rot}},
\end{equation}

\noindent where $P_{\rm rot}$ is the stellar equatorial rotation period. However, directly using Eq.~(\ref{eq:sinis}) to generate an $i_\star$ distribution is hazardous, because $v \sin i_\star$ and $P_{\rm rot}$ are correlated quantities \citep{Masuda2020}. To avoid such interdependence biases, we used the same procedure as \citet{Attia2023} to derive the stellar inclination. We fitted for $v \sin i_\star$ with a MCMC \citep[using the \texttt{emcee} package,][]{Foreman2013} based on the following formulation of Eq.~(\ref{eq:sinis}),
\begin{equation}
\centering
\label{eq:vsini}
    v \sin i_\star = \frac{2 \pi R_\star}{P_\mathrm{rot}} \sqrt{1 - \cos^2 i_\star},
\end{equation}

\noindent and using a semi-Gaussian distribution for $v \sin i_\star$ (since we only have an upper limit) as a constraint for the computation of the likelihood. We set $R_\star$, $P_{\rm rot}$, and $\cos i_\star$ as independent jump parameters with measurement-informed Gaussian priors on $R_\star$ and $P_{\rm rot}$ (Table \ref{tab:Stellar parameters}), and a uniform prior on $\cos i_\star$ (i.e.,~assuming an isotropic stellar inclination distribution). The number of walkers and the burn-in phase are adjusted to ensure the convergence of the chains. The ensuing probability distribution function (PDF) yields $i_\star = 32^{+15}_{-23}~^\circ$, the posterior estimate corresponding to the median value and the error bars to the 68\% highest density interval. This result indicates that the star is spinning faster ($v = 2.5 \pm 0.2$ \kms) but we are observing it close to pole-on.

Additionally, we used the TESS photometric data to constrain the fraction of stellar surface covered with spots. For this, we used the \texttt{SAGE}\footnote{\url{https://github.com/chakrah/sage}} tool to obtain flux variability models for a given spot configuration. We fitted the observed variability using the same procedure as described in \citet{Chakraborty2023}, with uniform priors on latitude, longitude and size of active regions. The wide priors allow spots to be present over the entire stellar surface and the spot sizes vary from small Sun-like spots (1$^{\circ}$) to large M-dwarf-type spots (15$^{\circ}$). To limit the degeneracy on constraining the parameters of active regions, we fixed the stellar inclination (to 32$^{\circ}$) and spot contrast in the TESS bandpass (to 0.2691). We iteratively increased the number of spots from 0 to 4 and found that a 1-spot model is most probable and has the lowest BIC. We conclude that at least 0.26$^{+0.20}_{-0.05}$ \% of the stellar surface is covered with spots to match the observed photometric variability. We also estimated the impact of these variations on the radius determination of the planets to be below 0.31\%.

\begin{table}
\small
\caption{Stellar parameters for TOI-815.}
\centering
\renewcommand{\arraystretch}{1.2}
\begin{tabular}{lcc}
\hline\hline
  & TOI-815  & Source \\
    \hline
    Identifiers& &\\
    TIC ID  & 102840239 & TICv8\\
    2MASS ID & J10232924-4350059 &  2MASS\\
    \textit{Gaia} ID & 5415648821879172096 & \textit{Gaia} DR3\\
    TYC &  7721-01451-1 & Tycho-2\\
    \hline
    Astrometric parameters & &\\
    Right ascension (J2016),  $\alpha$ & 10$^{\mathrm{h}}$ 23$^{\mathrm{m}}$ 29.25$^{\mathrm{s}}$ &\textit{Gaia} DR3\\
    Declination (J2016), $\delta$ & $-43^{\circ}$ 50$^{\prime}$ 05.84$^{\prime \prime}$&\textit{Gaia} DR3\\
    Parallax (mas) & 16.825 $\pm$ 0.013 &\textit{Gaia} DR3\\
    Distance (pc) & 59.437 $\pm$ 0.045 &\textit{Gaia} DR3\\
    $\mu_{\rm{R.A.}}$ (mas yr$^{-1}$) & 8.898 $\pm$ 0.011 &\textit{Gaia} DR3\\
    $\mu_{\rm{Dec}}$ (mas yr$^{-1}$) &7.248 $\pm$ 0.013 &\textit{Gaia} DR3\\
    \hline
    Photometric parameters & &\\
    TESS (mag) & 9.360 $\pm$ 0.006 & TICv8\\
    B (mag) &11.145 $\pm$ 0.061 & Tycho-2\\
    V (mag) &10.217 $\pm$ 0.005& Tycho-2\\
    G (mag) &9.9398 $\pm$ 0.0004&\textit{Gaia} DR3\\
    J (mag) &8.531 $\pm$ 0.024 & 2MASS\\
    H (mag) &8.066 $\pm$ 0.029 & 2MASS\\
    K (mag) &7.999 $\pm$ 0.029 & 2MASS\\
    \hline
    Bulk parameters & &\\
    Spectral type & K3 & Sec. $\ref{sec:Stellar parameters}$\\
    \teff\,(K) & 4869 $\pm$ 77& Sec. $\ref{sec:Stellar parameters}$\\
    $R_\star$  (\rsol)  & 0.770 $\pm$ 0.009 & Sec. $\ref{sec:age}$\\
    $M_\star$  (\msol)  & 0.776 $\pm$ 0.036 &Sec. $\ref{sec:age}$\\
    \feh (dex) & $-0.09$ $\pm$ 0.05&Sec. $\ref{sec:Stellar parameters}$\\
    \sih (dex) & $-0.03$ $\pm$ 0.06&Sec. $\ref{sec:Stellar parameters}$\\
    \mgh (dex) & $-0.03$ $\pm$ 0.09&Sec. $\ref{sec:Stellar parameters}$\\
    $\log g_*$ (cm\,s$^{-2}$) & 4.51 $\pm$ 0.04 & Sec. $\ref{sec:Stellar parameters}$\\
    $v \sin i_{*}$ (km\,s$^{-1}$) & $<$ 1 &Sec. $\ref{sec:Stellar parameters}$\\
    $P_{\rm rot}$ (days) &15.3 $\pm$ 1.2 &Sec. $\ref{sec: Stellar rotation period}$\\
    $i_\star$ ($^\circ$) & 32$^{+15}_{-23}$ &Sec. $\ref{sec: Stellar rotation period}$\\
    Age	(Myr) & $200 ^{+400} _{-200}$&Sec. $\ref{sec:age}$\\
    A (Li) (dex) & $1.15 \pm 0.05$&Sec. $\ref{sec:age}$\\
   \hline
\end{tabular}
\\
\begin{tablenotes}
\item Sources: 1) TICv8 \citep{Stassun2019} 2) 2MASS \citep{Skrutskie2006} 3) \textit{Gaia} DR3 \citep{GAIADR3} 4) Tycho-2 \citep{Tycho}.
\end{tablenotes}
\label{tab:Stellar parameters}
\end{table}
\begin{table}[h]
\tiny
\caption{Fitted and derived parameters for TOI-815b and TOI-815c.}
\centering
\renewcommand{\arraystretch}{1.5}
\setlength{\tabcolsep}{1pt}
\begin{center}
\begin{tabular}{lcc}
\hline\hline
\textbf{Parameter} &\textbf{TOI-815b} & \textbf{TOI-815c} \\
 \hline
     Orbital period, $\it{P_{\mathrm{orb}}}$ (days)\dotfill  & $ 11.197259^{+0.000018} _{-0.000017}$&$ 34.976145^{+0.000099} _{-0.000097}$ \\
    Planet radius, $\it{R_{\mathrm{P}}}$ (\re)\dotfill  & $2.94 ^{+0.05} _{-0.05}$&$ 2.62^{+0.10} _{-0.09}$ \\
    Planet mass, $\it{M_{\mathrm{P}}}$ (\me)\dotfill  & $7.6 ^{+1.5} _{-1.4}$  &$23.5 ^{+2.4} _{-2.4}$  \\
    Planet density, $\rho_\mathrm{P}$ (g~cm$^{-3}$)\dotfill  & 1.64$ ^{+0.33} _{-0.31}$ & 7.2$ ^{+1.1} _{-1.0}$\\
    Orbital inclination, $i$ (degrees)\dotfill & $89.36 ^{+0.17} _{-0.14}$  &$89.107 ^{+0.026} _{-0.026}$  \\
    Impact parameter, $\textit{b}$\dotfill  &$ 0.279^{+0.058} _{-0.073}$ & $ 0.839^{+0.015} _{-0.016}$ \\
    Semi-major axis, $\it{a}$ (au)\dotfill  & $0.0903 ^{+0.0018} _{-0.0019}$&$0.193 ^{+0.004} _{-0.004}$\\
    Transit duration, $T_{\mathrm{14}}$ (hours)\dotfill  & $3.379 ^{+0.023} _{-0.023}$&$2.97 ^{+0.09} _{-0.10}$ \\
    Scaled semi-major axis, $\it{a/\rstar}$\dotfill  & 25.2$^{+0.6} _{-0.6}$&53.9$^{+1.3} _{-1.3}$\\
    Equilibrium temperature$^{(a)}$, $T_{\mathrm{eq}}$ (K)\dotfill  & $686 ^{+13} _{-14}$ &$469 ^{+9} _{-9}$\\
    Atmospheric scale height$^{(b)}$, $H$ (km)\dotfill  & 288$ ^{+58} _{-54}$& $ 51^{+7} _{-6}$\\
    Insolation$^{(c)}$, $\it{S_{\mathrm{p}}}$ (\se)\dotfill  &  36.9$ ^{+2.9} _{-2.9}$& 8.03$ ^{+0.64} _{-0.64}$\\
    Restricted Jeans escape param.$^{(d)}$, $\Lambda$ \dotfill  &  28.6$ ^{+5.7} _{-5.3}$& 145$ ^{+16} _{-16}$\\
      \hline
\end{tabular}
\begin{tablenotes}
\item
\textbf{Notes:} $^{(a)}$ The equilibrium temperature is calculated using $T_{\mathrm{eq}}$ = \teff(1-A)$^{1/4}\sqrt{\frac{R_{*}}{2\textit{a}}}$, assuming a Bond albedo of A = 0. $^{(b)}$ The scale height is calculated using $H$ = $\frac{k_{b}T_{\mathrm{eq}}}{\mu g}$ assuming a mean molecular weight of 2.3 g/mol. $^{(c)}$ Insolation is calculated using $\it{S_{\mathrm{p}}}$ = $\frac{L_{*}}{4\pi\textit{a}^{2}}$. $^{(d)}$ The restricted Jeans escape parameter is calculated using $\Lambda=\frac{G M_{\mathrm{p}} m_{\mathrm{H}}}{k_{\mathrm{B}} T_{\mathrm{eq}} R_{\mathrm{p}}}$ (\citealt{Fossati2017}).
\end{tablenotes}
\label{tab:planetaryparams}
\end{center}
\end{table}

\section{Photometric and RV analysis}\label{sec:Joint transit and RV analysis}
\subsection{TESS-only modeling}\label{sec:TESSonly}
\begin{figure}
  \centering
  \includegraphics[width=0.5\textwidth]{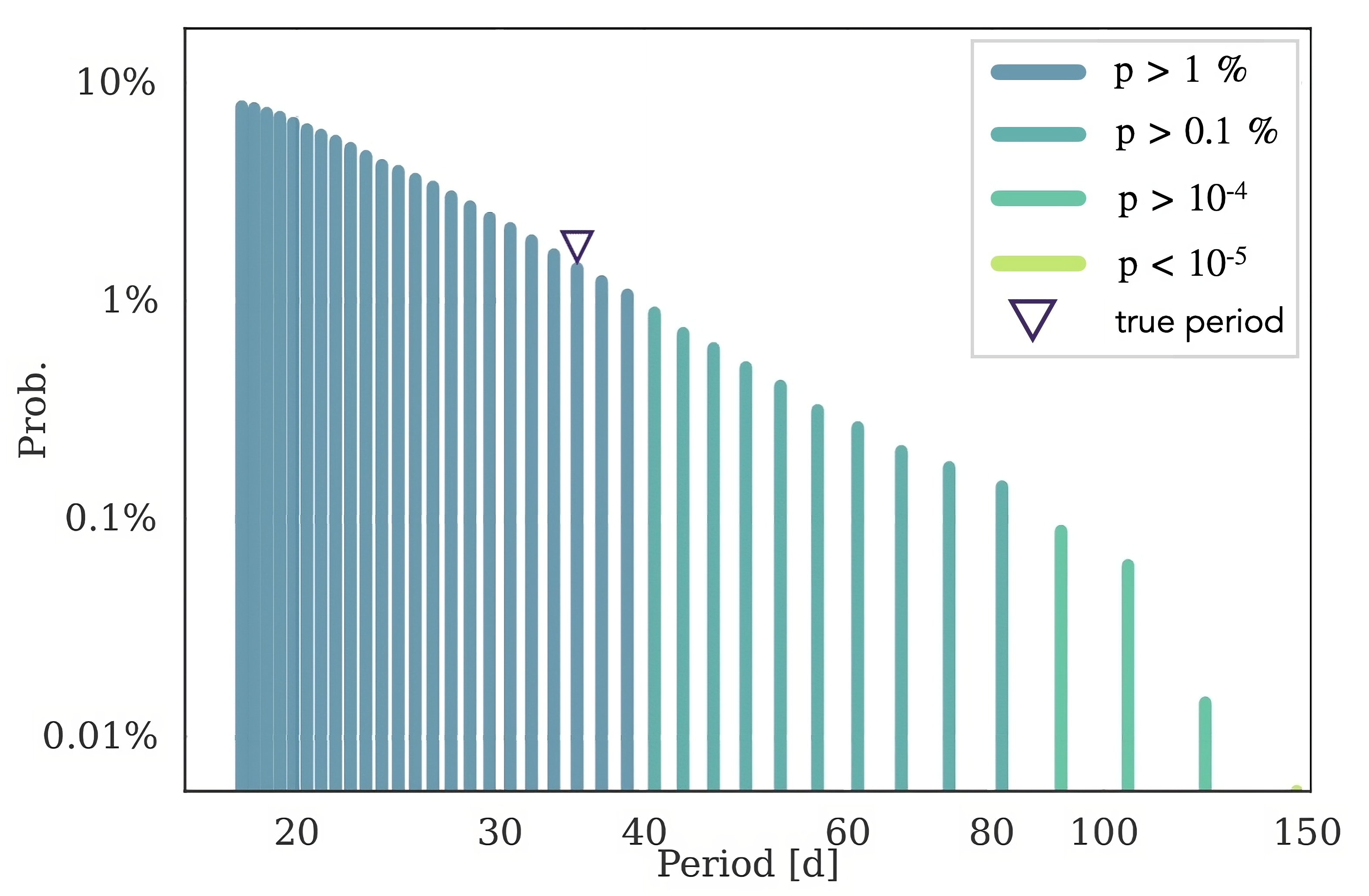}
  \caption{Marginalized probabilities for each of the period aliases of TOI-815c as calculated by \MonoTools from TESS-data only. The triangle indicates the true period of the planet.} 
  \label{fig:monotools}
\end{figure}

As mentioned in Section $\ref{sec:TESS_photometry}$, visual inspection of the light curves of TOI-815 in Sectors 9 and 36 revealed two similar single transits incompatible with the known P = 11.2~days inner planet. The two transits were separated by a gap of 734.5~days, which initially led to 41 separate period aliases between 17.49~days ($P_{\rm min}$) and 734.5~days ($P_{\rm max}$), where period on aliases shorter than this value would have produced a detectable third transit in TESS photometry. In order to determine the most likely periods, we modeled the system using \MonoTools\footnote{\url{https://github.com/hposborn/MonoTools}} \citep{Osborn2022b}. This uses a period-agnostic fit of the two transit events, combined with priors from period, eccentricity and system stability to determine a marginalized posterior probability for each period. In the case of TOI-815c, we found that the innermost aliases are most likely, with the 23 periods shorter than P = 40~days having probabilities greater than 1$\%$ (see Figure \ref{fig:monotools}).

\subsection{CHEOPS-only modeling}\label{sec:CHEOPSonly}
\begin{figure*}[ht]
  \centering
    \includegraphics[width=0.323\textwidth]{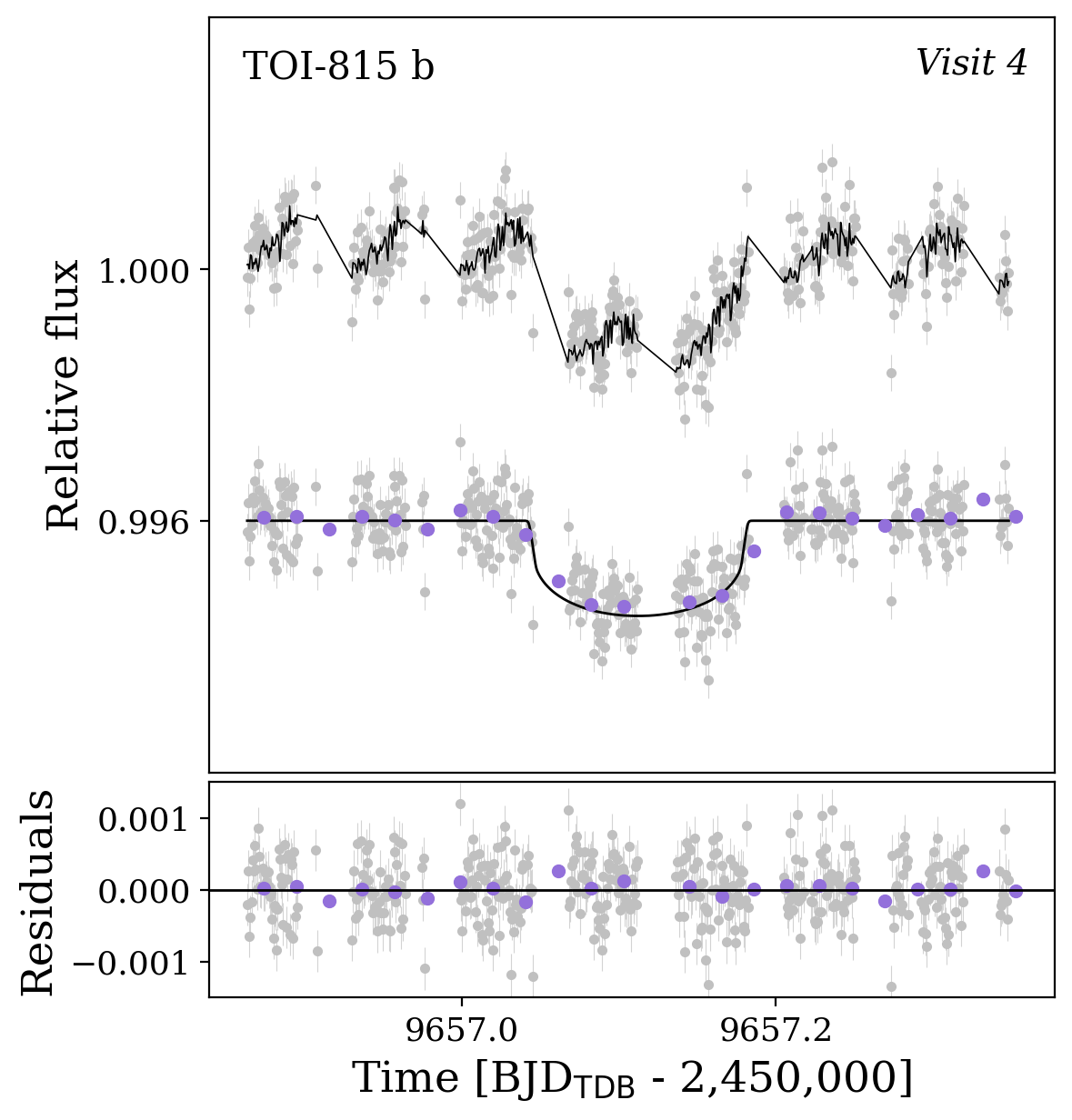}
    \includegraphics[width=0.33\textwidth]{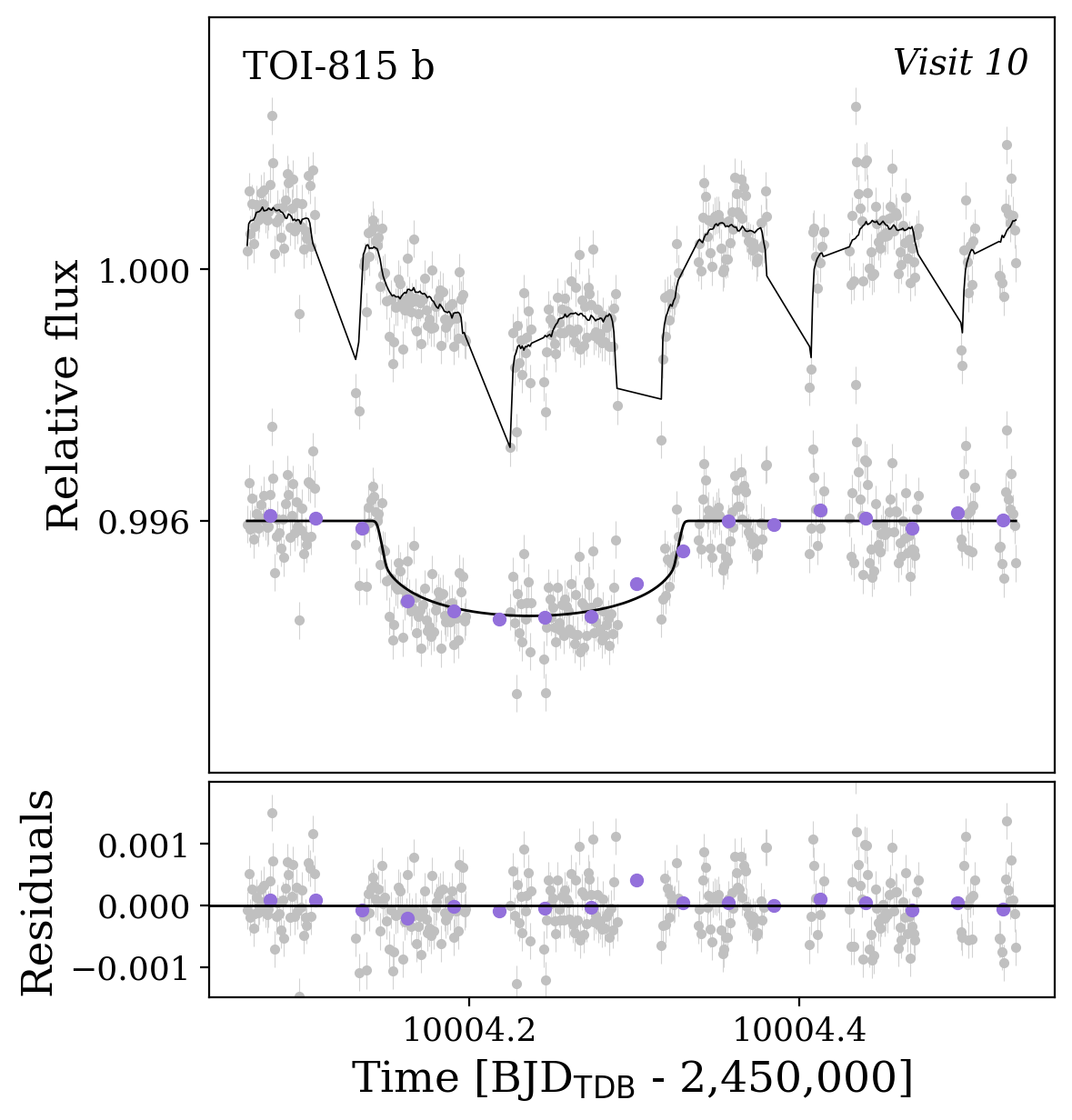}
    \includegraphics[width=0.338\textwidth]{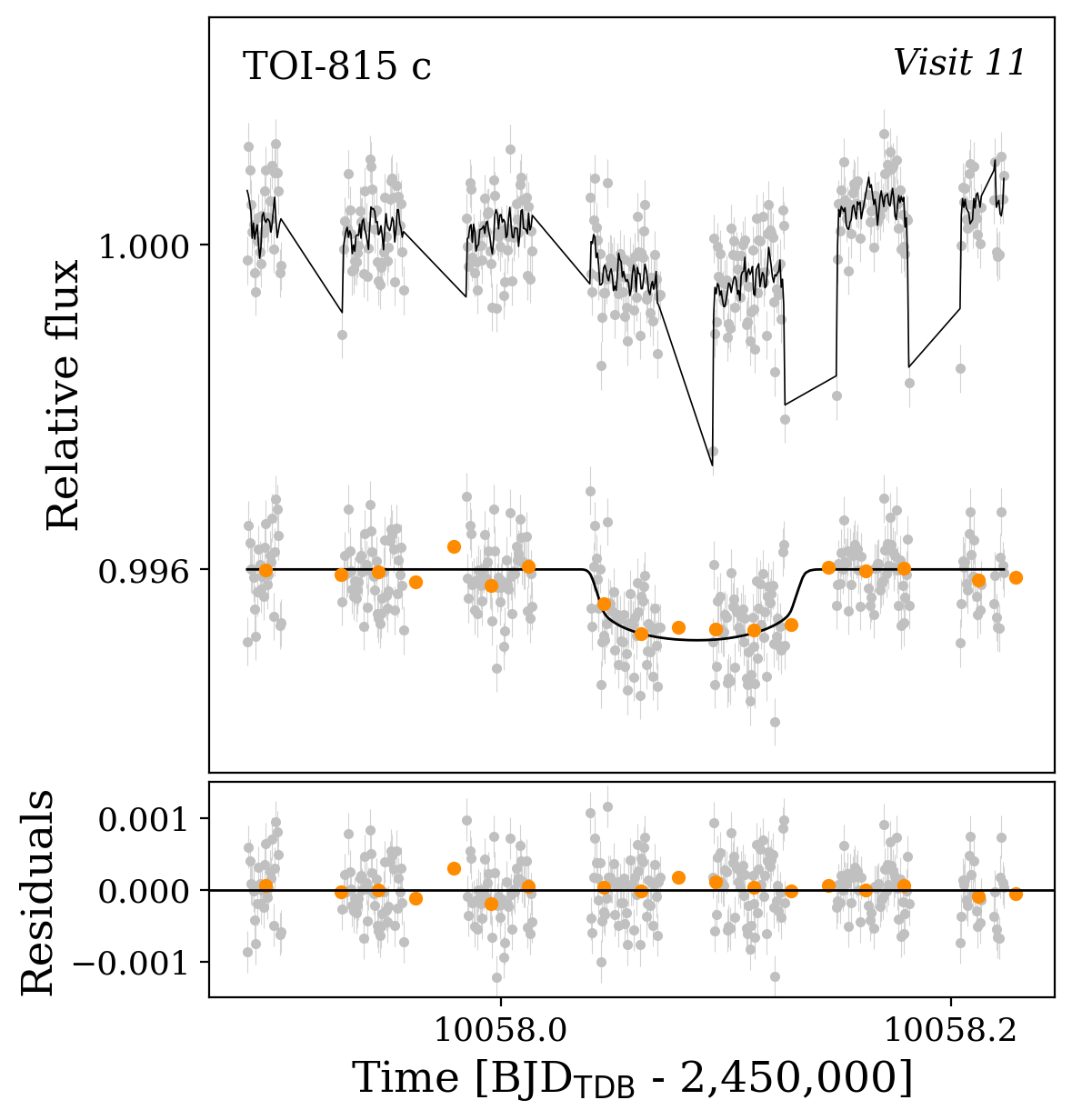}
  \caption{CHEOPS light curves of TOI-815b (visits 4 and 10) and TOI-815c (visit 11). The top light curves in the upper panel present the raw flux and the best-fit de-correlation model. The detrending terms are presented in Table \ref{tab:cheopsobservations}. The bottom light curves in the upper panel present the detrended flux, with the best-fit transit model overlaid in black. The residuals of the models are shown in the lower panel. The data are binned to 30 minutes (purple and orange circles).} 
  \label{fig:CHEOPS} 
\end{figure*}

We initially scheduled 13 of the most-likely period aliases of TOI-815c (all P < 25~days) with CHEOPS at varying priorities. This resulted in five observations in 2022, light curves for which are noted in Table \ref{tab:cheopsobservations} and can be seen in Figure \ref{fig:ALLCHEOPS}. The most important observation occurred on 2022 March 17 (visit 3); with this observation, due to its position exactly 367.25~days after the last transit (i.e., $t_{1} + P_{\rm max}/2$), we were able to rule out half of all aliases. The CHESS program (PR110048) also observed a single observation of TOI-815b, which showed little to no measurable TTVs. Observations resumed in 2023, with a further seven aliases scheduled, of which five were observed. One such observation, on 2023 February 28 (visit 10), was serendipitously during the transit of planet b, adding a second CHEOPS transit. On 2023 March 19, TESS re-observed a third transit of TOI-815c. However, this occurred exactly 734.5~days after the previously observed transit (i.e., $t_{2} = t_{1} + P_{\rm max}$), and therefore no period information was gained from the observation of this transit. 
We extracted a quick-look light curve using TICA \citep{fausnaugh2020calibrated} in order to verify the position of this third transit and updated the ephemeris of scheduled aliases. Thanks to these observations, very few aliases with P $<$ 40~days remained at this point. On 2023 April 23 (visit 11), a transit of TOI-815c at a unique transit time for the 34.97~days alias was finally observed with CHEOPS. Once a transit had been found all other observations were stopped. The CHEOPS light curves of TOI-815b and c are presented in Figure \ref{fig:CHEOPS}.
\subsection{ESPRESSO-only modeling}\label{sec:ESPRESSOonly}
As discussed in Section \ref{sec:Spectroscopic follow-up}, we conducted a spectroscopic follow-up of TOI-815 using ESPRESSO over a period of 20 months. In Figure \ref{fig:gls} we show that based on these observations we detected TOI-815c with a statistically significant signal at 35~days (FAP < 1\%) and marginally detected TOI-815b at 11.2~days (FAP $\sim$ 1\%). After fixing the transit ephemerides, the signal of TOI-815c and b was detected at a 10$\sigma$ and 5$\sigma$ level, respectively.

\subsection{Joint analysis}\label{sec:joint}
We performed an analysis of the planetary system around TOI-815 by combining photometric and RV data. To jointly model the datasets, we used the \juliet software package, which is a Bayesian framework capable of modeling multi-planet systems. \Juliet is an openly accessible tool that enables the simultaneous fitting of both transit models (using the batman package, \citealt{batman}) and RVs (via the radvel package, \citealt{radvel}). It incorporates GPs to model noise (via the \celerite package; \citealt{celerite}). To explore the parameter space, the code uses nested sampling algorithms with dynesty (\citealt{dynesty}) and conducts model comparison through Bayesian evidence (lnZ) using MultiNest (\citealt{multinest}) in conjunction with the PyMultiNest (\citealt{pymultinest}) Python software package.

\Juliet employs an efficient parametrization technique for the transit model by fitting the parameters r1 and r2, which are a combination of the planet-to-star ratio (p) and impact parameter (b). This approach ensures a uniform exploration of the (p, b) space. Furthermore, \juliet estimates the stellar density $\rho_{*}$, which, when combined with the orbital period, determines the scaled semimajor axis ($\it{a/\rstar}$).

To account for TESS photometric variability, we included a GP for Sectors 36 and 63 using a Mat\'ern 3/2 kernel with hyperparameters amplitude ($\sigma_\mathrm{GP}$) and timescale ($\rho_\mathrm{GP}$) as shown in Figure $\ref{fig:fullTESSLCs}$. The CHEOPS light curves exhibit correlations with certain instrumental parameters like contamination by background stars (bg), centroid movements (x, y) and the change in temperature of the telescope tube $\Delta$T. Additionally, periodic noise features occur once per CHEOPS orbit due to the satellite's rolling motion around its pointing direction. To account for these periodic instrumental effects, we detrended the light curves by considering the sine or cosine of the roll angle ($\phi$). In our global analysis, we incorporated linear models to account for both instrumental trends and stellar variability. However, it was essential to select only the relevant detrending parameters for each specific CHEOPS observation. To achieve this, we employed the \pycheops package (\citealt{pycheops}) that provided us with the optimal combination of detrending terms for each visit. The detrending terms for each CHEOPS visit are listed in Table \ref{tab:cheopsobservations}.

The stellar density was fitted using a Gaussian prior based on the stellar spectroscopic analysis conducted in Section \ref{sec:Stellar characterization}. We derived the quadratic stellar limb-darkening coefficients and their uncertainties for each photometric filter used using the \ldcu$\footnote{\url{https://github.com/delinea/LDCU}}$ code (\citealt{Deline2022}). The \ldcu code computes the limb-darkening coefficients and their corresponding uncertainties by considering a set of stellar intensity profiles and accounting for the uncertainties on the stellar parameters. In our global analysis, we utilized the determined limb-darkening coefficients from \ldcu as Gaussian priors for all TESS, CHEOPS, LCOGT, and ASTEP data. 

The SPOC crowding metric calculated for TOI-815 in Sectors 36 and 63 is $\sim$0.85. This indicates that, according to SPOC modeling after eliminating background noise, $\sim$85$\%$ of the flux detected within the photometric aperture originated from the target star. The remaining 15$\%$ was attributed to various other sources, with source 2 (TIC 102840237; Section \ref{sec:binary}) being particularly significant. To compensate for contamination by these additional sources, we fixed the dilution factor for all the light curves to 0.85. 

In order to prevent any potential Lucy-Sweeney bias in the eccentricity measurement (\citealt{Lucy1971}), we fixed the orbital eccentricity to zero. However, to explore the possibility of noncircular orbits, we ran a separate analysis without any constraints on the eccentricity. With the eccentricity fitted, we found a value of e = $ 0.110^{+0.072} _{-0.065}$ for TOI-815b and $ 0.035^{+0.064} _{-0.016}$ for TOI-815c. Therefore, the condition $e$ $>$ 2.45$\sigma_\mathrm{e}$ is not satisfied, which suggests that the RV data are compatible with a circular orbit. Additionally, in Table \ref{tab:posteriors} we show that the 3$\sigma$ upper limit on the eccentricity for TOI-815b and TOI-815c is 0.20 and 0.22, respectively.

\section{Discussion}\label{sec:Discussion}
\begin{figure*}[ht]
  \centering
  \includegraphics[width=0.9\textwidth]{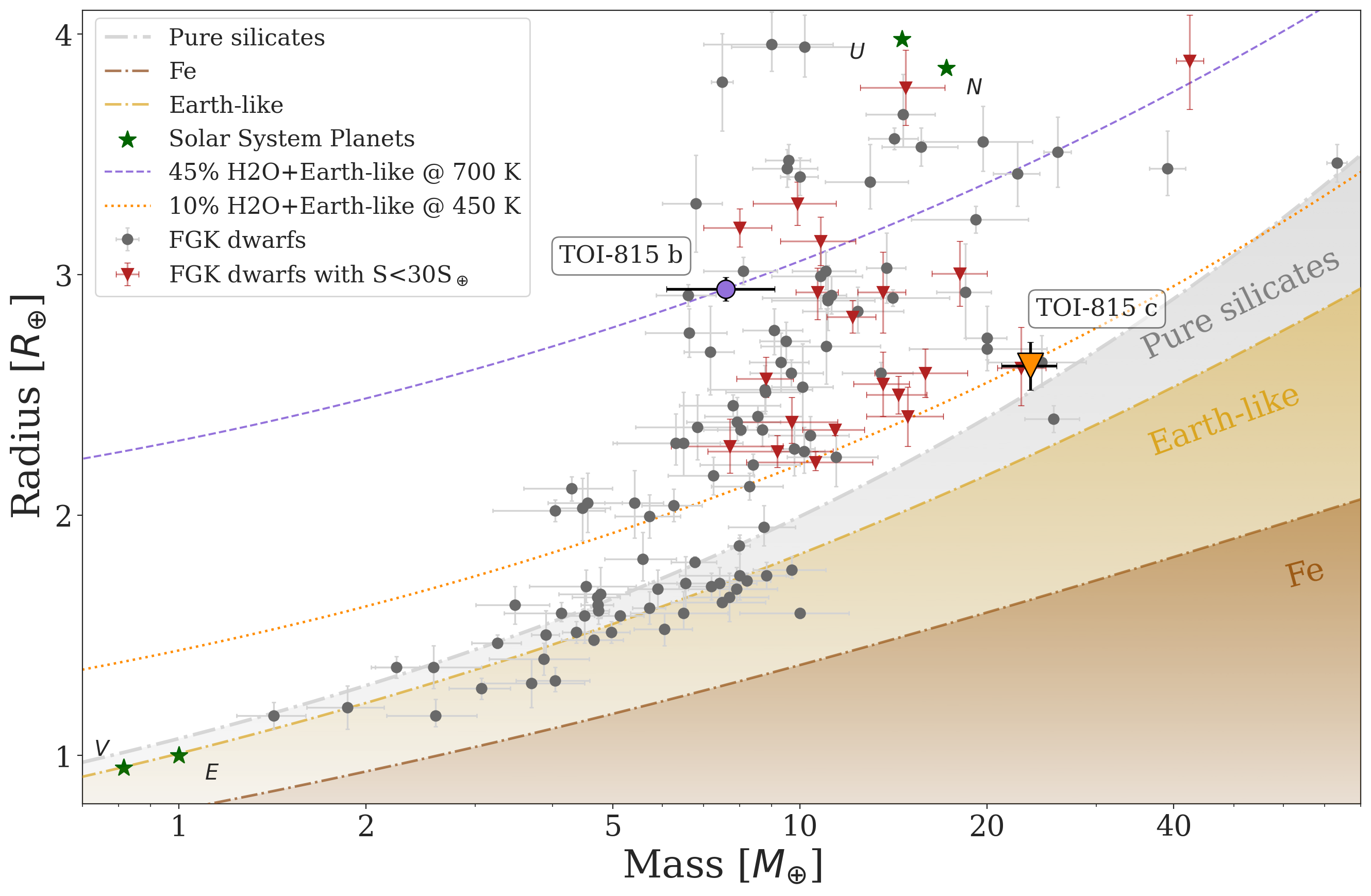}
  \caption{Mass-radius diagram of small exoplanets (with radii ranging from 2 to 4 \re) with precise densities\textsuperscript{\ref{precisedensity}} from the PlanetS catalog\textsuperscript{\ref{dacefootnote}} orbiting FGK stars. The red triangles correspond to exoplanets with stellar irradiation lower than 30 \se. The composition lines of pure silicates (gray) from \citet{Zeng2016}, Earth-like planets (light-brown), and pure iron (brown) from \citet{Dorn2015} are displayed. The purple and orange points represent TOI-815b and c, respectively. Two composition lines (\citealt{Aguichine2021}) that incorporate both water and terrestrial elements matching the equilibrium temperatures of the two planets are plotted as purple dashed and orange dotted lines.} 
  \label{fig:radiusmass}
\end{figure*}
We report the discovery and characterization of two transiting mini-Neptune planets around TOI-815, a young K3V star. TOI-815b is a 11.2~days period planet with a radius of 2.94 $\pm$ 0.05 \re and a mass of 7.6 $\pm$ 1.5 \me. The outermost planet, TOI-815c, has an orbital period of 35~days, a radius of 2.62 $\pm$ 0.10 \re and a mass of 23.5 $\pm$ 2.4 \me. Based on the stellar parameters (Table \ref{tab:Stellar parameters}), we find that the semimajor axis of TOI-815b and TOI-815c are 0.0903 $\pm$ 0.0019 au and 0.193 $\pm$ 0.004 au, respectively. Assuming a zero albedo and full heat redistribution ($f$ = 1), the equilibrium temperature of the inner planet is $686 ^{+13} _{-14}$~K and its stellar insolation is 36.9 $\pm$ 2.9 \se. The outer planet has an equilibrium temperature of $469 ^{+9} _{-9}$~K and receives a total amount of insolation equal to 8.03 $\pm$ 0.64 \se. 

Both planets lie within the population of sub-Neptunes that are transiting FGK stars, with 87 well-characterized exoplanet detections\textsuperscript{\ref{precisedensity}} (including TOI-815b and TOI-815c) from the PlanetS catalog\textsuperscript{\ref{dacefootnote}}. TESS has contributed in this sample with 46 discoveries during the past five years (\citealt{Naponiello2023}). Moreover, TOI-815c adds to the population of warm mini-Neptune-sized planets with low stellar irradiation (less than 30 \se); with its inclusion, the total number of well-characterized detections is now 21. Figure \ref{fig:radiusmass} places TOI-815b and TOI-815c in the mass-radius diagram of small exoplanets (2-4~\re) with precise densities orbiting FGK stars (gray points). The red triangles correspond to exoplanets with stellar irradiation lower than 30~\se. The composition lines of pure silicates (\citealt{Zeng2016}), Earth-like planets, and pure iron (\citealt{Dorn2015}) are displayed in gray, light-brown, and brown, respectively. We also plot two other composition lines from \citet{Aguichine2021} as dashed and dotted black lines. For TOI-815b, we plot a line at 700~K, corresponding to its equilibrium temperature, with a composition of 45\% of water and the rest as an Earth-like composition (32.5\% iron-67.5\% silicates). Since TOI-815c is denser and less irradiated, we can see that it lies on the composition lines with 10\% water and an Earth-like composition at 450~K, its equilibrium temperature. Finally, we can say that if TOI-815b is rich in water for the model under consideration, its atmosphere is inflated by its irradiation, giving it a large radius and therefore a low density for a quantity of water of around 45\%. TOI-815c, on the other hand, would have a lower water content of less than 10\%, and its lower irradiation would not have as much influence as its neighbor on its radius, which would explain its higher density. Of course, this remains one possible model, and it is worth remembering that the position of a planet in the mass-radius diagram does not uniquely determine its composition. These planets could also have atmospheres rich in H-He or heavier materials. A more detailed discussion of the internal structures can be found in Section \ref{sec:internalstructure}.
\subsection{Density discrepancy of lowly and highly irradiated mini-Neptunes}\label{sec:densities}
\begin{figure}
\centering
  \includegraphics[width=0.45\textwidth]{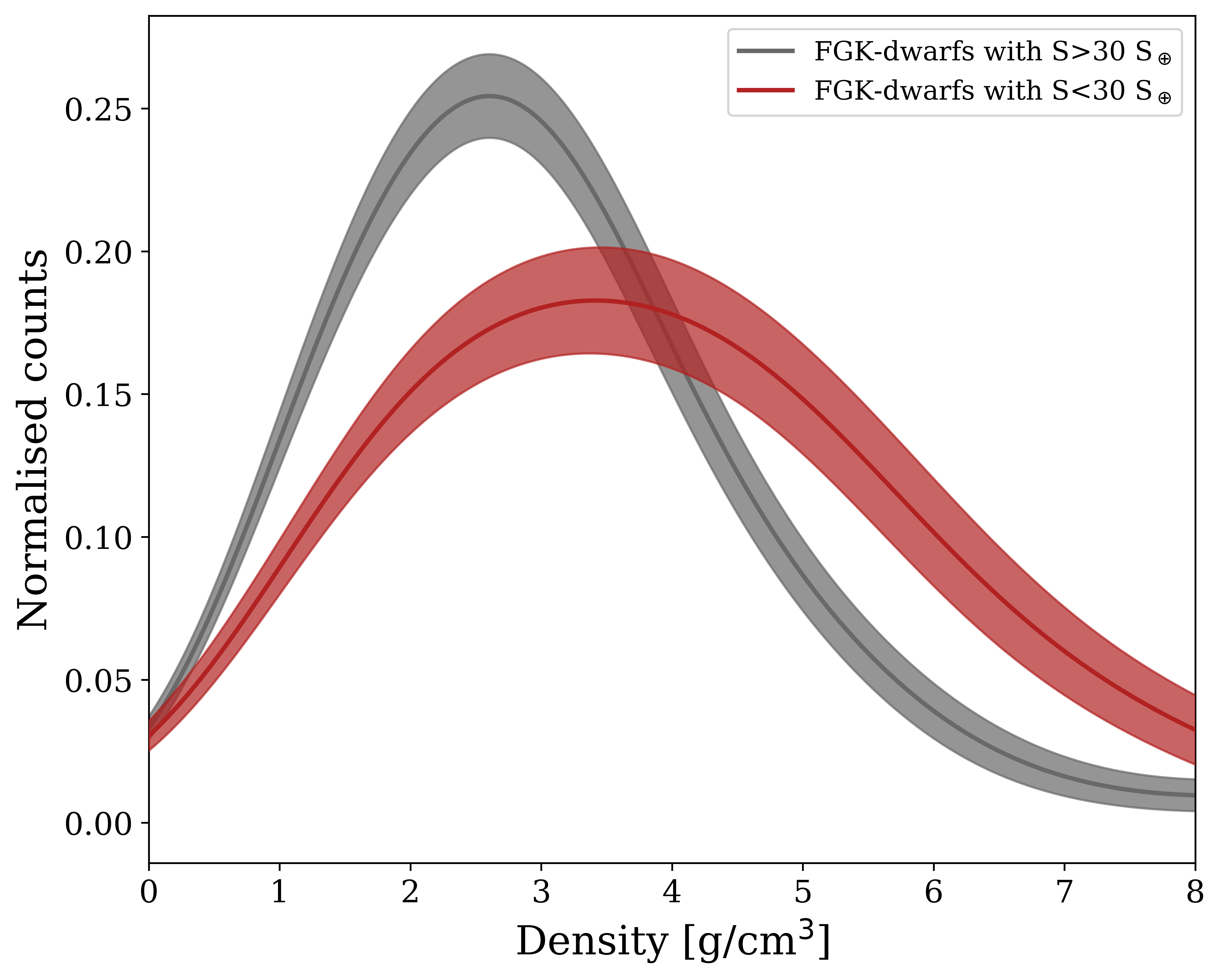}
  \caption{KDE for the two groups of sub-Neptunes (2-4 \re): with an irradiation above 30 \se (gray) and below 30 \se (red). The solid line represents the mean of the KDEs calculated with the 10, 000 samples drawn, and the transparent region represents the standard deviation.} 
  \label{fig:kde}
\end{figure}

 TOI-815c, with a bulk density of 7.2 $\pm$ 1.1~\gccc, becomes the densest well-characterized warm mini-Neptune (2 - 4 \re and $\leq$ 30 \se), with GJ 143b ($\rho_\mathrm{P}$ = 7.0 $\pm$ 1.6~\gccc and $\it{S_{\mathrm{p}}}$ = 6.0 $\pm$ 0.4 \se; \citealt{Dragomir2019}) being the second. Also, TOI-815c is the third densest well-characterized mini-Neptune (2 - 4 \re), the other two being K2-292b ($\rho_\mathrm{P}$ = 7.4 $\pm$ 1.6~\gccc and $\it{S_{\mathrm{p}}}$ = 72 $\pm$ 11 \se; \citealt{Luque2019}) and Kepler-411b ($\rho_\mathrm{P}$ = 10.3 $\pm$ 1.3~\gccc and $\it{S_{\mathrm{p}}}$ = 196 $\pm$ 9 \se; \citealt{Sun2019}). We analyzed the differences in bulk density of planets with 2 - 4 \re and $\leq$ 30 \se ("warm" sub-Neptunes) and planets of the same size but with $\geq$ 30 \se ("hot" sub-Neptunes). In the first group, we have 21 planets with a median density of 3.9~\gccc. In the second, we have 65 planets with a median density of 2.9~\gccc. We performed a resampling method to account for uncertainties in density measurements: a large number of new datasets (N = 10, 000 samples) are drawn, which lie within the normal distribution of the density measurements of the planets. For each of these new datasets, the kernel density estimation (KDE) is calculated. Similar to histograms, its aim is to estimate an unknown probability density function from a sample of data but in a more continuous way. We then calculated the mean and standard deviation of all the KDEs obtained. The result for the two groups is shown in Figure \ref{fig:kde}. There is evidence suggesting that sub-Neptune-size planets orbiting FGK stars with low stellar irradiation ("warm" sub-Neptunes) exhibit a slightly higher bulk density compared to those exposed to strong irradiation. 
 To test whether the difference in densities of planets in these two irradiation conditions is statistically significant, we utilized a two-sample Kolmogorov-Smirnov test, which studies whether the two data samples come from the same distribution. With a p-value of 0.026, we find that the densities of hot and warm sub-Neptunes belong to different distributions with over 95\% confidence. The inclusion of TOI-815b and c in the samples plays an important role in the statistical significance of this result, as they reduce the p-value from 0.051 to 0.026, enabling us to reach the 95\% confidence interval.  This difference in density could potentially be attributed to the cooler atmospheres of these less irradiated planets, which are more compressed and denser. However, such a pattern could be also linked to observational bias since more massive planets at long periods are easier to characterize using RVs. Therefore, further precise measurements of the mass of similar exoplanets using precise RVs are necessary.
\subsection{Similarity in the density of multi-planet systems}\label{sec:similarity}
Well-characterized exoplanets in multi-planet systems are especially interesting since they allow the system's architecture to be investigated. \citet{Otegi2022} studied the similarity of planets in radii, masses, densities, and period ratios within a multi-planet system and revealed that planets tend to be more similar in radius than in mass. This trend could be linked to the significant influence of the planetary radius on density compared to mass, and consequently, on the overall planetary composition. Interestingly, in their work, they observed a robust correlation between densities of adjacent planets with no clear evidence of bi-modality. 

\begin{figure}
  \includegraphics[width=0.5\textwidth]{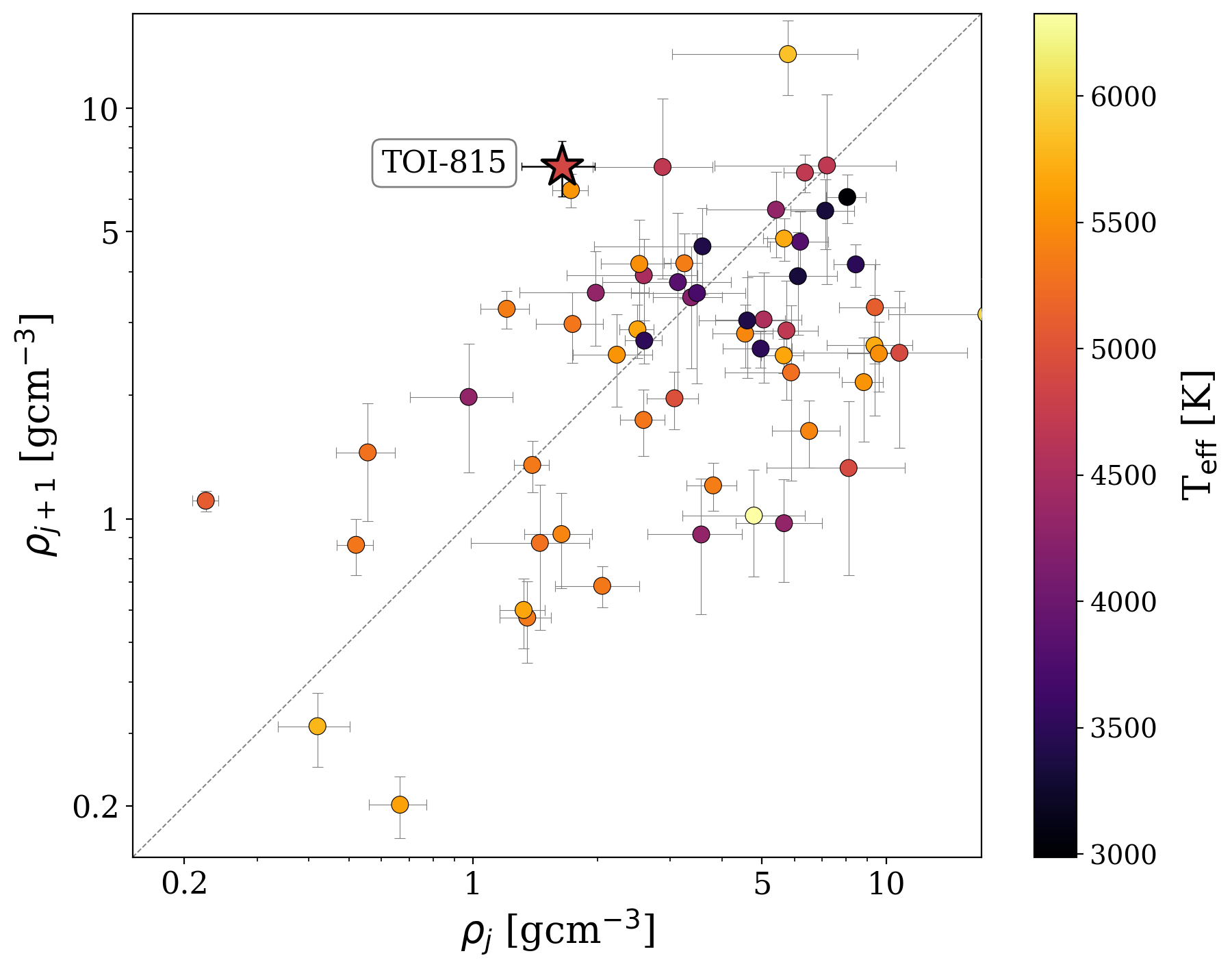}
  \caption{Density of a planet (\textit{j}) against the density of the next planet (\textit{j}+1; farther from the star). The TOI-815 system is indicated with a star symbol.} 
  \label{fig:density}
\end{figure}

In our study we used the NASA Exoplanet Archive\protect\footnote{\url{https://exoplanetarchive.ipac.caltech.edu/}} (\citealt{Akeson2013}) and narrowed down our selection to planets exhibiting mass and radius uncertainties below $\sigma_{M}/M$ $\leq$ 50$\%$ and $\sigma_{R}/R$ $\leq$16$\%$. For planets with masses characterized by TTVs we used the catalog presented by \citet{Hadden2017} and discarded the less robust (Kepler-25b and c, \citealt{Marcy2014}; Kepler-18c, \citealt{Cochran2011}). Our final sample consists of 116 planets within 39 multi-planet systems. In Figure \ref{fig:density} we plot the density of a planet (\textit{j}) within our sample against the density of the next planet (\textit{j+1}) farther from the star. The points are color-coded by the host effective temperature. The Pearson correlation test resulted in a p-value  = 0.04 that suggests a statistically significant linear correlation between densities. It is worth noting that planets orbiting M-dwarf stars tend to be closer to the 1:1 line compared to FGK stars with a p-value = 0.03. However, it is important to bear in mind that this observation is based on a limited sample of 22 planets within nine multi-planet systems.
We calculated the similarity metric for the TOI-815 system using Eq.~(2) of \citet{Millholland2017}:

\begin{equation}
\centering
    \mathcal{D}_X=\sum_{i=1}^{N_{\mathrm{sys}}} \sum_{\substack{j=1 \\ P_j<P_{j+1}}}^{N_{\mathrm{pl}}-1}\left|\log \frac{X_{j+1}}{X_j}\right|,
\end{equation}

\noindent with $X$ corresponding to the planetary radius or planetary mass, to be $\mathcal{D}_R$ = 0.05 and $\mathcal{D}_M$ = 0.49. The metrics indicate that the planets in the TOI-815 system are very similar in radii but rather different in mass. Placing the TOI-815 system in Figure \ref{fig:density}, we clearly see that TOI-815 is one of the least uniform systems in density. Similar to TOI-815, the Kepler-107 system (\citealt{Bonomo2023}) consists of two planets, Kepler-107b and c, with nearly identical radii (1.536 $\pm$ 0.025 \re and 1.597 $\pm$ 0.026 \re) but significantly different densities ($5.8 ^{+2.7} _{-2.6}$ \gccc and $13.5 ^{+2.9} _{-2.8}$ \gccc). \citet{Bonomo2019} showed that this density discrepancy can be explained by a giant impact on Kepler-107c. As presented in Section \ref{sec:atmosphericevolution}, such a mechanism can explain the large density of TOI-815c, which suggests that the two planets have formed in a similar way.

\subsection{Planetary composition and internal structure}\label{sec:internalstructure}
As shown in Figure \ref{fig:radiusmass}, the TOI-815 planetary system consists of two exoplanets that are located above an Earth-like composition line in the M-R diagram. This already indicates that both planets include a considerable amount of volatile materials, and for TOI-815b, a sizable atmosphere.

\begin{figure*}
\centering
  \includegraphics[width=0.45\textwidth]{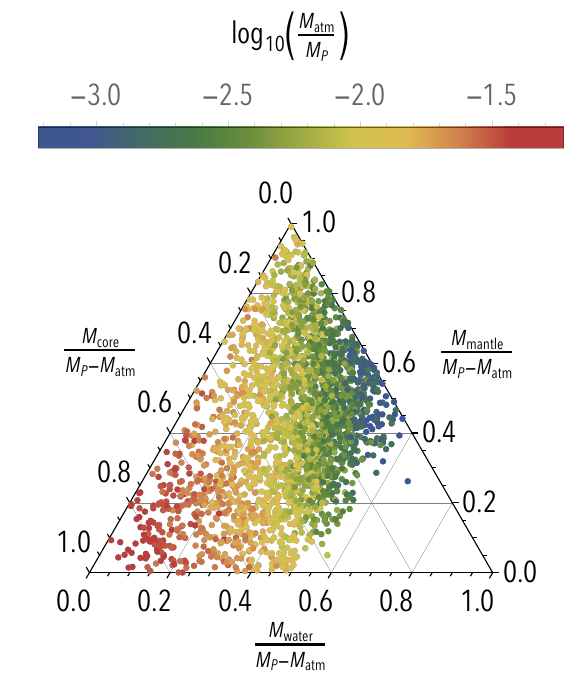}
  \includegraphics[width=0.45\textwidth]{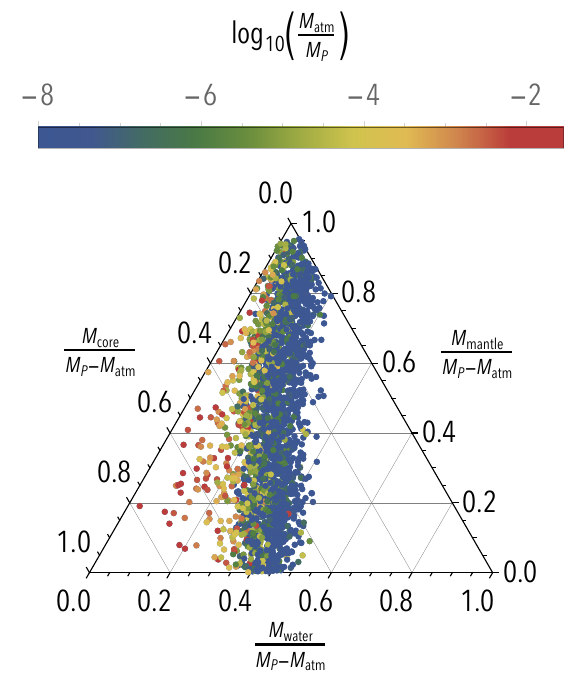}
  \caption{Ternary plots of the nested sampling results for the free model of TOI-815b (\textit{left}) and TOI-815c (\textit{right}). The axes are normalized with respect to $M_P - M_\mathrm{atm}$, and the color of the points indicates the atmospheric mass fraction in logarithmic scale.} 
    \label{fig:ternary_interior_structure}
\end{figure*}

To explore the range of possible interior structures, we used a nested sampling algorithm \citep{pymultinest} on the four-layered interior model \citep{Dorn2017}. This model contains an iron core, a rocky mantle, a water layer, and a H-He atmosphere. For both planets, we ran two bracketing interior models: one where we put no constraints on the layer masses (the free model), and one where we set the water mass fraction to zero in order to put an upper limit on the  atmospheric mass fraction  (the no-water model). Additionally, for TOI-815c, we included a model where we set the atmospheric mass fraction to zero (the no-atmosphere model). Such a model is not possible for TOI-815b, as its radius is too large to have no atmosphere. For all models, we varied the elemental ratios of [Fe/Si] and [Mg/Si] in the mantle and the planetary age, and we capped the water mass fraction at 50\%. As a proxy for the planet's age, we used the age constraints of the host star. The results are summarized in Table \ref{tab:interior_structure} and Figure \ref{fig:ternary_interior_structure}.

\begin{table}[h]
    \tiny
    \centering
    \renewcommand{\arraystretch}{1.}
    \setlength{\tabcolsep}{5pt}
    \caption{Nested sampling results for the interior structure of the free model (top), the no-water model (middle), and the no-atmosphere model (bottom).}
    \begin{tabularx}{\linewidth}{lcccc}
        \hline
        \hline
        \noalign{\smallskip}
         Planet (model) & $M_\mathrm{core}/M_p$ & $M_\mathrm{mantle}/M_p$ & $M_\mathrm{water}/M_p$ & $\log (M_\mathrm{atm}/M_p)$\\
         \noalign{\smallskip}
        \hline
        \noalign{\smallskip}
        free: & & & \\
         TOI-815b & $0.29 \pm 0.20$ & $0.42 \pm 0.22$ & $0.28 \pm 0.13$ & $-2.16\pm 0.40$\\
         TOI-815c & $0.32 \pm 0.16$ & $0.44 \pm 0.26$ & $0.23 \pm 0.12$ & $-7.70\pm 2.69$\\
         \noalign{\smallskip}
         \hline
         \noalign{\smallskip}
        no-water: & & & \\
         TOI-815b& $0.40 \pm 0.25$ & $0.58 \pm 0.26$ & $0$ & $-1.64\pm 0.22$\\
         TOI-815c& $0.30 \pm 0.23$ & $0.70 \pm 0.24$ & $0$ & $-2.78\pm 0.75$\\
         \noalign{\smallskip}
         \hline
         \noalign{\smallskip}
        no-atmosphere: & & & \\
         TOI-815c& $0.31 \pm 0.15$ & $0.43 \pm 0.25$ & $0.26 \pm 0.11$ & N/A\\
         \hline
    \end{tabularx}
    \label{tab:interior_structure}
\end{table}

As expected, we find that a plethora of interior structures are consistent with the observational constraints of TOI-815b and TOI-815c. These range from water-rich structures with little to no atmosphere in the case of TOI-815c, to structures devoid of water but with a non-negligible H-He atmosphere. However, while the internal structure of TOI-815b is rather poorly constrained, it is clear that the observational constraints can only be satisfied with an H-He atmosphere. This is in alignment with \citet{Hakim2018} who calculated the mass-radius diagram with different core and mantle compositions for planets with Moon-like, Earth-like, or Mercury-like internal structures. Even though they obtained a broad distribution of planetary radii for planets of the same mass, none of the models without an atmosphere can reproduce the large radius of TOI-815b. This further supports the conclusion that it is impossible to construct a model of TOI-815b without a H-He atmosphere.

We emphasize that our interior structure model does not include the pollution of the H-He envelope with heavier elements, which leads to contraction of the atmosphere (e.g., \citealt{Lozovsky_2018}). Furthermore, water might not only be present in an ocean layer, but also dissolved within the mantle of the planet \citep[e.g.,][]{Dorn_2021} or be present in the atmosphere \citep[e.g.,][]{2022NatAs...6..819M}. In addition, the assumption of distinct planetary layers might break down at high planetary masses \citep{Helled_2017, Bodenheimer_2018}. Nonetheless, in face of the observational constraints currently available, these additions are unlikely to  affect the main conclusions  of this analysis.


\subsection{Planetary atmospheric evolution}\label{sec:atmosphericevolution}
To model the atmospheric evolution of the planets in the TOI-815 system and constrain their primordial parameters, we present modeling conducted using the \texttt{P}lanetary \texttt{A}tmospheres and \texttt{S}tellar Ro\texttt{T}ation R\texttt{A}tes~\citep[\texttt{PASTA};][]{bonfanti2021_pasta} code, which is a planetary atmospheric evolution code based on the original code presented by \citet{Kubyshkina2019a,Kubyshkina2019b}. As stars can be born with different initial rotation periods, their $L_{\rm XUV}$ values can spread over a wide range during the first $\sim1$\,Gyr following protoplanetary disk dispersal \citep[e.g.,][]{tu2015}. To account for that, \texttt{PASTA} simultaneously constrains the evolution of planetary atmospheres and of the stellar rotation rate by combining a model predicting planetary atmospheric escape rates based on hydrodynamic simulations~\citep[this has the advantage over other commonly used analytical estimates in that it accounts for both extreme-UV-driven and core-powered mass loss;][]{kubyshkina2018grid}, a model of the evolution of the stellar extreme-UV (XUV) flux \citep{bonfanti2021_pasta}, a model relating planetary parameters and atmospheric mass~\citep{Johnstone2015}, and stellar evolutionary tracks~\citep{Choi2016}.

\texttt{PASTA} works under two main assumptions: (1) planet migration did not occur after the dispersal of the protoplanetary disk; and (2) the planets hosted at some point in the past or still host a hydrogen-dominated atmosphere. The free parameters (i.e., subject to uniform priors) are the planetary initial atmospheric mass fractions at the time of the dispersal of the protoplanetary disk ($f_{\rm atm}^{\rm start}$), which we assumed occurs at an age of 5\,Myr \citep[see for example][]{Alexander2014,Kimura2016,Gorti2016}, and the stellar rotation period at 150 Myr, which is used as a proxy for the stellar XUV emission. The code returns constraints on the free parameters and on their uncertainties by implementing the atmospheric evolution algorithm in a Bayesian framework (through the MC3 code of~\citealt{Cubillos2017}), using the system parameters with their uncertainties as input priors. Details of the algorithm can be found in~\citet{bonfanti2021_pasta}. 

\begin{figure}
  \centering
\includegraphics[width=0.42\textwidth]{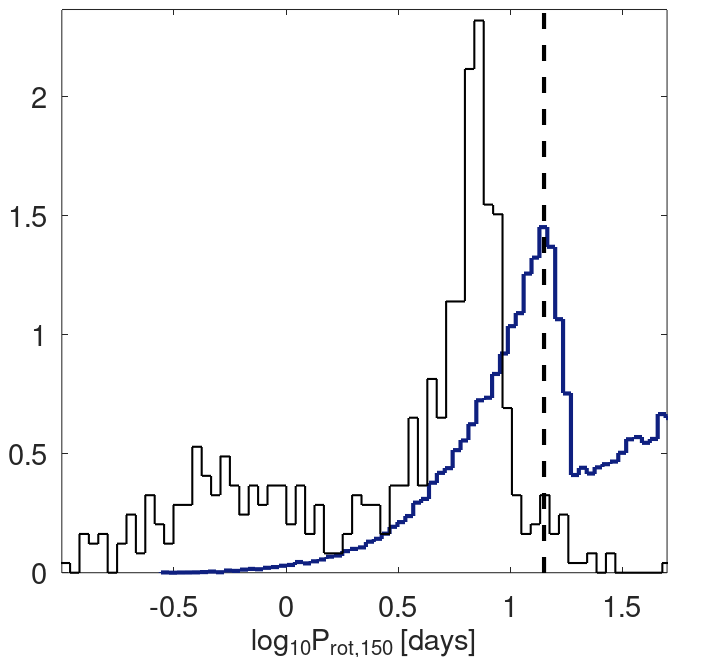}
  \caption{Posterior distribution (dark blue) of the stellar rotation rate of TOI-815 after 150 Myr derived by \texttt{PASTA}. The solid black line represents the distribution of the stellar rotation rate of young open cluster stars with masses comparable to that of TOI-815 based on the collection of data provided by \citet{johnstone2015_stII}. The dashed black line indicates the observed present-day stellar rotation rate.}
  \label{fig:atmosphericEvolutionStellarRotation}
\end{figure}

As a proxy for the evolution of the stellar rotation period, and thus XUV emission, Figure ~\ref{fig:atmosphericEvolutionStellarRotation} displays the posterior distribution of the stellar rotation period at an age of 150\,Myr ($P_{\mathrm{rot,150}}$). This distribution is then compared to that of stars member of young open clusters that are of comparable mass, as obtained from~\citet{johnstone2015_stII}. Because the system is still very young ($200^{+400}_{-200}$ Myr), the posterior distribution of $P_{\mathrm{rot,150}}$ is mostly dominated by the measurement of the present-day rotation rate ($15.3~\pm~1.2$ days). The distribution is, however, flatter than the prior on $P_{\mathrm{rot}}$, because of the large relative uncertainty of the stellar age. The result indicates that the star could not be a fast rotator (i.e., $P_{\mathrm{rot,150}} < 1 \, \rm days$), as constrained by the slow present-day rotation period. 

\begin{figure*}
  \includegraphics[width=\linewidth]{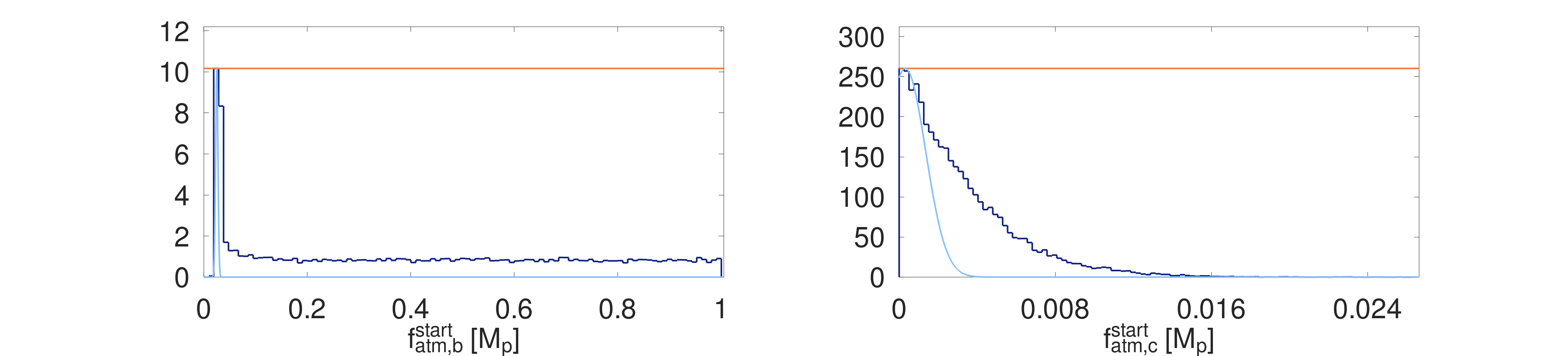}
  \caption{Posterior distribution (dark blue) for the mass of the planetary atmosphere of TOI-815b (left) and TOI-815c (right) at the time of the dispersal of the protoplanetary disk. The light blue line represents the distribution of the estimated present-day atmospheric mass fraction derived by \texttt{PASTA}. The orange horizontal line indicates the uninformative prior distribution.}
 \label{fig:atmosphericEvolutionMassFraction}
\end{figure*}

Figure~\ref{fig:atmosphericEvolutionMassFraction} shows the posterior distribution of the initial atmospheric mass fraction for TOI-815b (left panel) and TOI-815c (right panel) in comparison to the present-day atmospheric mass fraction. For planet b we find that \texttt{PASTA} clearly prefers evolutionary scenarios in which the evolution of the atmosphere was mostly unaffected by hydrodynamic escape. This indicates that the planet most likely still retains the majority of its primordial atmosphere. However, it must be noted that given the uncertainties of the observed present-day parameters \texttt{PASTA} is not able to completely exclude scenarios with significant escape.

In the case of TOI-815c, \texttt{PASTA} concludes that, independently of the evolution of the stellar rotation rate, the planet has not lost a significant amount of atmosphere via hydrodynamic escape. Given its present-day orbital position and relatively large core mass, it would be expected for the planet to have accreted an extensive hydrogen atmosphere if it had formed within the protoplanetary disk \citep{Ikoma2012,Lee2015,mordasini2020,Venturini2017}. However, such an envelope is not observed today and according to \texttt{PASTA} the planet is unable to have lost it due to hydrodynamic escape over the course of its evolution. One possible explanation would be formation within a gas-poor environment (e.g., \citealt{Lee_2022}). Another explanation is that the planet suffered from at least one giant impact toward the end of the disk's life \citep{Inamdar15}. This also aligns with the fact that the planets of the system are not in mean motion resonance and the expected spin-orbit misalignment. In both cases the results retrieved by \texttt{PASTA} are compatible with conclusions that can be drawn by the more simplistic approach of only evaluating the restricted Jeans parameter $\Lambda$ \citep{Fossati2017}. For TOI-815c the high value of $\Lambda = 145 ^{+16} _{-16}$ would imply a stable atmosphere, while for TOI-815b the intermediate value of $\Lambda = 28.6 ^{+5.7} _{-5.3}$ would favor a stable atmosphere but not exclude the possibility of a hydrodynamically escaping atmosphere.
We note that in principle \texttt{PASTA} is not set up to simulate the atmospheric evolution of planets with mean planetary densities larger than that of Earth. This is due to the fact that the code always assumes for the rocky core of a planet to have a mean density equal to that of the Earth. In practice, \texttt{PASTA} will underestimate the planetary mass of TOI-815c as its observed density is $1.30 \pm 0.20 \, \rho_\mathrm{\oplus}$. The planetary mass estimate for planet c returned by \texttt{PASTA} is $M_\mathrm{P,c} = 19.9 \pm 1.9 \, M_\mathrm{\oplus}$, which implies a bare rocky-core with a mean planetary density of $1 \, \rho_\mathrm{\oplus}$. A larger planetary mass, however, would just further hinder hydrodynamic escape, as escaping particles would need even more energy to escape the gravitational well of the planet. Therefore, since hydrodynamic escape affects only marginally the evolution of the planetary atmosphere, the impact of the assumption on the core density has a negligible impact on the results.

\subsection{Search for transit timing variations}\label{sec:TTVs}
\begin{figure}[h]
\centering
  \includegraphics[width=0.47\textwidth]{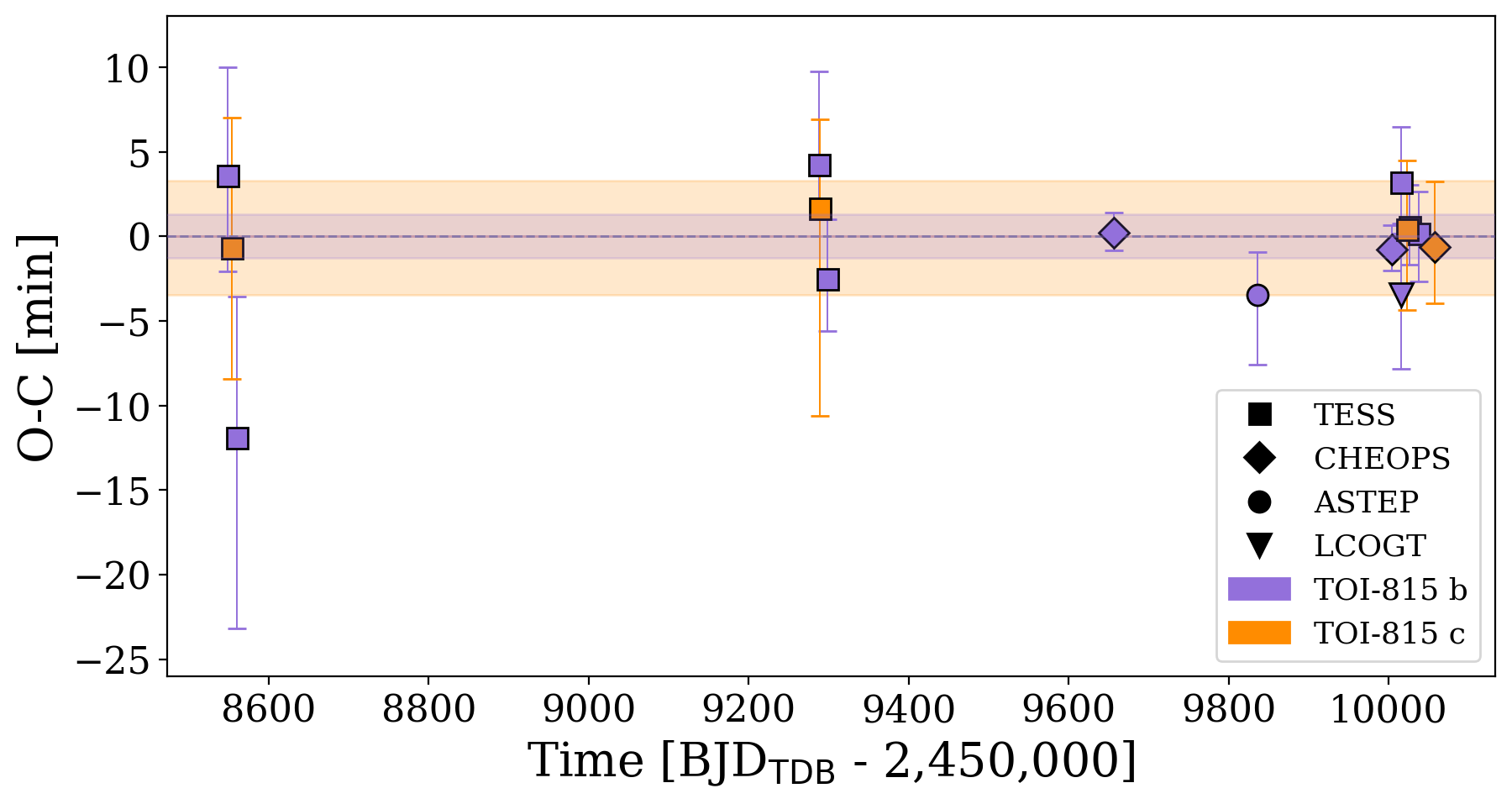}
  \caption{Transit timings for each transit of
TOI-815b (purple) and TOI-815c (orange) measured from the TESS and CHEOPS light curves. The purple and orange regions represent the 1$\sigma$ uncertainties of the inner and outer planet ephemeris reported in Table \ref{tab:posteriors}.} 
  \label{fig:ttv}
\end{figure}

Although the ratio of the orbital period of TOI-815 and c ($P_{\mathrm{orb,c}}$/$P_{\mathrm{orb,b}}$ $\simeq$ 3.12) lies exterior to the 3:1 resonance, we conducted a TTV analysis to search for additional, non-transiting planets. Using \juliet, we performed the analysis by utilizing all available photometric datasets from TESS, CHEOPS, ASTEP, and LCOGT. Instead of fitting a single period ($P_{\mathrm{orb}}$) and time-of-transit center ($T_0$), \juliet employs a method that seeks individual transit times. This approach involves fitting each transit independently and determining one transit time for each one, resulting in a more consistent and coherent analysis. Figure \ref{fig:ttv} illustrates the results of our analysis, which compares the observed transit times with the calculated linear ephemeris derived from all the transits. Notably, no significant variations were detected within the data. 

\subsection{Potential for atmospheric and Rossiter-McLaughlin studies}
We used the metric proposed by \citet{Kempton2018} to assess the potential of atmospheric characterization for the planets in the TOI-815 system  with the \textit{James Webb Space Telescope} (\textit{JWST}; \citealt{JWST}). To calculate the transmission spectroscopy metric for all exoplanets in the PlanetS catalog\textsuperscript{\ref{dacefootnote}} with a radii from 2 to 4 \re, we utilized the scale factors provided in Table 1 by \citet{Kempton2018} instead of the suggested value of 0.167 for temperate planets. Notably, the transmission spectroscopy metric values for TOI-815b and c were found to be 97 $\pm$ 20 and 15 $\pm$ 2, respectively. Putting these values into context, TOI-815b is located in the top 10$\%$ of well-characterized exoplanets in this regime and therefore stands out as an excellent candidate for atmospheric characterization via transmission spectroscopy. 

We also explored the possibility of studying the spin--orbit of both planets via  Rossiter-McLaughlin (RM) observations (\citealt{Rossiter1924}; \citealt{McLaughlin1924}). According to Eq.~(40) by \citet{Winn2010}, we expect the maximum amplitude of the RM effect to be $\sim$1.13 \ms and 0.52 \ms for TOI-815b and TOI-815c, respectively. Using the classical RM technique, the signals are small and challenging, but the sky-projected spin--orbit angle $\lambda$ should be detectable using newer methods such as the RM effect revolutions technique (\citealt{Bourrier2021}). Moreover, since the star is seen close to pole-on and since the planets are transiting, the two orbits are probably misaligned. This can be seen using the following equation linking the 3D spin--orbit angle $\psi$ to other observable quantities \citep[e.g.,][]{Winn2007}:

\begin{equation}
\label{eq:cospsi}
\cos \psi = \sin i_\star \sin i \cos \lambda + \cos i_\star \cos i,
\end{equation}

\noindent yielding a small value for $\cos \psi$ for any value of $\lambda$ we measure. The case that would disfavor the orbital misalignment picture the most is if we observe an aligned sky-projected spin--orbit angle $\lambda \sim 0^\circ$. Sampling a $\psi$ distribution from Eq.~(\ref{eq:cospsi}) with a Dirac distribution at $0^\circ$ for $\lambda$ and measurement-informed Gaussians for the other parameters, we find $\psi$ = 58$^{+18}_{-19}$, which still represents a misaligned orbit at a 3.2$\sigma$ level. Given the young nature of the system, tidal realignment \citep[e.g.,][]{Barker2010,Attia2023} may have not had time to occur. Hence, TOI-815 offers us a unique possibility to capture a peculiar orbital architecture, especially given the scarcity of misaligned orbits in multi-planet systems \citep[e.g.,][]{Albrecht2022}.

\section{Conclusions}\label{sec:Conclusions}
We have reported the detection and characterization of two sub-Neptune planets that are transiting TOI-815. TOI-815 is a young ($200 ^{+400} _{-200}$~Myr) K dwarf in a K-M binary system. The planets were initially detected by the TESS mission and subsequently confirmed with CHEOPS observations as well as precise RV measurements obtained with ESPRESSO, which enabled the determination of their masses. Through the combined use of TESS, CHEOPS, LCOGT, and ASTEP photometry and the ESPRESSO spectroscopic data, we precisely characterized both planets: TOI-815b has a period of 11.2 days, a radius of 2.94 $\re$, and a mass of 7.6 $\me$, while TOI-815c has a period of 35 days, a radius of 2.62 $\re$, and a mass of 23.5 $\me$. The combination of these facilities allowed us to determine the planetary radius and mass with precision better than 2$\%$ and 20$\%$ for the inner planet and 4$\%$ and 10$\%$ for the outer planet, respectively. 

TOI-815c, with a bulk density of 7.2 $\pm$ 1.1 \gccc, stands as the densest well-characterized warm mini-Neptune. Our study reveals a statistically significant difference between the bulk densities of warm sub-Neptunes and hot sub-Neptunes, with TOI-815b and c being instrumental in establishing a 95\% confidence. Further exoplanet detections with precise mass measurements are needed to confirm this trend. Our analysis of the TOI-815 system using similarity metrics demonstrates that while the planets in this system have remarkably similar radii, they exhibit significantly different masses, making TOI-815 one of the least uniform systems in terms of planetary density.

Internal structure modeling that includes four layers (an iron core, a rocky mantle, a water layer, and a H-He atmosphere) suggests that TOI-815b likely has a small H-He atmosphere (on the order of a percent of the planet's mass) if water is included. If no water is assumed, the atmospheric mass fraction can be an order of magnitude higher. On the contrary, the radius of TOI-815c can be reproduced without any atmosphere. This object is interesting because it is rather massive and yet seems to have only a tiny (or no) atmosphere. This is a challenge for planet formation theories since at these planetary masses strong accretion of H-He is expected (for comparison, Uranus and Neptune have lower masses but consist of $\sim$10\% of H-He). A possible explanation would be for the planet to have lost the majority of its primordial atmosphere. However, our atmospheric evolution model shows that such an extensive mass loss cannot be caused by XUV-driven hydrodynamic escape. Thus, the most likely scenario is that planet c, similar to Kepler-107c (\citealt{Bonomo2019}), suffered from at least one giant impact (which removed a substantial amount of primordial gas) during the unstable dynamical phase of disk dispersal. This is consistent with the planets of this system not being in resonance. Alternatively, a formation within a gas-poor environment cannot be ruled out either.


Finally, the combination of the stellar rotation period ($P_{\rm rot}$ = 15.3 $\pm$ 1.2 days) constrained from the WASP and TESS photometry and the very low spectroscopically measured projected rotational velocity ($v \sin i_{*}$ $<$ 1~km\,s$^{-1}$) suggest that we are observing the star close to pole-on. Given that the planets are on orbits with inclinations close to $\sim$90~$\degree$, we expect a spin-orbit misalignment at a 3.2$\sigma$ level. Future RM observations will be able to confirm this.

\begin{acknowledgements}
We thank the Swiss National Science Foundation (SNSF) and the Geneva University for their continuous support to our planet low-mass companion search programs. This work has been carried out within the framework of the National Centre of Competence in Research PlanetS supported by the Swiss National Science Foundation under grants 51NF40$\_$182901 and 51NF40$\_$205606. OA and VB have received funding from the European Research Council (ERC) under the European Union's Horizon 2020 research and innovation programme (project {\sc Spice Dune}, grant agreement No 947634; grant agreement No 730890). This material reflects only the authors' views and the Commission is not liable for any use that may be made of the information contained therein. The authors acknowledge the financial support of the SNSF. This publication makes use of The Data \& Analysis Center for Exoplanets (DACE), which is a facility based at the University of Geneva (CH) dedicated to extrasolar planet data visualization, exchange, and analysis. DACE is a platform of NCCR $PlanetS$ and is available at https://dace.unige.ch. This paper includes data collected by the TESS mission. Funding for the TESS mission is provided by the NASA Explorer Program. JSJ acknowledges support by FONDECYT grant 1201371 and partial support from the ANID Basal project FB210003. Funding for the TESS mission is provided by NASA's Science Mission Directorate. KAC and CNW acknowledge support from the TESS mission via subaward s3449 from MIT. This research has made use of the Exoplanet Follow-up Observation Program (ExoFOP; DOI: 10.26134/ExoFOP5) website, which is operated by the California Institute of Technology, under contract with the National Aeronautics and Space Administration under the Exoplanet Exploration Program. 
This paper made use of data collected by the TESS mission and are publicly available from the Mikulski Archive for Space Telescopes (MAST) operated by the Space Telescope Science Institute (STScI). We acknowledge the use of public TESS data from pipelines at the TESS Science Office and at the TESS Science Processing Operations Center. Resources supporting this work were provided by the NASA High-End Computing (HEC) Program through the NASA Advanced Supercomputing (NAS) Division at Ames Research Center for the production of the SPOC data products.

CHEOPS is an ESA mission in partnership with Switzerland with important contributions to the payload and the ground segment from Austria, Belgium, France, Germany, Hungary, Italy, Portugal, Spain, Sweden, and the United Kingdom. The CHEOPS Consortium would like to gratefully acknowledge the support received by all the agencies, offices, universities, and industries involved. Their flexibility and willingness to explore new approaches were essential to the success of this mission. CHEOPS data analyzed in this article will be made available in the CHEOPS mission archive (\url{https://cheops.unige.ch/archive_browser/}). This work makes use of observations from the ASTEP telescope. ASTEP benefited from the support of the French and Italian polar agencies IPEV and PNRA in the framework of the Concordia station program, from OCA, INSU, Idex UCAJEDI (ANR- 15-IDEX-01) and ESA through the Science Faculty of the European Space Research and Technology Centre (ESTEC). This work makes use of observations from the LCOGT network. Part of the LCOGT telescope time was granted by NOIRLab through the Mid-Scale Innovations Program (MSIP). MSIP is funded by NSF. Co-funded by the European Union (ERC, FIERCE, 101052347). Views and opinions expressed are however those of the author(s) only and do not necessarily reflect those of the European Union or the European Research Council. Neither the European Union nor the granting authority can be held responsible for them. C.B. acknowledges support from the Swiss Space Office through the ESA PRODEX program. This work has been carried out within the framework of the NCCR PlanetS supported by the Swiss National Science Foundation under grants 51NF40\_182901 and 51NF40\_205606. ML acknowledges support of the Swiss National Science Foundation under grant number PCEFP2\_194576. S.G.S. acknowledge support from FCT through FCT contract nr. CEECIND/00826/2018 and POPH/FSE (EC). YAl acknowledges support from the Swiss National Science Foundation (SNSF) under grant 200020\_192038. ABr was supported by the SNSA. DG gratefully acknowledges financial support from the CRT foundation under Grant No. 2018.2323 ``Gaseousor rocky? Unveiling the nature of small worlds''. JV acknowledges support from the Swiss National Science Foundation (SNSF) under grant PZ00P2\_208945.
MNG is the ESA CHEOPS Project Scientist and Mission Representative, and as such also responsible for the Guest Observers (GO) Programme. MNG does not relay proprietary information between the GO and Guaranteed Time Observation (GTO) Programmes, and does not decide on the definition and target selection of the GTO Programme.
KGI is the ESA CHEOPS Project Scientist and is responsible for the ESA CHEOPS Guest Observers Programme. She does not participate in, or contribute to, the definition of the Guaranteed Time Programme of the CHEOPS mission through which observations described in this paper have been taken, nor to any aspect of target selection for the programme. We acknowledge financial support from the Agencia Estatal de Investigación of the Ministerio de Ciencia e Innovación MCIN/AEI/10.13039/501100011033 and the ERDF “A way of making Europe” through projects PID2019-107061GB-C61, PID2019-107061GB-C66, PID2021-125627OB-C31, and PID2021-125627OB-C32, from the Centre of Excellence “Severo Ochoa'' award to the Instituto de Astrofísica de Canarias (CEX2019-000920-S), from the Centre of Excellence “María de Maeztu” award to the Institut de Ciències de l’Espai (CEX2020-001058-M), and from the Generalitat de Catalunya/CERCA programme. S.C.C.B. acknowledges support from FCT through FCT contracts nr. IF/01312/2014/CP1215/CT0004. XB, SC, DG, MF and JL acknowledge their role as ESA-appointed CHEOPS science team members. LBo, GBr, VNa, IPa, GPi, RRa, GSc, VSi, and TZi acknowledge support from CHEOPS ASI-INAF agreement n. 2019-29-HH.0. ACC acknowledges support from STFC consolidated grant numbers ST/R000824/1 and ST/V000861/1, and UKSA grant number ST/R003203/1. P.E.C. is funded by the Austrian Science Fund (FWF) Erwin Schroedinger Fellowship, program J4595-N. This project was supported by the CNES. The Belgian participation to CHEOPS has been supported by the Belgian Federal Science Policy Office (BELSPO) in the framework of the PRODEX Program, and by the University of Liège through an ARC grant for Concerted Research Actions financed by the Wallonia-Brussels Federation; L.D. is an F.R.S.-FNRS Postdoctoral Researcher. This work was supported by FCT - Fundação para a Ciência e a Tecnologia through national funds and by FEDER through COMPETE2020 - Programa Operacional Competitividade e Internacionalização by these grants: UIDB/04434/2020; UIDP/04434/2020; 2022.06962.PTDC. O.D.S.D. is supported in the form of work contract (DL 57/2016/CP1364/CT0004) funded by national funds through FCT. S.G.S acknowledges the support from FCT through Investigador FCT contract nr. CEECIND/00826/2018 and  POPH/FSE (EC). B.-O. D. acknowledges support from the Swiss State Secretariat for Education, Research and Innovation (SERI) under contract number MB22.00046. This project has received funding from the Swiss National Science Foundation for project 200021\_200726. It has also been carried out within the framework of the National Centre of Competence in Research PlanetS supported by the Swiss National Science Foundation under grant 51NF40\_205606. The authors acknowledge the financial support of the SNSF. MF and CMP gratefully acknowledge the support of the Swedish National Space Agency (DNR 65/19, 174/18). T.D. acknowledges support by the McDonnell Center for the Space Sciences at Washington University in St. Louis. M.G. is an F.R.S.-FNRS Senior Research Associate. CHe acknowledges support from the European Union H2020-MSCA-ITN-2019 under Grant Agreement no. 860470 (CHAMELEON). SH gratefully acknowledges CNES funding through the grant 837319. K.W.F.L. was supported by Deutsche Forschungsgemeinschaft grants RA714/14-1 within the DFG Schwerpunkt SPP 1992, Exploring the Diversity of Extrasolar Planets. This work was granted access to the HPC resources of MesoPSL financed by the Region Ile de France and the project Equip@Meso (reference ANR-10-EQPX-29-01) of the programme Investissements d'Avenir supervised by the Agence Nationale pour la Recherche. PM acknowledges support from STFC research grant number ST/M001040/1. This work was also partially supported by a grant from the Simons Foundation (PI Queloz, grant number 327127). A. S. acknowledges support from the Swiss Space Office through the ESA PRODEX program. TWi acknowledges support from the UKSA and the University of Warwick. NAW acknowledges UKSA grant ST/R004838/1. V.V.G. is an F.R.S-FNRS Research Associate. C.D. acknowledges support from the Swiss National Science Foundation under grant TMSGI2\_211313. GyMSz acknowledges the support of the Hungarian National Research, Development and Innovation Office (NKFIH) grant K-125015, a a PRODEX Experiment Agreement No. 4000137122, the Lend\"ulet LP2018-7/2021 grant of the Hungarian Academy of Science and the support of the city of Szombathely. DR was supported by NASA under award number NNA16BD14C for NASA Academic Mission Services. R.L. acknowledges funding from University of La Laguna through the Margarita Salas Fellowship from the Spanish Ministry of Universities ref. UNI/551/2021-May 26, and under the EU Next Generation funds. R. A. is a Trottier Postdoctoral Fellow and acknowledges support from the Trottier Family Foundation. This work was supported in part through a grant from the Fonds de Recherche du Qu\'ebec - Nature et Technologies (FRQNT). This work was funded by the Institut Trottier de Recherche sur les Exoplan\'etes (iREx). This work made use of \texttt{tpfplotter} by J. Lillo-Box (publicly available in www.github.com/jlillo/tpfplotter), which also made use of the python packages \texttt{astropy}, \texttt{lightkurve}, \texttt{matplotlib} and \texttt{numpy}.

\end{acknowledgements}
\bibliographystyle{aa}
\bibliography{bib}
\begin{appendix}
\appendix
\onecolumn
\section{TESS light curves}
\begin{figure*}[h]
  \centering
  \includegraphics[width=1.\textwidth]{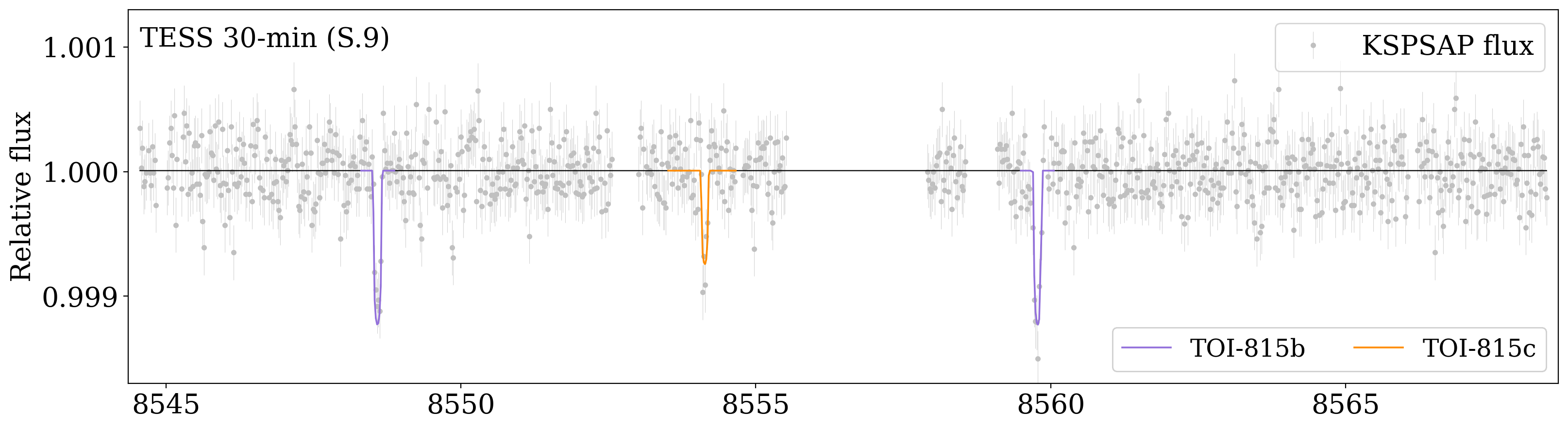}
  \includegraphics[width=1.\textwidth]{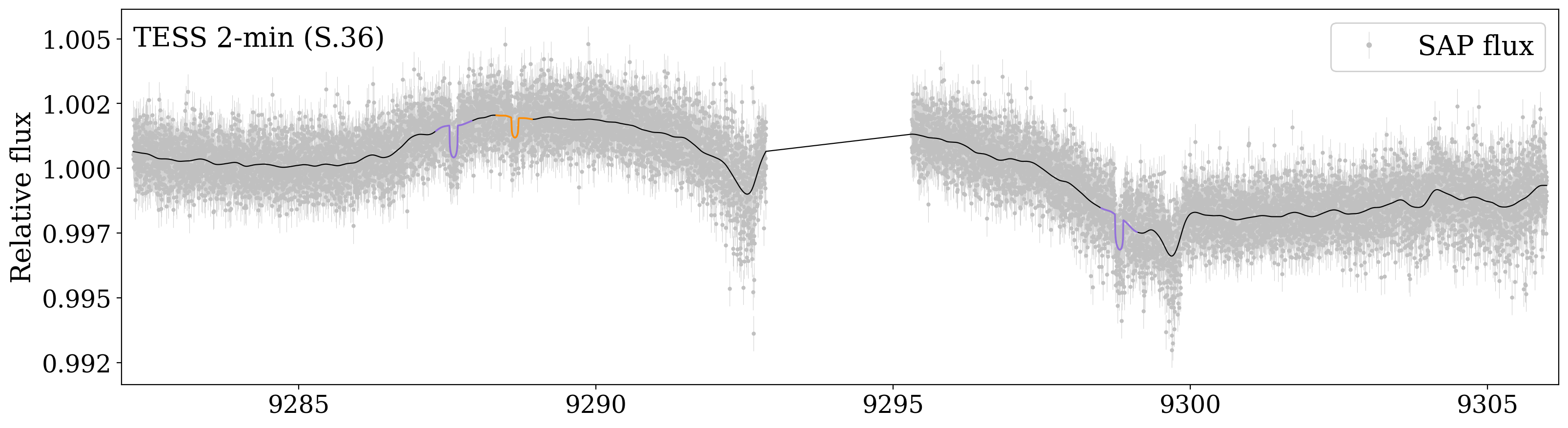}
  \includegraphics[width=1.\textwidth]{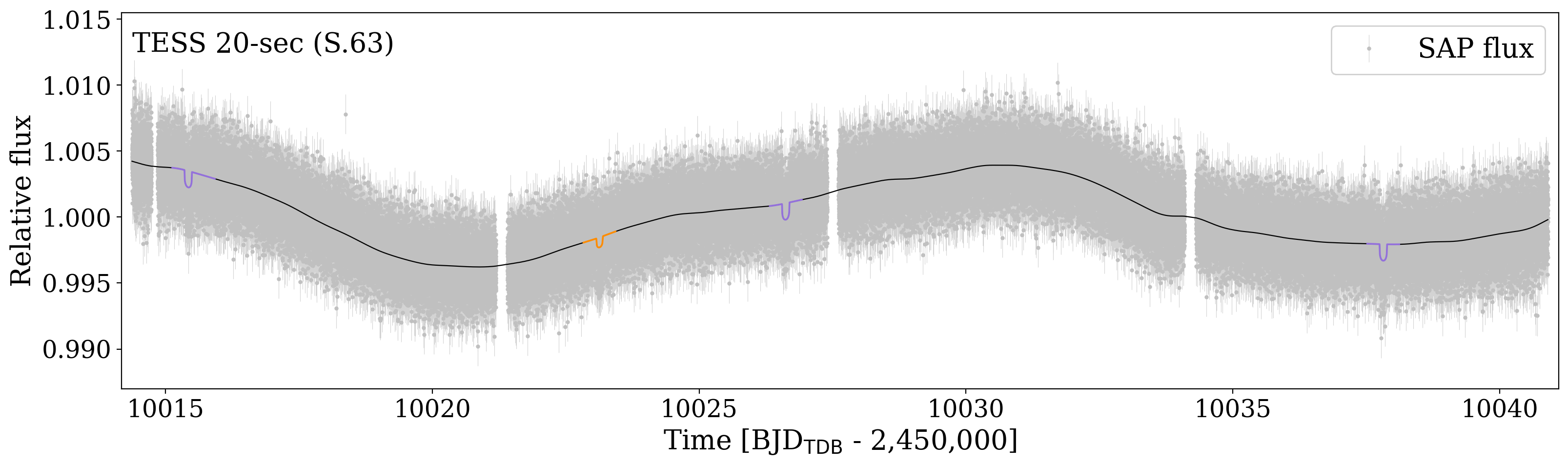}
  \caption{TESS SAP flux light curves with the best-fit \juliet model shown with the black line (see Sect. \ref{sec:Joint transit and RV analysis} for details on the modeling). The transits of TOI-815b and TOI-815c are indicated with purple and orange, respectively.} 
  \label{fig:fullTESSLCs}
\end{figure*}
\newpage
\section{RVs with ESPRESSO}

\begin{table*}[h]
\caption{ESPRESSO RV observations for TOI-815, along with activity indicator measurements, i.e. FWHM, bisector, and  logR$'_{\mathrm{HK}}$.}
\centering
\renewcommand{\arraystretch}{1.1}
\setlength{\tabcolsep}{19pt}
\begin{center}
\begin{tabular}{lcccccc}
\hline\hline
 Time & RV & $\sigma_{\mathrm{RV}}$ & FWHM$^{(2)}$ &  BIS$^{(2)}$ & logR$'_{\mathrm{HK}}$$^{(3)}$ & t$_{\mathrm{exp}}$ \\

[BJD-2 450 000] & [\ms]  & [\ms] & [\ms] & & & [s]\\
 \hline
9308.60433	&	-2973.3950	&	0.6145	&	6474.143	&	-33.038	&	-4.7084	&	600	\\
9310.60679	&	-2970.1704	&	0.7233	&	6470.941	&	-33.052	&	-4.7338	&	600	\\
9313.73506$^{(1)}$	&	-2962.6373	&	0.5712	&	6492.966	&	-30.291	&	-4.6757	&	600	\\
9316.52823	&	-2967.0427	&	0.7459	&	6494.718	&	-23.625	&	-4.7107	&	600	\\
9318.53033	&	-2968.9726	&	0.5863	&	6478.690	&	-29.060	&	-4.6762	&	600	\\
9324.54484	&	-2979.7547	&	0.8117	&	6513.799	&	-28.181	&	-4.6859	&	600	\\
9327.50777$^{(1)}$	&	-2965.7154	&	0.4397	&	6482.962	&	-32.938	&	-4.6229	&	600	\\
9329.52592$^{(1)}$	&	-2971.0105	&	0.9842	&	6543.347	&	-29.283	&	-4.7529	&	600	\\
9332.57219	&	-2979.7266	&	0.6474	&	6491.382	&	-26.738	&	-4.6824	&	600	\\
9336.49956$^{(1)}$	&	-2982.4246	&	1.3115	&	6515.936	&	-40.000	&	-4.9931	&	600	\\
9339.48187	&	-2973.6498	&	0.5988	&	6468.715	&	-33.267	&	-4.7189	&	600	\\
9341.51194$^{(1)}$	&	-2966.5835	&	0.9108	&	6513.090	&	-35.131	&	-4.7516	&	600	\\
9346.54064	&	-2972.2321	&	0.5959	&	6482.420	&	-24.888	&	-4.6930	&	600	\\
9373.49668	&	-2973.5189	&	0.5467	&	6462.905	&	-33.991	&	-4.6367	&	600	\\
9375.51475	&	-2971.6794	&	0.6513	&	6471.909	&	-32.600	&	-4.6265	&	600	\\
9406.50596	&	-2975.2108	&	0.7212	&	6492.234	&	-32.504	&	-4.6149	&	600	\\
9410.48054	&	-2971.8872	&	0.6759	&	6499.208	&	-30.202	&	-4.6978	&	600	\\
9415.50797	&	-2971.5355	&	0.8107	&	6491.252	&	-26.700	&	-4.6620	&	600	\\
9673.54348	&	-2977.4158	&	0.5684	&	6488.390	&	-34.257	&	-4.6894	&	700	\\
9675.51515	&	-2973.7947	&	0.9406	&	6487.917	&	-25.694	&	-4.6418	&	700	\\
9677.65231	&	-2972.6960	&	0.8151	&	6492.405	&	-31.643	&	-4.6580	&	700	\\
9679.55323$^{(1)}$	&	-2973.4334	&	0.5445	&	6497.504	&	-32.766	&	-4.6080	&	700	\\
9683.53185	&	-2981.6090	&	0.4694	&	6497.328	&	-29.483	&	-4.6329	&	700	\\
9685.55122	&	-2980.4065	&	0.7458	&	6496.394	&	-24.517	&	-4.6203	&	700	\\
9687.52944	&	-2972.0070	&	0.4487	&	6495.493	&	-29.672	&	-4.6432	&	700	\\
9695.52289	&	-2967.3200	&	0.9486	&	6512.740	&	-29.566	&	-4.7299	&	700	\\
9697.58163	&	-2965.9727	&	0.5759	&	6511.966	&	-26.858	&	-4.6611	&	700	\\
9784.49262	&	-2977.5647	&	0.6128	&	6525.410	&	-22.486	&	-4.6248	&	700	\\
9904.83330	&	-2969.5919	&	0.4342	&	6487.460	&	-35.249	&	-4.6440	&	700	\\
9905.77809	&	-2970.1154	&	0.6533	&	6491.720	&	-34.630	&	-4.6797	&	700	\\
9906.80910	&	-2970.1976	&	0.4551	&	6492.162	&	-33.028	&	-4.6399	&	700	\\
9907.79772	&	-2969.7455	&	1.3028	&	6534.337	&	-38.113	&	-4.8075	&	700	\\
9910.75047	&	-2966.0686	&	0.7246	&	6503.361	&	-25.248	&	-4.6902	&	700	\\
9917.72905	&	-2978.0604	&	0.6064	&	6481.170	&	-33.996	&	-4.6673	&	700	\\
\hline
\end{tabular}
\begin{tablenotes}
\item
\textbf{Notes:}
\item[a] $^{(1)}$ These data did not pass the ESPRESSO flux quality control and therefore were excluded from the analysis.
\item[a] $^{(2)}$ The uncertainties of FWHM and BIS are twice the uncertainties of the RVs (2$\sigma_{\mathrm{RV}}$).
\item[a] $^{(3)}$ The median uncertainty of logR$'_{\mathrm{HK}}$$^{(3)}$ is 0.03.
\end{tablenotes}
\label{tab:rvs}
\end{center}
\end{table*}

\newpage
\section{Spectral synthesis}
\begin{figure}[hb]
  \centering
    \includegraphics[width=0.8\textwidth]{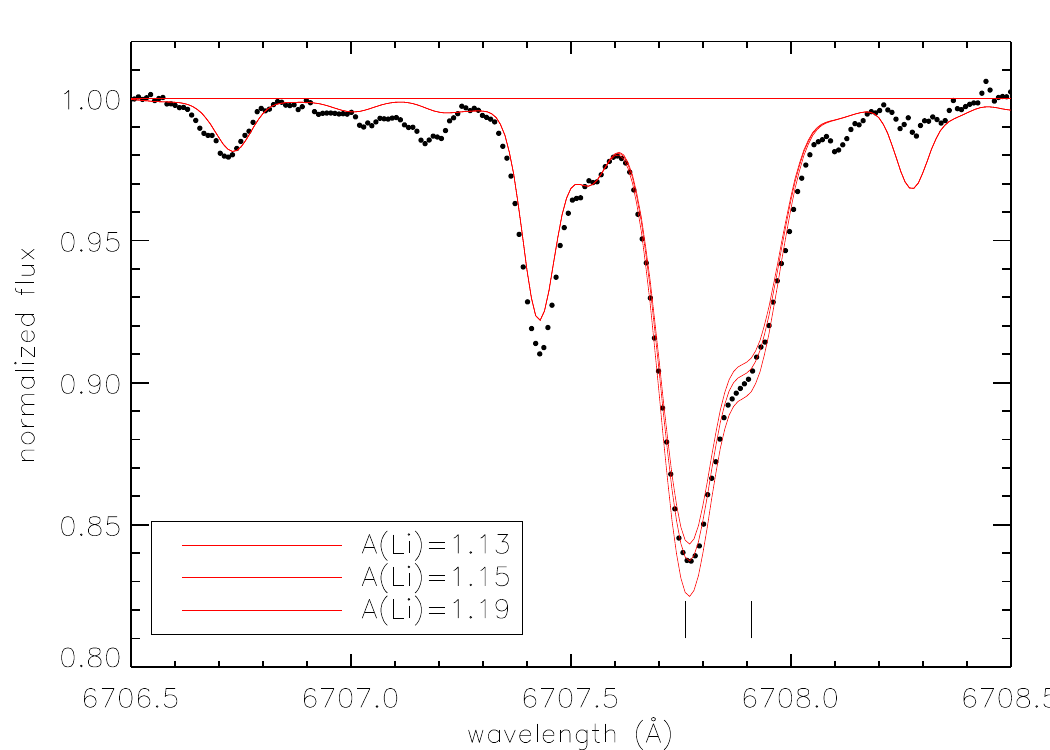}
  \caption{Spectral synthesis around the Li doublet. The vertical lines mark the position of the lines.} 
  \label{fig:spectralsynthesis} 
\end{figure}
\newpage
\section{Generalized Lomb-Scargle periodograms}
\begin{figure}[hb]
  \centering
    \includegraphics[width=0.8\textwidth]{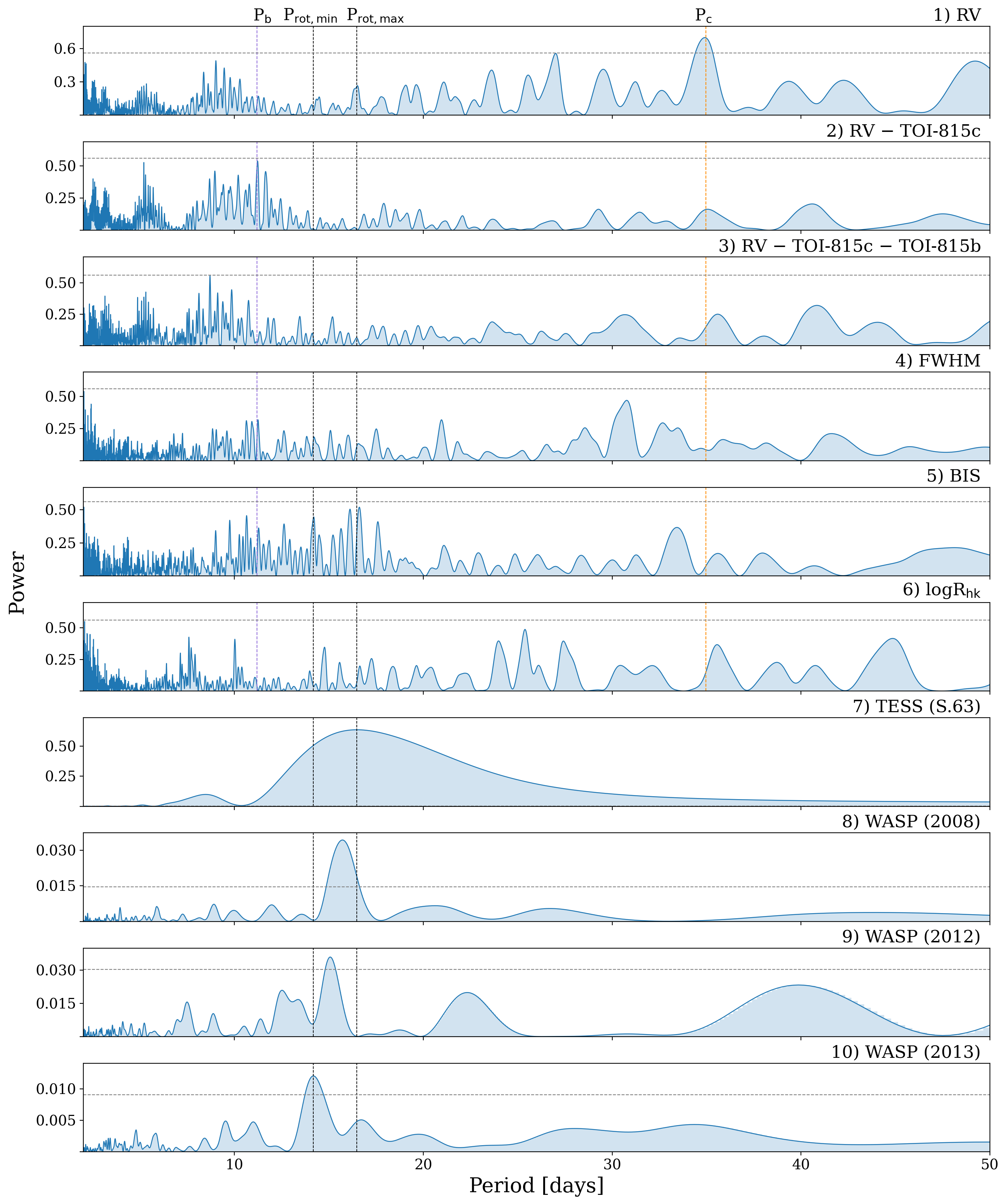}
  \caption{GLS periodogram of the TOI-815 spectroscopic and photometric time series. More specifically: (1) RV ESPRESSO measurements, (2) RV residuals, after the subtraction of TOI-815c signal, (2) RV residuals, after the subtraction of TOI-815b and c signals, (3)-(5) FWHM, bisector, and Ca\hspace{0.04cm}II\hspace{0.04cm}H\hspace{0.04cm}\&\hspace{0.04cm}K line activity indicator (logR$'_{\mathrm{HK}}$), (7) TESS Sector 63 photometric data, and (8)-(10) WASP photometric data (2008, 2012, and 2023). The two black vertical lines correspond to the minimum and maximum rotation periods (P$_\mathrm{rot,min}$ $\sim$ 14.2 days and P$_\mathrm{rot,max}$ $\sim$ 16.5 days). The purple and orange vertical lines correspond to the orbital periods of planets b (P$_\mathrm{b}$ $\sim$ 11.2 days) and c (P$_\mathrm{c}$ $\sim$ 35 days), respectively. The gray horizontal lines indicating the 1$\%$ FAP.} 
  \label{fig:gls} 
\end{figure}

\newpage
\section{Priors}

\begin{table*}[h]
\tiny
\caption{Maximum values and 68$\%$ confidence intervals of the posterior distributions from joint modeling with \juliet.}
\centering
\renewcommand{\arraystretch}{1.1}
\setlength{\tabcolsep}{38pt}
\begin{center}
\begin{tabular}{lcc}
\hline\hline
\textbf{Parameter} & \textbf{Prior$^{(a)}$}& \textbf{Value$^{(b)}$} \\
 \hline
    \textbf{Jump stellar parameters} &   &\\
    Stellar density, $\rho_*$ (\denssol)\dotfill  &  $\mathcal{N}(2396, 139)$ &  $ 2431^{+122} _{-125}$ \\  
    \textbf{Jump parameters for TOI-815b} &  \\
    Orbital period, $\it{P_{orb}}$ (days)\dotfill  & $\mathcal{U}(10, 12)$ & $11.197259 ^{+0.000018} _{-0.000017}$\\
    Transit epoch, $T_0$ (BJD)\dotfill & $\mathcal{N}(2459287.6, 0.1)$&$2459287.6028 ^{+0.0009} _{-0.0009}$\\
    Scaled planetary radius, $R_{P}$/\rstar\dotfill &$\mathcal{U}(0, 1)$&$0.0349 ^{+0.0004} _{-0.0004}$  \\
    Impact parameter, $\textit{b}$\dotfill  &$\mathcal{U}(0, 1)$&$0.279 ^{+0.058} _{-0.073}$ \\
    Eccentricity, $e$\dotfill  & $\mathcal{F}(0)$& 0 (adopted, 3$\sigma$ $<$ 0.20)\\
    Argument of periastron, $\omega_*$ (deg)\dotfill  & $\mathcal{F}(90)$& -\\
    RV semi-amplitude, $\textit{K}$ (m\,s$^{-1}$)\dotfill  & $\mathcal{U}(0, 100)$&$2.6 ^{+0.5} _{-0.5}$     \\
    \textbf{Jump parameters for TOI-815c} &  \\
    Orbital period, $\it{P_{orb}}$ (days)\dotfill  & $\mathcal{U}(34, 36)$ &$34.976145 ^{+0.000099} _{-0.000097}$\\
    Transit epoch, $T_0$ (BJD)\dotfill & $\mathcal{N}(2459288.6, 0.1)$ & $2459288.6331 ^{+0.0023} _{-0.0024}$\\
    Scaled planetary radius, $R_{P}$/\rstar\dotfill &$\mathcal{U}(0, 1)$&$0.0312 ^{+0.0011} _{-0.0010}$ \\
    Impact parameter, $\textit{b}$\dotfill  &$\mathcal{U}(0, 1)$&$0.839 ^{+0.015} _{-0.016}$ \\
    Eccentricity, $e$\dotfill  & $\mathcal{F}(0)$& 0 (adopted, 3$\sigma$ $<$ 0.22)\\
    Argument of periastron, $\omega_*$ (deg)\dotfill  & $\mathcal{F}(90)$& -\\
    RV semi-amplitude, $\textit{K}$ (m\,s$^{-1}$)\dotfill  & $\mathcal{U}(0, 100)$&$5.5 ^{+0.5} _{-0.5}$\\
    \textbf{Instrumental photometric parameters} &  \\
    Offset relative flux, $M_{\mathrm{TESS}_{S9}}$ (10$^{-6}$) \dotfill &$\mathcal{N}(0,0.1)$ & $-8.6 ^{+8.0} _{-8.2}$\\
    Jitter, $\sigma_{w,\mathrm{TESS}_{S9}}$ (ppm)\dotfill & log$\mathcal{U}(0.01,1000)$ & $0.5 ^{+6.5} _{-0.5}$ \\  
    Offset relative flux, $M_{\mathrm{TESS}_{S36}}$ (10$^{-6}$)\dotfill &$\mathcal{N}(0,0.1)$ & $-72 ^{+411} _{-417}$ \\
    Jitter, $\sigma_{w,\mathrm{TESS}_{S36}}$ (ppm)\dotfill & log$\mathcal{U}(0.01,1000)$ & $466 ^{+8} _{-8}$ \\  
    Offset relative flux, $M_{\mathrm{TESS}_{S63}}$ (10$^{-6}$)\dotfill &$\mathcal{N}(0,0.1)$ & $-3594 ^{+3821} _{-6137}$\\
    Jitter, $\sigma_{w,\mathrm{TESS}_{S63}}$ (ppm)\dotfill & log$\mathcal{U}(0.01,1000)$ & $457 ^{+11} _{-11}$\\  
    Offset relative flux, $M_{\mathrm{CHEOPS}_{1}}$ (10$^{-6}$)\dotfill &$\mathcal{N}(0,0.1)$ & $-197^{+46} _{-44}$\\
    Jitter, $\sigma_{w,\mathrm{CHEOPS}_{1}}$ (ppm)\dotfill & log$\mathcal{U}(0.01,1000)$ & $299^{+23} _{-22}$\\  
    Offset relative flux, $M_{\mathrm{CHEOPS}_{2}}$ (10$^{-6}$)\dotfill &$\mathcal{N}(0,0.1)$ & $-474^{+48} _{-48}$\\
    Jitter, $\sigma_{w,\mathrm{CHEOPS}_{2}}$ (ppm)\dotfill & log$\mathcal{U}(0.01,1000)$ & $404^{+24} _{-23}$\\  
    Offset relative flux, $M_{\mathrm{CHEOPS}_{3}}$ (10$^{-6}$)\dotfill &$\mathcal{N}(0,0.1)$ & $-322^{+34} _{-35}$\\
    Jitter, $\sigma_{w,\mathrm{CHEOPS}_{3}}$ (ppm)\dotfill & log$\mathcal{U}(0.01,1000)$ & $293^{+19} _{-19}$\\  
    Offset relative flux, $M_{\mathrm{LCO}}$ (10$^{-6}$)\dotfill &$\mathcal{N}(0,0.1)$ & 1089$^{+88} _{-88}$\\
    Jitter, $\sigma_{w,\mathrm{LCO}}$ (ppm)\dotfill & log$\mathcal{U}(0.01,5000)$ & 986$^{+72} _{-68}$\\  
    Offset relative flux, $M_{\mathrm{ASTEP}}$ (10$^{-6}$)\dotfill &$\mathcal{N}(0,0.1)$ & 761$^{+133} _{-131}$\\
    Jitter, $\sigma_{w,\mathrm{ASTEP}}$ (ppm)\dotfill & log$\mathcal{U}(0.01,5000)$ & 3369$^{+89} _{-85}$\\   
    \textbf{GP parameters} &  \\
    $\sigma_{GP,\mathrm{TESS}_{S9}}$ (relative flux)\dotfill & $\mathcal{F}(0)$& -\\  
    $\rho_{GP,\mathrm{TESS}_{S9}}$ (days)\dotfill & $\mathcal{F}(0)$& -\\  
    $\sigma_{GP,\mathrm{TESS}_{S36}}$ (10$^{-4}$ relative flux)\dotfill &   log$\mathcal{U}$(10$^{-6}$, 100) & $12 ^{+2} _{-2}$ \\  
    $\rho_{GP,\mathrm{TESS}_{S36}}$ (days)\dotfill & log$\mathcal{U}(0.001, 100)$ & $0.94 ^{+0.13} _{-0.11}$ \\  
    $\sigma_{GP,\mathrm{TESS}_{S63}}$ (10$^{-4}$ relative flux)\dotfill &   log$\mathcal{U}$(10$^{-6}$, 100) & $57 ^{+44} _{22}$\\  
    $\rho_{GP,\mathrm{TESS}_{S63}}$ (days)\dotfill & log$\mathcal{U}(0.001, 100)$ & $9.11 ^{+4.37} _{-2.77}$\\  
    \textbf{Instrumental RV parameters} &  & \\  
    Jitter, $\sigma_{w,\mathrm{ESPRESSO}}$ ($\ms$)\dotfill &log$\mathcal{U}(0.001,100)$ & $1.88 ^{+0.29} _{-0.25}$  \\
    Systemic RV, $\mu_{\mathrm{ESPRESSO}}$ ($\ms$)\dotfill &$\mathcal{U}(-3500,-1000)$ & $-2973.61 ^{+0.36} _{-0.36}$\\    
    \textbf{Limb darkening parameters} &  \\
    $q_{1,\mathrm{ESPRESSO}}$\dotfill  & $\mathcal{N}(0.408,0.017)$ & 0.407$^{+0.015} _{-0.015}$\\
    $q_{2,\mathrm{ESPRESSO}}$\dotfill  & $\mathcal{N}(0.349,0.022)$ & 0.345$^{+0.020} _{-0.019}$\\
    $q_{1,\mathrm{CHEOPS}}$\dotfill  & $\mathcal{N}(0.519,0.012)$ & 0.518$^{+0.011} _{-0.011}$\\
    $q_{2,\mathrm{CHEOPS}}$\dotfill  & $\mathcal{N}(0.408,0.012)$ & 0.406$^{+0.011} _{-0.011}$\\
     $q_{1,\mathrm{LCO}}$\dotfill  & $\mathcal{N}(0.324,0.008)$ & 0.324$^{+0.007} _{-0.007}$\\
    $q_{2,\mathrm{LCO}}$\dotfill  & $\mathcal{N}(0.314,0.014)$ & 0.314$^{+0.012} _{-0.012}$\\
    $q_{1,\mathrm{ASTEP}}$\dotfill  & $\mathcal{N}(0.492,0.016)$ & 0.492$^{+0.014} _{-0.014}$\\
    $q_{2,\mathrm{ASTEP}}$\dotfill  & $\mathcal{N}(0.376,0.017)$ & 0.378$^{+0.015} _{-0.016}$\\
    \hline
\end{tabular}
\begin{tablenotes}
\item
\textbf{Notes:} $^{(a)}$For the priors, $\mathcal{N}$($\mu$, $\sigma^{2}$) indicates a normal distribution with mean $\mu$ and variance $\sigma^{2}$, $\mathcal{U}$(\textit{a}, \textit{b}) a uniform distribution between \textit{a} and \textit{b}, log$\mathcal{U}$(\textit{a}, \textit{b}) a log-uniform distribution between \textit{a} and \textit{b} and $\mathcal{F}$(\textit{a}) a parameter fixed to value \textit{a}.
$^{(b)}$The posterior estimate indicates the median value and then error bars for the 68 $\%$ credibility intervals.
\end{tablenotes}
\label{tab:posteriors}
\end{center}
\end{table*}

\newpage
\section{CHEOPS alias hunting}

\begin{table*}[h]
\tiny
\caption{CHEOPS photometric observations of TOI-815. }
\centering
\renewcommand{\arraystretch}{1.1}
\setlength{\tabcolsep}{3pt}
\begin{center}
\begin{tabular}{lccccccccccc}
\hline\hline
 Visit & Expected planet & Start time & Duration  & Eff. & RMS &Alias &Detection &File key & Detrending$^{(d)}$\\
  & & [UTC] &[h] & [$\%$] &ppm & [d] & & \\ 
 \hline
1 &c&2022-03-06 00:36& 8.34& 73&422 &22.26&no &CH\_PR110048\_TG018001\_V0300 & ${(e)}$\\
2 &c&2022-03-06 16:26& 8.37&60 & 461&20.99& no&CH\_PR110048\_TG018901\_V0300 & ${(e)}$\\
3 &c&2022-03-17 04:35 & 9.84&61 &417 & 22.95&no &CH\_PR110048\_TG018101\_V0300 & ${(e)}$\\
4$^{(a)}$ &b&2022-03-18 08:36& 11.64& 66& 514& 11.20&yes&CH\_PR100031\_TG050801\_V0300 & t, x, y, y$^{2}$, sin($\phi$), cos($\phi$), cos(2$\phi$)\\
5 &c&2022-04-13 01:20 & 8.47& 60&429 & 17.92&no &CH\_PR110048\_TG025201\_V0300 &${(e)}$\\
6 &c&2022-04-30 06:47& 8.34&51 &497 &29.38&no &CH\_PR110048\_TG027601\_V0300 & t (clear flux trend in data)\\
7 &c&2023-02-15 11:09& 8.29&69 &520 &31.93&no&CH\_PR110048\_TG036501\_V0300 & ${(e)}$\\
8 &c&2023-02-22 01:40& 9.04&68 &447 &25.33&no&CH\_PR110048\_TG036701\_V0300 &${(e)}$\\
9 &c&2023-02-27 14:48& 8.22&64& 580&19.85&no&CH\_PR110048\_TG036901\_V0300 & ${(e)}$\\
10$^{(b)}$ &b&2023-02-28 14:16& 8.39& 62&460 &18.83&yes&CH\_PR110048\_TG037001\_V0300 & t$^{2}$, x, y, y$^{2}$, bg, sin($\phi$), sin(2$\phi$), cos(2$\phi$)\\
11$^{(c)}$ &c&2023-04-23 08:31& 10.09&58 &462 &34.98&yes&CH\_PR110048\_TG042201\_V0300 & t, t$^{2}$, y, y$^{2}$, bg, cos($\phi$)\\

    \hline
\end{tabular}
\begin{tablenotes}
\item
\textbf{Notes:}
\item[a] $^{(a)}$ Figure \ref{fig:CHEOPS}, left.
\item[b] $^{(b)}$ Figure \ref{fig:CHEOPS}, center.
\item[c] $^{(c)}$ Figure \ref{fig:CHEOPS}, right.
\item[d] $^{(d)}$ The correlation terms were determined using \pycheops.
\item[e] $^{(e)}$ Detrended with: x, y, bg, cos($\phi$), cos(2$\phi$), cos(3$\phi$), $\Delta$T, sin($\phi$), sin(2$\phi$), sin(3$\phi$), bg$^{2}$, x$^{2}$, y$^{2}$
\end{tablenotes}
\label{tab:cheopsobservations}
\end{center}
\end{table*}

\begin{figure}[hb]
  \centering
    \includegraphics[width=1\textwidth]{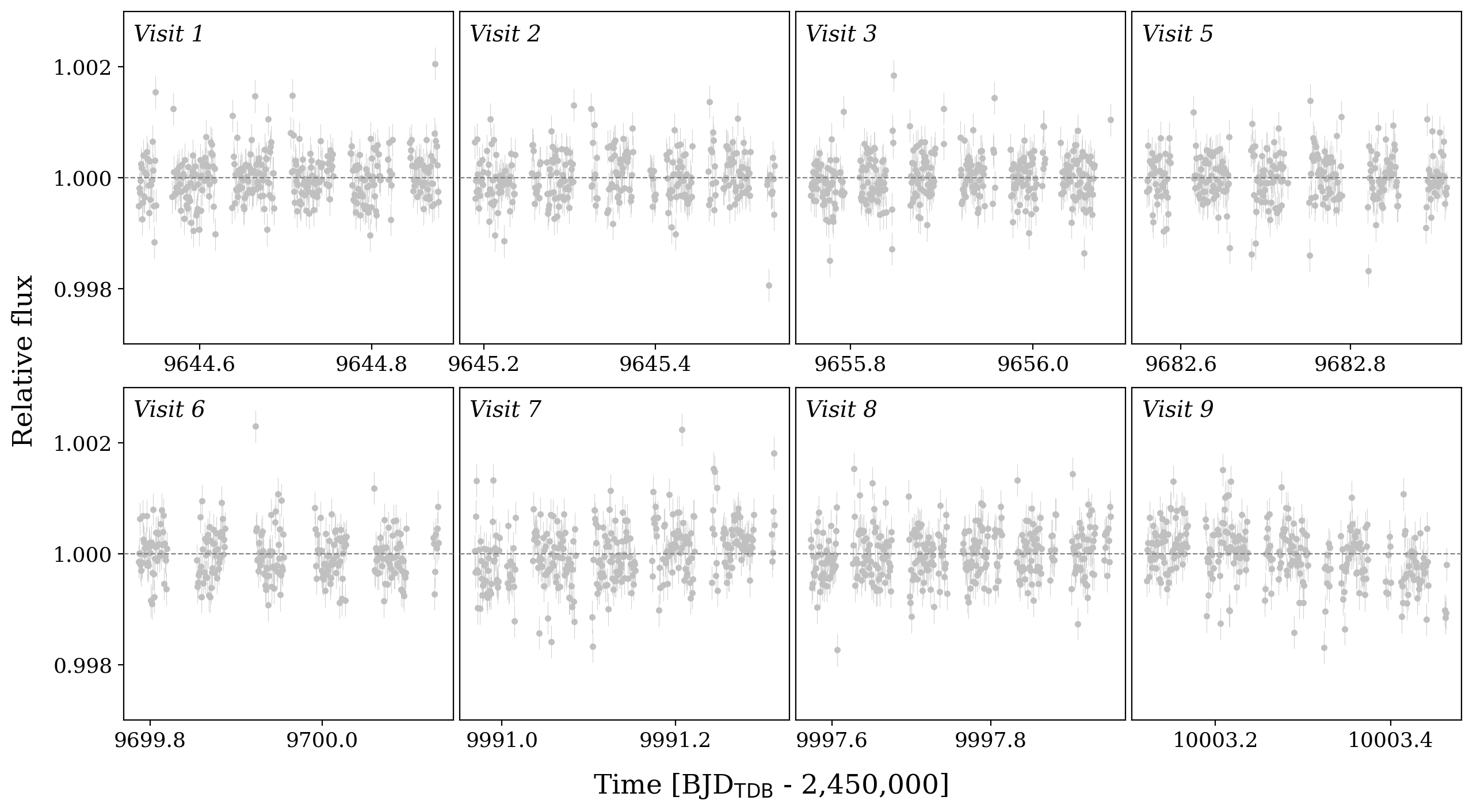}
  \caption{CHEOPS unsuccessful transit observations that ruled out the most likely period aliases of TOI-815c. The data have been detrended using the detrending terms presented in Table \ref{tab:cheopsobservations}.} 
  \label{fig:ALLCHEOPS} 
\end{figure}
\end{appendix}
\end{document}